\documentclass{jfm}
\usepackage{amssymb,psfrag}
\usepackage[applemac]{inputenc}
\usepackage{pgfplots}

\usepackage{amsmath,url}
\usepackage[titletoc]{appendix}
\usepackage{amssymb}
\usepackage{lscape,amstext,mathrsfs}
\usepackage{graphicx}
\usepackage{epstopdf}
\usepackage{tikz,pmat}
\usepackage{afterpage}
\usepackage{rotating}
\usepackage[graphicx]{realboxes}
\usepackage{pgfgantt}
\usepackage[europeanresistors,cuteinductors,emptydiodes]{circuitikz}
\usepackage{graphics,array} 
\usepackage{epsfig} 
\usepackage{epstopdf}
\usepackage{epstopdf}
\usepackage{mhchem,bm}
\usepackage{caption}
\usepackage{subcaption}
\usepackage{hyperref}
\usetikzlibrary{shapes,arrows}
\usetikzlibrary{arrows, decorations.markings}

\newtheorem{thm}{Theorem}[section]
\newtheorem{lemma}[thm]{Lemma}
\newtheorem{prop}[thm]{Proposition}
\newtheorem{corollary}[thm]{Corollary}

\newtheorem{defi}[thm]{Definition}

\newenvironment{pf}{\paragraph{Proof:}}{\hfill$\square$}

\shorttitle{A Framework for Input-Output Analysis of Wall-Bounded Shear  Flows}
\shortauthor{M. Ahmadi, G. Valmorbida, D. Gayme, and  A. Papachristodoulou}

\title{A Framework for Input-Output Analysis of Wall-Bounded Shear  Flows}

\author{Mohamadreza Ahmadi\aff{1}
  \corresp{\email{mrahmadi@utexas.edu}},
  Giorgio Valmorbida\aff{2},\\
    Dennice Gayme\aff{3}, and
  Antonis Papachristodoulou\aff{4}
 }

\affiliation{\aff{1}Institute for Computational Engineering and Sciences (ICES), University of Texas at Austin, 201 E 24th St, Austin, TX 78712, USA.
\aff{2} L2S, CentraleSupelec, University of Paris, Plateau de Moulon, 3 rue Joliot-Curie, F-91192 Gif-sur-Yvette Cedex, France.
\aff{3} Department of Mechanical Engineering, Johns Hopkins University, Lathrobe Hall 223, 3400 North Charles St, Baltimore, MD 21218, USA.
\aff{4}Department of Engineering Science, University of Oxford,  Parks Rd, Oxford OX1 3PJ, UK.

}

\begin{document}

\maketitle

\begin{abstract}

We propose a framework  to understand input-output amplification properties of nonlinear partial differential equation (PDE) models of wall-bounded shear flows, which are spatially invariant in one  coordinate (e.g., streamwise-constant plane Couette flow). Our methodology is based on the notion of dissipation inequalities in control theory.  In particular, we consider flows with body and other forcings, for which we study the input-to-output properties, including energy growth, worst-case disturbance amplification, and stability to persistent  disturbances. The proposed method can be applied  to a large class of  flow configurations as long as the base flow is described by a polynomial. This includes many examples in  both channel flows and pipe flows, e.g., plane Couette flow, and Hagen-Poiseuille flow. The methodology we use is numerically implemented as the solution of a (convex) optimization problem. We use the framework to study input-output amplification mechanisms in 
 rotating Couette flow, plane Couette flow, plane Poiseuille flow, and Hagen-Poiseuille flow. In addition to showing that the application of the proposed framework leads to results that are consistent with theoretical and experimental amplification scalings obtained in the literature through linearization around the base flow, we demonstrate that the stability bounds to persistent forcings can be used as a means to predict  transition to turbulence in wall-bounded shear flows. \end{abstract}


\section{Introduction}

\subsection{Literature Review}
The dynamics of incompressible fluid flows are described by a set of nonlinear PDEs known as the Navier-Stokes equations. The properties of such flows are then characterized in terms of a dimensionless parameter $Re$, the Reynolds number. Experiments show that many wall-bounded shear flows have a critical Reynolds number $Re_C$ {  below which the flow is stable
with respect to disturbances of any amplitude.} However, spectrum analysis of the linearized Navier-Stokes equations, considering only infinitesimal perturbations, 
predicts a linear stability limit $Re_L$ which upper-bounds $Re_C$~(\cite{DR81}). On the other hand, the bounds using energy methods $Re_E$, the limiting value for which the energy of arbitrary large perturbations decreases monotonically, are much below $Re_C$~(\cite{J76}). For Couette flow, for instance, $Re_E =32.6$ was computed by~\cite{Se59} using the energy functional, $Re_L = \infty$ using spectrum analysis~(\cite{Romanov73}), and $Re_C \approx 350$ was estimated empirically by~\cite{TA92}. \\

Conventional hydrodynamic stability methods usually  involve linearization of the Navier-Stokes equations around a base flow followed by spectrum analysis, revealing the Reynolds number estimate for when this solution becomes unstable.  The discrepancy between $Re_L$ and $Re_C$ has long been attributed to the eigenvalues analysis approach of the linearized Navier-Stokes operator~(\cite{Trefethen30071993}). Other theoretical methods for studying stability of flows are often based on spectral truncation of the  Navier-Stokes equations into an ODE system. This method is fettered by  truncation errors and by the mismatch between the dynamics of the truncated model and the Navier-Stokes PDE. To alleviate this drawback,  recently in~(\cite{GC12,CGHP14}) a method was proposed based on keeping  a number of modes from the Galerkin expansion of the nonlinear Navier-Stokes equations and bounding the energy of the remaining modes. It was shown in~(\cite{Huang20150622}) that, in the case of rotating Couette flow, this method can  find  a global stability limit, which is better than the energy method but not as good as the linear stability limit\footnote{Recall that the linear stability  and the global stability limits coincide  for the Taylor-Couette flow (\cite{Taylor289}).}.\\

 In fact, even in the seminal paper by~\cite{Rey83}, it was observed that external excitations and body forces play an important role in flow instabilities.  Mechanisms such as energy amplification of external excitations and body forcings  have shown to be crucial in understanding transition to turbulence as highlighted by~\cite{J76}. Therefore, instead of studying stability, researchers began to focus on growth and  were able to uncover additional flow properties through the new paradigm of input-output analysis. A phenomenon called \textit{transient growth} is known as the culprit for flow instability; \textit{i.e.}, although the perturbations to the linearized Navier-Stokes equation are stable (and the eigenvalues have negative real parts), they undergo high amplitude transient amplifications that steer the trajectories out of the region of linearization. The root cause of the transient growth phenomenon is the non-normality of the stable Navier-Stokes operator
that has been linearized about a base flow. This phenomenon has led to studying the resolvent operator or $\varepsilon$-pseudospectra to uncover when transition occurs, based on the general solution to the linearized Navier-Stokes equations (\cite{Sch07}). In particular, (\cite{mckeon_sharma_2010}) used resolvent analysis to study the amplification scalings from an input composed of nonlinear terms and periodic forcings for turbulent pipe flows.  \\

The input-output properties can be characterized based on the class of forcings (noise vs square integrable signals)  and the flow model (linear vs nonlinear or finite-dimensional vs infinite dimensional) one considers.  For stochastic forcings (Gaussian noise), energy amplification   to the linearized Navier-Stokes equations in wall-bounded shear flows was  studied by~\cite{FI93}. In a similar vein,~(\cite{BD01}), using the stochastically forced linearized Navier-Stokes equation, showed analytically through the calculation of traces of operator Lyapunov equations, that  the input-output $\mathcal{H}^2$-norm from streamwise constant excitations to perturbation velocities   in channel flows is proportional to $Re^3$. The  amplification scaling of the linearized Navier-Stokes equation was further characterized in~(\cite{JB05}) and (\cite{mjphd04}), where the authors studied the influence of each component of the body forces  in terms of the input-output $\mathcal{H}^2$-norm. For square integrable forcings, \cite[Chapter 9]{mjphd04} and~(\cite{JB05})  provided worst-case amplification mechanisms for incompressible viscous channel flows based on the linearized Navier-Stokes equations. \\

 \subsection{Contribution}
 
 Our work extends the rich input-output analysis paradigm. We propose a method based on dissipation inequalities (\cite{W72}) to study input-output amplification in wall-bounded shear flows (described by the \emph{nonlinear} Navier-Stokes PDE, rather than finite-dimensional ODE approximations or linearizations) that are invariant in one of the spatial directions. 
   Here, a dissipation inequality  establishes  a relation between the rate of change of the
weighted kinetic energy of the flow perturbations (characterized by a storage functional), the energy supplied  from the body forces, and the energy dissipated via viscosity  (characterized by a supply rate). This approach exploits our previous work (\cite{AVPaut}) wherein dissipation inequalities for nonlinear PDEs were formulated. \\
 
  Based on these dissipation inequalities, we study three flow properties. We start by studying energy growth from initial perturbations, which is tantamount to the notion of transient growth (\cite{Trefethen30071993}). Note that the definition of transient growth requires a linear approximation of the dynamics; whereas, the concept of energy growth used in this study is applied directly to nonlinear dynamics. Additionally, we consider body forcings and external excitations that are square integrable and we study worst-case amplification  mechanisms. In addition to square integrable forcings, we provide a mathematical framework to consider a new class of forcings, in particular those that are constrained only in terms of either their  maximum or their absolute value  for all time. This is the first time that input-output response of wall bounded shear flows under  persistent forcings is being investigated. \\
  
  
Furthermore, for flows with streamwise constant perturbations described by the nonlinear Navier-Stokes equations, we find a weighted kinetic energy form as the storage functional that converts the dissipation inequalities into integral inequalities with quadratic integrands in perturbation velocities and their spatial derivatives. Then, using these functionals, we propose conditions based on matrix inequalities that can be checked via convex optimization using available MATLAB software. One strength of the method is that the results can be directly extended to more complex flow geometries as long as they can be described by semi-algebraic sets.  A precise characterization of this condition is provided in Section~\ref{sec:convex}. \\


Our proposed methodology allows us to  study multiple input-output aspects, such as energy growth, worst-case disturbance amplification, and stability to persistent  disturbances of a broad class of shear flows within a single framework. We evaluate the performance of the proposed method by several examples from both channel and pipe flows, namely rotating Couette flow, plane Couette flow, plane Poiseuille flow, and Hagen-Poiseuille flow. We demonstrate that our results tally with the transient growth results in the literature.  For channel flows, we show the results obtained using our method are consistent with the results in~(\cite{JB05}) and~\cite[Chapter 9]{mjphd04} in terms of worst-case disturbance amplification and we show that our framework can be used to study pipe flows, as well. Moreover, we observe an intriguing correspondence between the stability bounds  to persistent forcings  and the experimental Reynolds numbers for transition to turbulence, which provides a theoretical tool to predict transition.\\

Preliminary mathematical results on this work were presented in~(\cite{7403365}). The current paper is different from~(\cite{7403365}) in several aspects. From a theoretical standpoint, the current paper provides a method for energy growth analysis and extends the formulation to both flows between parallel plates and flows in pipes. In addition, it presents the mathematical proofs of the input-output analysis framework and the formulation based on convex optimization. From the examples standpoint, in addition to an extended study of the rotating Couette flow, we applied the framework to investigate the input-output properties of plane Couette flow, plane Poiseuille, and the  Hagen-Poiseuille flow. Furthermore, the current version includes a comparison with previous results in the literature and  an examination of flow structures corresponding to maximum input-output amplifications. 

\subsection{Organization}

In the next section, we briefly describe the flow model studied in the paper. In Section~\ref{sec:dissiInq}, we propose the flow input-output analysis framework  based on dissipation inequalities. 
In Section~\ref{sec:convex}, we show how the input-output analysis can be computationally implemented as the solution to a convex optimization problem. In Section~\ref{sec:NR}, we demonstrate the effectiveness of the proposed framework by applying it to study input-output properties of rotating Couette flow, plane Couette flow, plane Poiseuille flow, and Hagen-Poiseuille flow. Finally, in Section~\ref{sec:conclusions}, we present some concluding remarks and provide directions for future research.

\section{The Flow Perturbation Model}

 Let $I$ be an index set corresponding to the spatial coordinates.
  The  dynamics of forced incompressible shear flows are described by the Navier-Stokes equations, given~by
\begin{eqnarray} \label{eq:NS}
\partial_t \boldsymbol{\bar{u}} &=& \frac{1}{Re} \nabla^2  \boldsymbol{\bar{u}}-  \boldsymbol{\bar{u}}\cdot \nabla  \boldsymbol{\bar{u}}- \nabla {\bar{p}} + F  \boldsymbol{\bar{u}}+\boldsymbol{d}, \nonumber \\
0 &=& \nabla \cdot \bar{\boldsymbol{u}},
\end{eqnarray}

\noindent where $t>0$, $F \in \mathbb{R}^{3\times3}$ represents terms coming from rotation,~{$\mathrm{x} \in \Omega = \Omega_{i} \times \Omega_j \subset \mathbb{R} \times \mathbb{R}$} with $i \neq j$, $i,j \in I$ are  spatial coordinates and $\partial_s(\cdot) = \frac{\partial (\cdot)}{\partial s}$. 
The dependent variable $\boldsymbol{d}: \mathbb{R}_{\ge 0} \times \Omega \to \mathbb{R}^3$ is the input vector representing exogenous excitations or body forces, $\boldsymbol{\bar{u}} : \mathbb{R}_{\ge 0} \times \Omega \to \mathbb{R}^3$ is the velocity vector, and ${\bar{p}}:\mathbb{R}_{\ge 0} \times \Omega \to \mathbb{R}$ is the pressure. $\nabla^2$ is the Laplacian operator, 
$\nabla$ denotes the gradient, and $\nabla \cdot \boldsymbol{u}$ denotes the divergence of $\boldsymbol{u}$.

We consider perturbations $(\boldsymbol{u},{p})$ to the { steady solution} $(\boldsymbol{U},{P})$, which are spatially invariant  in one of the directions, say $x_m$, $m \in I$, \textit{i.e.,}~\mbox{$\partial_{x_m} =0$}. 
Let $I_0= I-\{m\}$. The velocity field can be decomposed as
\begin{equation} \label{eq:perturbsub}
\boldsymbol{\bar{u}} = \boldsymbol{u} + \boldsymbol{U},~{\bar{p}} = {p} + {P},
\end{equation}
where $(\boldsymbol{U},P)$ are divergence free steady state solutions, \textit{i.e.},
\begin{eqnarray} \label{eq:eU}
0 &=& \frac{1}{Re} \nabla^2 \boldsymbol{U} -\boldsymbol{U} \cdot \nabla \boldsymbol{U} - \nabla P + F\boldsymbol{U}.
\end{eqnarray}
Substituting~\eqref{eq:perturbsub} in~\eqref{eq:NS} and using~\eqref{eq:eU}, we  obtain the perturbation  dynamics
\begin{eqnarray} \label{eq:mainNS}
\partial_t \boldsymbol{u} &=& \frac{1}{Re} \nabla^2 \boldsymbol{u} - \boldsymbol{u} \cdot \nabla \boldsymbol{u} - \boldsymbol{U} \cdot \nabla \boldsymbol{u} - \boldsymbol{u} \cdot \nabla \boldsymbol{U}  -\nabla p + F \boldsymbol{u} + \boldsymbol{d}, \nonumber \\
0 &=& \nabla \cdot \boldsymbol{u}.
\end{eqnarray}
%
In the rest of this paper, we study the properties of~\eqref{eq:mainNS}. We concentrate on perturbations with no-slip boundary conditions $\boldsymbol{u}|_{\partial \Omega} \equiv 0$ (in the direction with solid boundaries) and periodic boundary conditions (in the spatially homogeneous direction). In a similar manner, we extend the results to pipe flows (cylindrical coordinates) as discussed in Appendix~\ref{app:cylindr}. Next, we introduce the input-output analysis method based on dissipativity theory. 

%
%

\section{Dissipation Theory and Dissipation Inequalities}\label{sec:dissiInq}

In systems and control theory, dissipativity  (\cite{W72,Willems2007134,hill1980dissipative})\footnote{Note that the notion of dissipativity used here should not be confused with dissipative operators in semigroup theory (\cite{lumer1961}). The latter is concerned with proving the existence of a contraction semigroups and used to prove well-posedness of solutions to PDEs, \cite{CZ95}; whereas, the dissipativity notion we use here is concerned with the input-output properties of a dynamical system.} establishes a relationship between the energy stored in the system represented by  a continuous, non-negative functional  $V(u)$, known as the storage functional, and the power supplied to the system $W(u,d,y)$, known as the supply rate, with $d$ and $y$ being the inputs and outputs of the system, respectively. This relationship is often given by a dissipation inequality (in differential form) as
\begin{equation} \label{eq:DisInq}
\frac{d V(u)}{dt} \le W(u,d,y).
\end{equation}
A system is called dissipative with respect to the supply rate $W(u,d,y)$, if there is a non-negative functional $V(u)$ that satisfies~\eqref{eq:DisInq}. Dissipativity theory has a close connection with Lyapunov stability theory~(\cite{khalil1996noninear}). In particular, dissipativity theory can be understood as a generalization of the Lyapunov stability theory  to systems with inputs and outputs.

%
%

Given the dissipation inequality~\eqref{eq:DisInq} with a fixed supply rate, the main challenge is to find a corresponding storage functional that satisfies the dissipation inequality along the solutions of the flow. In fact, kinetic energy was shown to be a candidate storage functional for some input-output properties. In the special case of an irrotational flow ($F=0$ in~\eqref{eq:mainNS})   under no-slip, stress-free or periodic boundary condition,  if we set $V$ to be the kinetic energy of the perturbations $V({\boldsymbol{u}}) = \int_\Omega | {\boldsymbol{u}} |^2~{d}\Omega$, we can show~\cite[p. 31]{DG95}  that the total kinetic energy of the perturbations satisfies the following equality
$$
\frac{d V(\boldsymbol{u})}{dt} = -\frac{1}{Re} \| \nabla \boldsymbol{u} \|_{\mathcal{L}^2_\Omega}^2 - \int_\Omega \boldsymbol{u} \cdot \nabla \boldsymbol{U} \cdot \boldsymbol{u}~d\Omega +  \int_\Omega {\boldsymbol{u}} \cdot \boldsymbol{d}~d\Omega.
$$
The above  equality implies that the kinetic energy of the perturbations in the flow  changes according to three effects:
the energy  dissipated by viscosity, the energy either injected or dissipated depending on the base flow,  and the energy  expended by the external force. 
Since the viscosity term $\frac{1}{Re} \| \nabla \boldsymbol{u} \|_{\mathcal{L}^2_\Omega}^2$ is always non-negative, we can  obtain the following  inequality 
$$
\frac{d V(\boldsymbol{u})}{dt} \le  - \int_\Omega \boldsymbol{u} \cdot \nabla \boldsymbol{U} \cdot \boldsymbol{u}~d\Omega +  \int_\Omega {\boldsymbol{u}} \cdot \boldsymbol{d}~d\Omega.
$$
If the base flow $\boldsymbol{U}$ is such that the term $\int_\Omega \boldsymbol{u} \cdot \nabla \boldsymbol{U} \cdot \boldsymbol{u}~d\Omega$ is non-negative, we can conclude that the  following dissipation inequality holds
$$
\frac{d V(\boldsymbol{u})}{dt} \le    \int_\Omega {\boldsymbol{u}} \cdot \boldsymbol{d}~d\Omega,
$$
where $W(\boldsymbol{u},\boldsymbol{d}) = \int_\Omega {\boldsymbol{u}} \cdot \boldsymbol{d}~d\Omega$ is the supply rate. This is a well-known dissipation inequality that corresponds to \textit{passivity}. Passivity has been used to study  finite-dimensional linear discretizations of the Navier-Stokes equation with the nonlinearity being modeled as an input~(\cite{3662449,HEINS2016348}).

The general dissipation inequality framework allows us to consider more general energy inequalities rather than only the passivity inequality. In particular, our formulation considers weighted kinetic energy as the storage functional and three different  supply rates. As  will be shown in Section~\ref{sec:convex}, for the class of fluid flows studied in this paper, we present an algorithmic way to find the storage functionals based on convex optimization.


\subsection{Input-Output Properties}

We now define the three types of input-output properties that we study within the dissipativity framework and discuss their relation to common notions in the literature.

%
%
The first property that we can study is bounds on the maximum \textit{energy growth} due to initial perturbation velocities for the nonlinear Navier-Stokes equation~\eqref{eq:mainNS}. In the context of linear systems, this corresponds to maximum transient growth,~\cite{858386, reddy_henningson_1993,gustavsson_1991}. 

\begin{defi} [Energy Growth]
Let \mbox{$\boldsymbol{d}\equiv 0$} in ~\eqref{eq:mainNS}. If there exists a constant $\gamma>0$ such that 
\begin{equation} \label{eq:engr}
\| \boldsymbol{u} \|_{\mathcal{L}^2_{[0,\infty),\Omega}} \le \gamma \| \boldsymbol{u}(0,\cdot) \|_{\mathcal{L}^2_{\Omega}},
\end{equation}
where $\| u \|_{\mathcal{L}^2_{[0,T),\Omega}} = \left ( \int_0^T \int_\Omega u^2(\tau,\theta)~\mathrm{d}\tau \mathrm{d}\theta  \right)^{\frac{1}{2}}$ and $\| u_0 \|_{\mathcal{L}^2_{\Omega} }= \left (  \int_\Omega u_0^2(\theta)~ \mathrm{d}\theta  \right)^{\frac{1}{2}}$, then we say that the flow perturbations have bounded energy growth.
\end{defi}

 The next property of interest is related to amplifications from square integrable  body forces or disturbances (see \cite[Chapter 9]{mjphd04} for results pertaining to a linearized model of channel flows).  The square integrable forcings are of special interest, because they can be interpreted as finite energy forcings. 
 
 We refer to this class of amplifications as \textit{worst-case disturbance amplification}.

\begin{defi} [Worst-Case Disturbance Amplification]
 If there exists  \mbox{$\eta_i >0$, $i\in I$}, such that 
\begin{equation} \label{eq:L2}
\| \boldsymbol{u} \|_{\mathcal{L}^2_{[0,\infty),\Omega}}^2 \le \sum_{i\in I}\eta_i^2 \| d_{i} \|_{\mathcal{L}^2_{[0,\infty),\Omega}}^2,
\end{equation}
subject to zero initial perturbations $\boldsymbol{u}(0,\mathrm{x})\equiv0,~\forall \mathrm{x} \in \Omega$, then we say that the flow has bounded worst-case disturbance amplification.
\end{defi}

The above property is equivalent to~the induced $\mathcal{L}^2$-norm in control theory (\cite{van2017l2}). In other words, each $\eta_i$ upper-bounds the peak amplification  of perturbation velocities from the forcing in the direction $i$, $d_i$, when the forcings in other directions are set to zero, i.e., $d_j = 0$,~$j \in I$, $i \neq j$. That is,
$$
\eta_i \le \sup_{\| {d}_i \|_{\mathcal{L}^2_{[0,\infty),\Omega}} \neq  0} \frac{\| \boldsymbol{u} \|_{\mathcal{L}^2_{[0,\infty),\Omega}}}{\| {d}_i \|_{\mathcal{L}^2_{[0,\infty),\Omega}}}.
$$

Due to nonlinear flow dynamics, the actual induced $\mathcal{L}^2$-norm of system~\eqref{eq:mainNS} is a nonlinear function of $\|\boldsymbol{d}\|_{\mathcal{L}^2_{[0,\infty),\Omega}}$ \cite[Example I]{AVPaut}. The quantities $\eta_i,~i\in I$ provide  upper-bounds on the actual  induced $\mathcal{L}^2$-norms. In this sense, minimizing $\eta_i >0$, $i\in I$, provides an upper bound to the worst-case disturbance amplification. 


From a practical perspective, global
stability of a base flow is often not very meaningful, because small disturbances may cause an unstable behavior. Hence, we require a notion of stability that relates disturbances to perturbation velocities. Besides, the definition of the worst-case disturbance amplification requires the forcings to be square integrable. This automatically leads to the exclusion of persistent forcings, e.g. constant  and sinusoidal forcings, that are defined for all time. To include these classes of forcings in a nonlinear context\footnote{ In the fluids literature, the ensemble average energy density or the $\mathcal{H}^2$-norm has been used to study amplifications from Gaussian stochastic forcings to the linearized flow dynamics (\cite{FI93,JB05}). The $\mathcal{H}^2$-norm is equivalent to the (root mean square) RMS-value of the linearized flow response to  white noise forcings. However, extension of $\mathcal{H}^2$ analysis to the nonlinear Navier-Stokes equations is an open problem.}, we employ the concept of input-to-state stability (\cite{So08})  to study the class of upper-bounded forcings. We refer to this extended notion of stability, as \textit{stability to persistent disturbances}. 

Prominent among the features of this property
are that forcings that are bounded, eventually
small, integrally small, or convergent should
lead to perturbation velocities with the respective property. Furthermore, this property quantifies in what
manner initial perturbation velocities affect transient behavior. Flows with this property
do not have unstable behavior for persistent (nonvanishing) forcings. 

To characterize this property, let us introduce a few comparison functions. $\mathcal{K}$ denote the class of nonnegative functions  that are strictly increasing and zero for zero argument, and $\mathcal{K}_\infty$ denote the class of functions that, in addition, become unbounded as their argument goes to infinity.

\begin{defi} [Stability to persistent disturbances]
If there exist  some scalar $\psi>0$, functions $\beta,\tilde{\beta},\chi \in \mathcal{K}_\infty$, and  $\sigma \in \mathcal{K}$, such that 
{
\begin{multline}\label{eq:iss}
\|\boldsymbol{u}(t,\cdot)\|_{\mathcal{L}^2_\Omega} \le \beta \left( e^{-\psi t} \chi \left(\|\boldsymbol{u}(0,\cdot)\|_{\mathcal{L}^2_\Omega} \right)  \right)  + \tilde{\beta} \left( \sup_{\tau \in [0,t)} \big( \int_\Omega \sigma \big(|\boldsymbol{d}(\tau,\mathrm{x})|\big) \,\, d\Omega \big)  \right),
\end{multline}}
\noindent for all $t>0$, then we call the flow stable to persistent disturbances.
\end{defi}

Property~\eqref{eq:iss} implies   convergence to the base flow $(\boldsymbol{U},P)$ in the $\mathcal{L}^2_\Omega$-norm (the norm corresponding to the space of square integrable functions over the spatial domain) when the disturbances are not present ($\boldsymbol{d} \equiv 0$). Indeed, the $\beta \left( e^{-\psi t} \chi \left(\|\boldsymbol{u}(0,\cdot)\|_{\mathcal{L}^2_\Omega} \right)  \right)$ term dominates for small $t$, and  this serves
to quantify the magnitude of the transient growth as a function of the size of the
initial state $\|\boldsymbol{u}(0,\cdot)\|_{\mathcal{L}^2_\Omega}$.

 Moreover, as $t \to \infty$, we obtain 
\begin{multline}
\lim_{t \to \infty}\| \boldsymbol{u}(t,\cdot) \|_{\mathcal{L}^2_\Omega} \le \tilde{\beta} \left(  \int_\Omega \|\sigma(|\boldsymbol{d}(\cdot,\mathrm{x})|) \|_{\mathcal{L}^\infty_{[0,\infty)} }\,\, d\Omega  \right)  
 \le  \tilde{\beta} \left(   \int_\Omega \sigma(\|\boldsymbol{d}(\cdot,\mathrm{x})\|_{\mathcal{L}^\infty_{[0,\infty)} }) \,\, d\Omega  \right),
\end{multline}
\noindent where, $\sigma, \beta \in \mathcal{K}$ and $\|f\|_{\mathcal{L}^\infty_{[0,\infty)} }=\sup_{\tau \in [0,\infty)} |f(\tau)|$. Hence, as long as the external excitations or body forces $\boldsymbol{d}$ are upper-bounded, the perturbation velocities  $\boldsymbol{u}$ are bounded in the  $\mathcal{L}^2_\Omega$-norm, meaning that they remain square integrable over the flow geometry. 

\begin{figure}

\centering{
                \includegraphics[width=9cm]{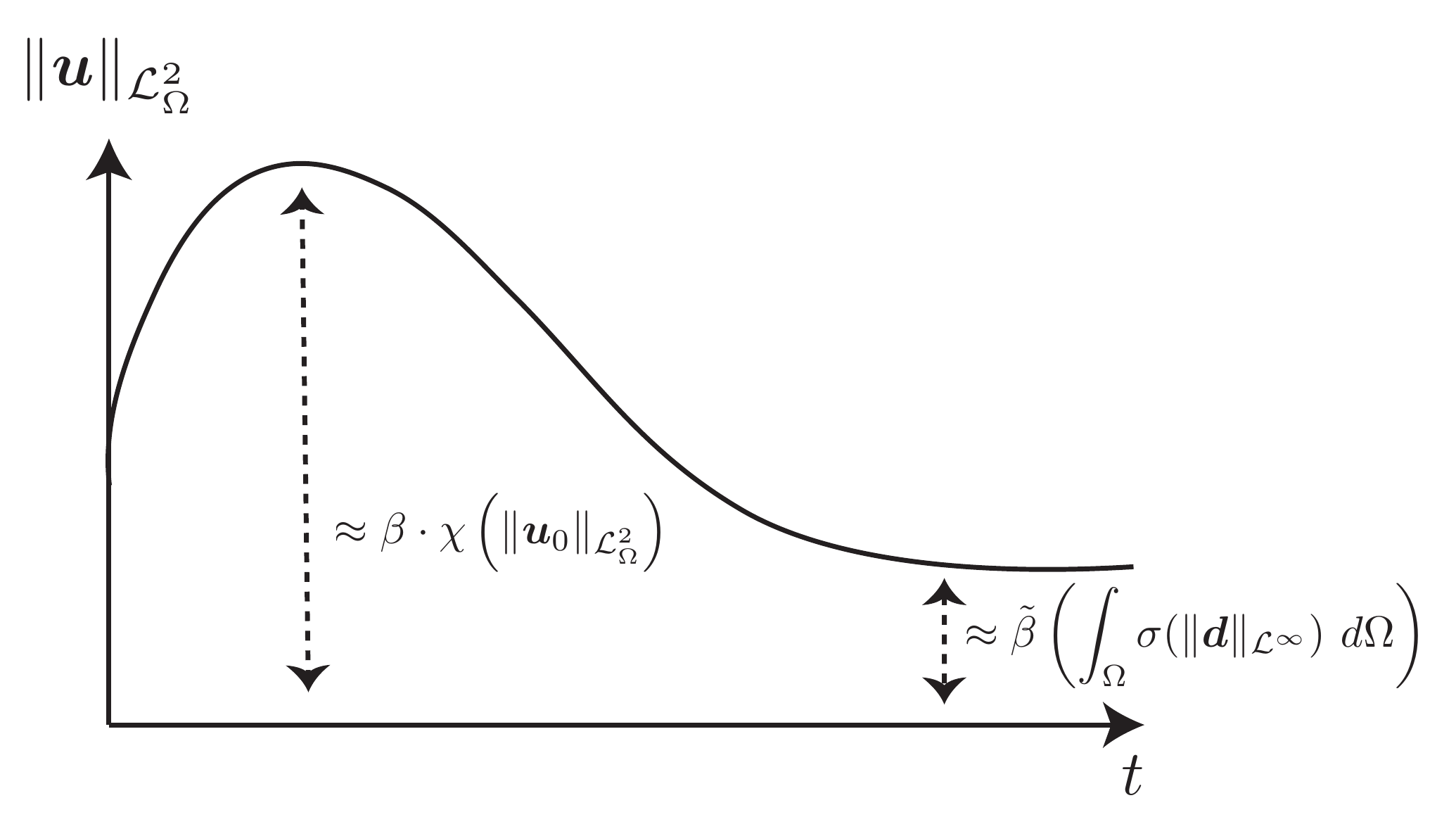}}

        \caption{ The stability to persistent disturbances property combines transient growth (overshoot)
and asymptotic behavior.}\label{figiss}
\end{figure}

In fact, by input-to-state superposition theorem~(\cite{Sontag2013}),
we can shows that stability to persistent disturbances is the conjunction of two
properties, one of them concerned  with asymptotic
bounds on the perturbation velocities, in the sense of $\|\boldsymbol{u}(t,\cdot)\|_{\mathcal{L}^2_\Omega}$,
 as a function of the magnitude of
the forcings, and the other one providing a transient
term obtained when we ignore forcings (see Figure~\ref{figiss}).

We now demonstrate how the problem of verifying the properties in Definitions 3.1-3.3 can be cast as verifying a set of dissipation inequalities. This result which can be derived from~\cite[Theorem~6]{AVPaut} allows for the extension of well known methods for stability, input/output, and optimal perturbation  analysis of linear systems to the full nonlinear Navier-Stokes equation. 
{
\begin{thm} \label{bigthmfluidfluid}
Consider the perturbation model~\eqref{eq:mainNS}. If there exist  a positive semidefinite  storage functional {$V(\boldsymbol{u})$}, positive scalars $\{\eta_i\}_{i\in I}$, $\psi$,  $\gamma$, and functions $\beta_1,\beta_2 \in \mathcal{K}_\infty$, $\sigma \in \mathcal{K}$, such that\\
\mbox{I)} when $\boldsymbol{d}\equiv 0$, 
\begin{equation} \label{fjffjhfjghfjfh}
V(\boldsymbol{u}) \le \gamma^2 {  \| \boldsymbol{u}(t,\cdot) \|^2_{\mathcal{L}^2_\Omega}},
\end{equation}
\begin{equation} \label{con:engr}
\frac{d V(\boldsymbol{u}(t,\mathrm{x}))}{dt}  \le - \int_\Omega \boldsymbol{u}^\prime (t,\mathrm{x})\boldsymbol{u}(t,\mathrm{x}) \,\, d\Omega,
\end{equation}
then it has bounded energy growth as given by~\eqref{eq:engr};\\
\mbox{II)}
\begin{eqnarray} \label{fluide6}
\frac{d V(\boldsymbol{u}(t,\mathrm{x}))}{dt}  \le - \int_\Omega \boldsymbol{u}^\prime (t,\mathrm{x})\boldsymbol{u}(t,\mathrm{x}) \,\, d\Omega
+  \int_\Omega \sum_{i \in I} \eta_i^2 {d}_i^2(t,\mathrm{x}) \,\, d\Omega,
\end{eqnarray}
then the perturbation velocities~\eqref{eq:mainNS} has  worst-case disturbance amplification upper-bounds  $\eta_i$, $i\in I$ as in~\eqref{eq:L2};\\
\mbox{III)}
\begin{equation} \label{fluidse11}
{ \beta_1(\|\boldsymbol{u}(t,\cdot)\|_{\mathcal{L}^2_\Omega}) \le V(\boldsymbol{u}) \le \beta_2(\|\boldsymbol{u}(t,\cdot)\|_{\mathcal{L}^2_\Omega})},
\end{equation}
\begin{equation} \label{fluidse12}
\frac{d V(\boldsymbol{u}(t,\mathrm{x}))}{dt}  \le -  \psi V(\boldsymbol{u}(t,\mathrm{x})) + \int_\Omega \sigma(|\boldsymbol{d}(t,\mathrm{x})|) \,\, d\Omega,
\end{equation}
then perturbation velocities described by~\eqref{eq:mainNS} are stabe to persistent disturbances as  given by \eqref{eq:iss} with $\chi = \beta_2$, $\beta=\beta_1^{-1}\circ 2$ and $\tilde{\beta}=\beta_1^{-1}\circ \frac{2}{\psi}$, where $\circ$ implies function composition.
%
%
%
%
%
%
\end{thm}
}

In the following, we derive  classes of storage functionals $V(\boldsymbol{u})$ suitable for the analysis of perturbation dynamics~\eqref{eq:mainNS} invariant in one of the three spatial coordinates. We consider two classes of flows, namely, channel flows  with perturbations that vary in two spatial dimensions and time discussed in Section~\ref{sec:channelwww} and  pipe flows  invariant in the axial direction discussed in Appendix~\ref{app:cylindr}.

\subsection{Flows Between Parallel Plates} \label{sec:channelwww}

In Cartesian coordinates, for a scalar function ${v}$, $\nabla {v} = \sum_i \partial_{x_i} v \overrightarrow{e}_{i}$  and $\nabla^2 {v} = \sum_i \partial_{x_i}^2 v$, {where $\overrightarrow{e}_{i}$ is the unit vector in the direction $x_i$.} For a vector valued function $\boldsymbol{w}=\sum_i w_i \overrightarrow{e}_{i}$, the divergence $\nabla \cdot \boldsymbol{w}$ is given by $\nabla \cdot \boldsymbol{w} = \sum_i \partial_{x_i} w_i$. 
In the following,  $\{x_1,x_2,x_3\}$  corresponds to $\{x,y,z\}$ (streamwise,  wall-normal, and spanwise directions) and $I=\{1,2,3\}$. Additionally,  we adopt Einstein's multi-index notation over index $j$, that is the sum over repeated indices $j$, e.g., $v_j \partial_{x_j} u_i = \sum_j v_j\partial_{x_j} u_i$.

The perturbation model \eqref{eq:mainNS}  can be re-written as
\begin{eqnarray} \label{eq:mainNSEin}
\partial_t u_i &=& \frac{1}{Re} \nabla^2 u_i - u_j  \partial_{x_j} u_i - U_j  \partial_{x_j} u_i  - u_j  \partial_{x_j} U_i - \partial_{x_i} p + F_{ij} u_j + d_i, \nonumber \\
0 &=& \partial_{x_j} u_j.
\end{eqnarray}
where $i,j \in I$ and $F_{ij}$ is the $(i,j)$ entry of $F$. {To simplify the exposition, without  loss of generality, we assume that the perturbations are invariant with respect to $x_1$. Since $x_i,~i=1,2,3$ are arbitrary, this does not affect the formulation.}


The next proposition states that, by choosing a suitable storage functional structure (weighted kinetic energy of the perturbation velocities), the time derivative of the storage functional turns out to be upper-bounded  by a quadratic form in the  velocity fields $\boldsymbol{u}$ and their spatial derivatives. This property paves the way for a convex optimization based method to check stability and input-output properties. Convex optimization is a subfield of optimization that studies the problem of minimizing convex functions over convex sets. The convexity makes optimization easier than the general case since local minimum must be a global minimum, and first-order conditions are sufficient conditions for optimality~(\cite{BV04}). Convex optimization problems can be solved efficiently by interior-point methods~(\cite{nesterov1994interior}). Convex optimization was used by~\cite{4876195} to obtain a low-order decomposition of the
Navier-Stokes equations  based on resolvent modes.

\begin{prop} \label{fluidsprop1}
Consider the perturbation model~\eqref{eq:mainNSEin} subject to  periodic or   no-slip boundary conditions $\boldsymbol{u}|_{\partial\Omega} =0$. Assume the velocity perturbations in~\eqref{eq:mainNSEin} are invariant with respect to $x_1$. Let $I_0 = \{2,3\}$ and
\begin{equation} \label{eq:Lyap}
V(\boldsymbol{u}) = \frac{1}{2}\int_\Omega \boldsymbol{u}^\prime Q \boldsymbol{u} \,\, d\Omega, 
\end{equation}
where $Q =\left[ \begin{smallmatrix} q_1 & 0 & 0 \\ 0 & q_i & 0 \\ 0 & 0 & q_j \end{smallmatrix} \right]>0$, $q_i = q_j$ for $i\neq j$, $i,j \in I_0$, be a candidate storage functional. Then, the time derivative of~\eqref{eq:Lyap} along the solutions to~\eqref{eq:mainNSEin} satisfies 
\begin{multline} \label{eq:Lyapmaindt}
\frac{dV(\boldsymbol{u})}{dt} \le  -\sum_{i\in I}q_i\int_\Omega  \bigg( \frac{ C}{Re} u_i^2   +  U_j  u_i \partial_{x_j} u_i   +u_j u_i  \partial_{x_j} U_i  -  u_i F_{ij} u_j -u_i d_i \bigg) \,\, d\Omega,
\end{multline}
where $C$ is a positive constant that only depends on the domain $\Omega$.
\end{prop}

The proof of this proposition is given in Appendix~\ref{appfdsfdfdsfcccvr}.

Remark that a special case of~\eqref{eq:Lyap} was used in~(\cite{JH71}) to study the stability of viscous flows (subject to streamwise constant perturbations) in pipes and between rotating cylinders. The authors referred to this structure as \emph{the two energy function}. In the formulation presented in this paper, assuming invariant perturbations in the $x_1$-direction, we can represent the two energy function as
$$
V(\boldsymbol{u}) = \frac{1}{2}\int_\Omega \boldsymbol{u}^\prime \left[ \begin{smallmatrix} q & 0 & 0\\ 0 & 1 & 0\\ 0 & 0 & 1\end{smallmatrix}  \right] \boldsymbol{u} \,\, d\Omega, 
$$
 where $q>0$ is a constant. The ``optimal" value for this constant was then calculated analytically for the pipe Poiseuille and the Taylor-Couette flow by~\cite{JH71}.

Note that in \eqref{eq:Lyapmaindt} the Poincar\'e constant, $C$, appears. There are several estimates for the optimal  Poincar\'e constant. The optimal constant (\cite{PW60}) we use in this paper is 
\begin{equation}
C(\Omega) = \frac{\pi^2}{D(\Omega)},
\end{equation}
where $D(\Omega)$ is the diameter of the domain $\Omega$.

Proposition~\ref{fluidsprop1} allows us to provide an algorithmic method for input-output analysis of fluid flows based on convex optimization. These convex optimization problems are in terms of linear matrix inequalities and polynomial matrix inequalities. This formulation is delineated in more detail in the next section.

  \vspace{-.85cm}

\section{Matrix Inequality Formulation for  Streamwise Constant Perturbations}
\label{sec:convex}

In this section,  we show that the input-output analysis problem  outlined in Section 3 for the class of streamwise constant perturbations can be converted into a set of matrix inequalities. These matrix inequalities can be solved by  convex optimization, provided that the base flow is a polynomial in the spatial coordinates and the flow geometry is described by a semi-algebraic set\footnote{
Let $\mathcal{R}[x]$ be the set of polynomials with real coefficients. A set is semi-algebraic if it can be described by a finite number of polynomial equalities and inequalities. That is,  $\mathcal{S} \subset \mathbb{R}^n$ for some closed field, say $\mathbb{R}$, is defined by a set of polynomial equalities and inequalities as follows
$
\mathcal{S} = \left\{ x \in \mathbb{R}^n \mid p_i(x) \ge 0,~i=1,2,\ldots, n_p,~ q_i(x)=0,~i=1,2,\ldots,n_q   \right\},
$
where $\{p_i \}_{i=1}^{n_p},\{q_i\}_{i=1}^{n_q} \in \mathcal{R}[x]$, where $\mathcal{R}[x]$ denotes the set of polynomials in the variable $x$ with real coefficients.
}. Examples are laminar base flows that are linear or parabolic, and turbulent flows that can be represented by polynomial fits  (or by piecewise polynomial functions).


 To present a convex method for checking the conditions in Theorem~\ref{bigthmfluidfluid} (also see Corollary~\ref{cor1} in Appendix B), we restrict our attention to streamwise constant perturbations in the $x_1$-direction with base flow $\boldsymbol{U} = U_m(x_2,x_3) \overrightarrow{e}_1$, where $\overrightarrow{e}_1$ denotes the unit vector in the $x_1$-direction.
 
In order to present the procedure, we first need to define the following notation. For a square matrix $M$,  $M \succcurlyeq 0$ ($M  \succ 0$) implies that the matrix is positive semidefinite (positive definite), i.e., all the eigenvalues of $M$ are non-negative (positive). Similarly, $M \preccurlyeq 0$ ($M  \prec 0$) signifies that $-M \succcurlyeq 0$ ($-M \succ  0$). By $\mathrm{I}_{n \times n}$, we denote the square matrix of dimension $n \times n$ with diagonal entries set to $1$.
  
 \begin{corollary} \label{LMIcor}
 Consider the perturbation dynamics given by~\eqref{eq:mainNSEin}, that are constant in the streamwise direction $x_1$ and with base flow $\boldsymbol{U} = U_m(\mathrm{x}) \overrightarrow{e}_1$, where $\mathrm{x}=(x_2,x_3)$. Let $I_0=\{2,3\}$. If there exist positive constants $\{q_l\}_{l\in I}$ with $q_i=q_j$, $i,j\in I_0$, $\{\eta_l\}_{l\in I}$,  $\{\psi_l\}_{l\in I}$, and functions $\{\sigma_l\}_{l\in I}$ such that \\
 \begin{multline} \label{eq:Mmat}
M(\mathrm{x}) =\\ \begin{bmatrix} 
 \left(\frac{C}{Re}-F_{11} \right)q_1 &  \frac{q_1(\partial_{x_j}U_m(\mathrm{x})-F_{1j})-q_jF{j1}}{2}  & \frac{q_1(\partial_{x_i}U_m(\mathrm{x})-F_{1i})-q_iF{i1}}{2}   \\  \frac{q_1(\partial_{x_j}U_m(\mathrm{x})-F_{1j})-q_jF{j1}}{2} &  \left(\frac{C}{Re}-F_{jj} \right)q_j & -\frac{q_j F_{j1}}{2} \\ \frac{q_1(\partial_{x_i}U_m(\mathrm{x})-F_{1i})-q_iF{i1}}{2} & -\frac{q_j F_{j1}}{2}  & \left(\frac{C}{Re}-F_{ii} \right)q_i
 \end{bmatrix},\\~i,j \in I_0, i\neq j.
 \end{multline}\\
  \mbox{I)}  when $\boldsymbol{d}\equiv 0$, 
  \begin{equation} \label{condengr}
  M\left(\mathrm{x}\right) - \mathrm{I}_{3 \times 3} \succcurlyeq 0,~\mathrm{x}\in \Omega,
  \end{equation} 
 \mbox{II)} 
 \begin{equation} \label{eq:Nmat}
N(\mathrm{x}) = \begin{pmat}[{..|}] 
~ & ~  & ~ & -\frac{q_1}{2} & 0 & 0\cr 
~ & M(\mathrm{x})-\mathrm{I}_{3\times 3} & ~ & 0 & -\frac{q_j}{2} & 0 \cr 
~ & ~& ~ & 0 & 0 & -\frac{q_i}{2} \cr\-
-\frac{q_1}{2} & 0 & 0  & \eta_{1}^2 & 0 & 0 \cr
0 & -\frac{q_j}{2} & 0  & 0 & \eta_{i}^2  & 0   \cr
0 & 0 & -\frac{q_i}{2}  & 0 & 0 & \eta_{j}^2  \cr
\end{pmat} \succcurlyeq 0,
\end{equation}
for $i,j \in I_0, i\neq j$ and $\mathrm{x}\in \Omega$,\\
\mbox{III)} $\sigma_l(\mathrm{x}) \ge 0,~\mathrm{x} \in \Omega$, $l\in I$ and 
\begin{equation}\label{eq:Pmat}
Z(\mathrm{x}) =\begin{pmat} [{..|}]  ~ & ~  & ~ & -\frac{q_1}{2} & 0 & 0 \cr
 ~ & M(\mathrm{x})-W & ~ & 0 & -\frac{q_j}{2} & 0 \cr
~ & ~ & ~ & 0 & 0 & -\frac{q_i}{2} \cr\-
 -\frac{q_1}{2} & 0 & 0  & \sigma_{1}(\mathrm{x}) & 0 & 0 \cr
0 & -\frac{q_j}{2} & 0  & 0 & \sigma_{j}(\mathrm{x})  & 0   \cr
0 & 0 & -\frac{q_i}{2}  & 0 & 0 & \sigma_{i}(\mathrm{x}) \cr \end{pmat} \succcurlyeq 0,
\end{equation}
for  $i,j \in I_0, i\neq j$ and $\mathrm{x}\in \Omega$, where $W=\left[\begin{smallmatrix}\psi_1q_1 & 0 & 0\\0&\psi_jq_j&0\\0&0&\psi_i q_i\end{smallmatrix} \right]$. Then, it follows that \\
\mbox{I)} the flow energy growth is bounded by $\gamma^2 = \max_{i \in I} q_i$ as described by~\eqref{eq:engr},\\
\mbox{II)}  the worst-case disturbance amplification (induced  $\mathcal{L}^2$ norm from the disturbances to perturbation velocities) is bounded by $\eta_i$, \mbox{$i\in  I$} as in~\eqref{eq:L2} when the initial perturbations have zero velocity,\\
\mbox{III)} the flow is stable to persistent disturbances  in the sense of~\eqref{eq:iss} with $\sigma(|\boldsymbol{d}|) =  \sum_{i\in I} \sigma_i(\mathrm{x}) d_i^2$.
  \end{corollary}

The proof of the above Corollary is given in Appendix~B. 

  
  When $U_m(\mathrm{x})$ is a polynomial function, inequalities~\eqref{eq:Mmat}-\eqref{eq:Pmat} are polynomial matrix inequalities that should be checked for all $\mathrm{x} \in \Omega$. If the set $\Omega$ is a semi-algebraic set,~\textit{i.e.,}
\begin{equation*}
  \Omega = \left\{ \mathrm{x} \in \mathbb{R}^2 \mid g_l(\mathrm{x})=0,~f_k(\mathrm{x})>0,~l = 1,2,\ldots,L,~k=1,2,\ldots,K  \right\},
\end{equation*}
 where $\{g_l\}_{l=1}^{L}$ and $\{f_k\}_{k=1}^{K}$ are polynomial functions,
 then these inequalities can be cast as a sum-of-squares program by applying Corollary~\ref{cor:psatz}. We  show in the next section that this assumption is indeed the case for several well-known flows. For a brief introduction to sum-of-squares programming refer to Appendix~\ref{app:sosp}. Note that once the input-output analysis problem is cast as a sum-of-squares program, it can be checked using available MATLAB toolboxes such as~SOSTOOLS (\cite{sostools}) and YALMIP~ (\cite{YALMIP}).
  
We can compute the bound on the maximum energy grown described in~\eqref{eq:engr} by solving an optimization problem. To this end, we solve 
\begin{eqnarray} \label{opprobengr}
& \underset{\{q_i\}_{i \in I}}{\min} \left( \underset{{i \in I}}{\max}~q_i \right)& \nonumber \\
&\text{subject~to}& \nonumber \\
&M(\mathrm{x}) - \mathrm{I}_{3 \times 3} \succcurlyeq 0,& \nonumber \\
&q_i>0,~i \in I.&
\end{eqnarray}

 In order to find upper-bounds on the worst-case disturbance amplification (the induced $\mathcal{L}^2$-norm) from the body forces or disturbances~$\boldsymbol{d}$ to the perturbation velocities $\boldsymbol{u}$ as described in~\eqref{eq:L2}, we solve the following optimization problem
\begin{eqnarray}
&\underset{\{q_i\}_{i \in I}}{\min}~\sum_{i\in I} \eta_i^2 & \nonumber \\
&\text{subject~to}& \nonumber \\
&N(\mathrm{x})\succcurlyeq 0,& \nonumber \\
&q_i>0,~i \in I.&
\end{eqnarray}

\section{Numerical Results} \label{sec:NR}

In this section, we illustrate the proposed method by analyzing four benchmark flows, namely,  plane Couette flow,  plane Poiseuille flow, rotating Couette flow (a simplified Taylor-Couette flow model),  and  Hagen-Poiseuille flow. For worst-case disturbance amplification, we carry out a comparative analysis of the influence of each of the disturbance components. For stability to persistent disturbances, we find the maximum Reynolds number for which stability to persistent disturbances holds.

\subsection{Plane Couette Flow} \label{example:planecouette}

\begin{figure}

\centering{
                \includegraphics[width=10cm]{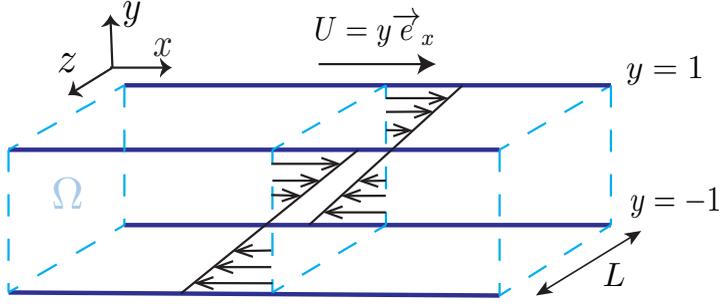}}

        \caption{Schematic of the plane Couette flow geometry.}\label{fig1}
\end{figure}

We consider the  flow of viscous fluid between two parallel plates, where the gap between the plates is much smaller than the length of the plates as illustrated in Figure~\ref{fig1}. 

We consider  no-slip boundary conditions \mbox{$\boldsymbol{u}|_{y=-1}^1 = 0$} in the wall-normal direction and  $\boldsymbol{u}(t,y,z)=\boldsymbol{u}(t,y,z+L)$ in the spanwise direction. The Poincar\'e constant is then given by $C=\frac{\pi^2}{\sqrt{L^2+2^2}}$.


We are interested in studying bounds on energy growth, worst-case amplification, and stability to persistent forcings. To this end, we consider the following storage functional 
\begin{equation} \label{ExampleCouette}
V(u) = \int_0^{L} \int_{-1}^1 \left[\begin{smallmatrix} u_x \\ u_y \\ u_z \end{smallmatrix}\right]^\prime \left[\begin{smallmatrix} q_x & 0 & 0 \\ 0 & q_y & 0 \\ 0 & 0 & q_z \end{smallmatrix} \right]\left[\begin{smallmatrix} u_x \\ u_y \\ u_z \end{smallmatrix} \right] \,\, dydz,
\end{equation}
with $q_y=q_z$, which is the same as storage functional~\eqref{eq:Lyap} considering invariance with respect to $x$. 

 For this flow ($m=x,j=y,i=z$), the $M$ matrix~\eqref{eq:Mmat} described as
\begin{equation} \label{sddfsdf}
M = \begin{bmatrix} \frac{q_xC}{Re} &    \frac{  q_x}{2} & 0 \\ \frac{ q_x}{2} & \frac{q_yC}{Re}  & 0 \\
0 & 0 & \frac{q_yC}{Re}
\end{bmatrix} 
\end{equation}

\begin{figure}

\centerline{
                \includegraphics[scale=.35]{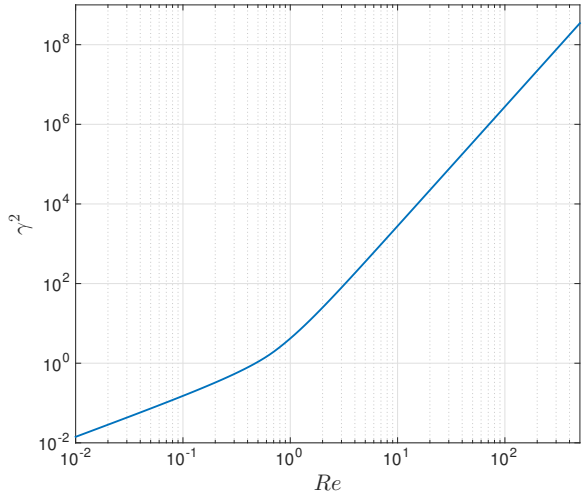}                           
}
        \caption{Upper bounds on the maximum energy growth for plane Couette flow in terms of Reynolds numbers.}\label{figeng}
\end{figure}

Let $L = \pi$. For energy growth analysis, we solve optimization problem~\eqref{opprobengr} with $M$  given by~\eqref{sddfsdf}. The results are depicted in Figure~\ref{figeng}. For small Reynolds numbers $\gamma^2 \propto O(Re)$, whereas for larger Reynolds numbers $\gamma^2 \propto O(Re^3)$. Therefore, it can be inferred that $\gamma^2 = c_0 Re + c_1 Re^3$ with $c_0,c_1>0$. This is consistent with the results by~\cite{BBD02} where the maximum energy growth of steamwise constant (nonlinear) plane Couette flow was calculated analytically. 


For worst-case amplification analysis, we apply inequality~\eqref{eq:Nmat} which for this particular flow is given by the following linear matrix inequality

\begin{eqnarray}
N = \begin{pmat}[{..|}] 
~ & ~  & ~ & -\frac{q_x}{2} & 0 & 0\cr 
~ & M-\mathrm{I}_{3\times 3} & ~ & 0 & -\frac{q_y}{2} & 0 \cr 
~ & ~& ~ & 0 & 0 & -\frac{q_y}{2} \cr\-
-\frac{q_x}{2} & 0 & 0  & \eta_{x}^2 & 0 & 0 \cr
0 & -\frac{q_y}{2} & 0  & 0 & \eta_{y}^2  & 0   \cr
0 & 0 & -\frac{q_y}{2}  & 0 & 0 & \eta_{z}^2  \cr
\end{pmat} \succcurlyeq 0 \nonumber
\end{eqnarray}
with $M$ as in~\eqref{sddfsdf}. 


\begin{figure}

\centerline{
                \includegraphics[scale=.35]{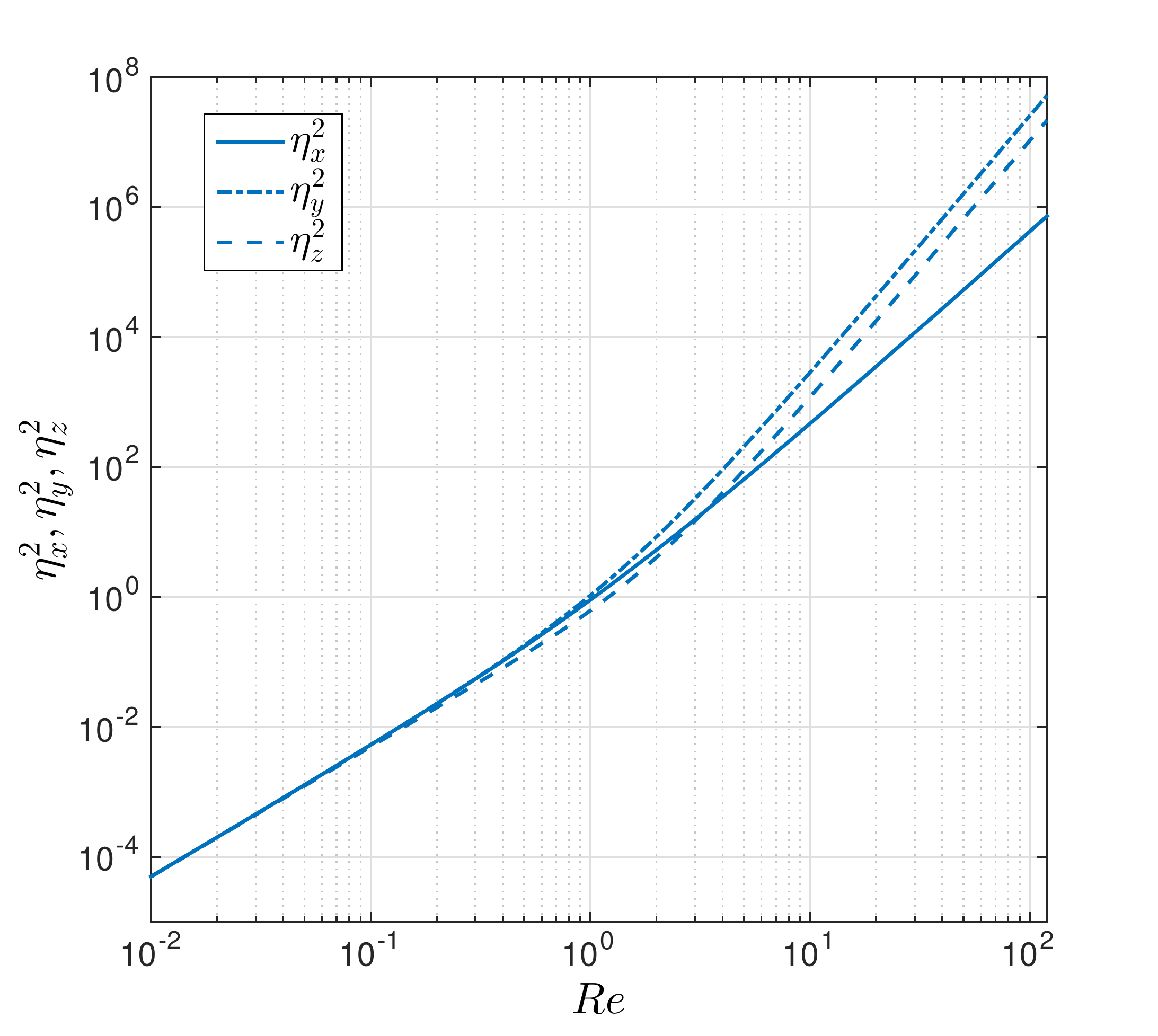}                           
}
        \caption{Upper bounds on the worst-case amplification for perturbation velocities of plane Couette flow for different Reynolds numbers.}\label{fig4}
\end{figure}

The obtained upper-bounds on the worst-case amplification for Couette flow are  given in Figure~\ref{fig4}. Since the flow is stable for all Reynolds numbers, the worst-case amplifications are increasing monotonically  with Reynolds number. The obtained upper-bounds depicted in Figure~\ref{fig4} imply $\eta_x^2 = a_0 Re^{2}+a_1Re^{3}$, $\eta_y^2 = b_0Re^2+b_1 Re^4$ and $\eta_z^2 = c_0 Re^2 +c_1 Re^4$ with \mbox{$a_0,a_1,b_0,b_1,c_0,c_1>0$}. This implies that worst-case amplification in all three components of disturbances grow with a $Re^2$ ratio for low Reynold numbers. For Reynolds numbers approximately greater than $1$, the streamwise disturbances are amplified proportional to $Re^3$; whereas, the wall-normal and spanwise disturbance components are amplified relative to $Re^4$. Therefore, for high Reynolds numbers, worst-case amplification from wall-normal and spanwise disturbance components are approximately $Re$ times larger than  the worst-case amplification from streamwise forcings.

The obtained upper-bounds depicted in Figure~\ref{fig4} can be compared with Corollary~\ref{app:corollary} (see Appendix~\ref{app:calculation}), wherein it was demonstrated that $\eta_x^2  = f_0 Re^2$,  $\eta_{y}^2=g_0 Re^2 + g_1Re^4$ and $\eta_{z}^2 = h_0 Re^2 + h_1Re^4$ for the linearized plane Couette flow with constants $ f_0,g_0,g_1,h_0,h_1>0$. 


Lastly, in order to check the stability to persistent forcings property, we check inequality~\eqref{eq:Pmat} from Corollary~\ref{LMIcor}  for the  Couette flow under study,~\textit{i.e.,}
\begin{equation}
Z =\begin{pmat} [{..|}]  ~ & ~  & ~ & -\frac{q_x}{2} & 0 & 0 \cr
 ~ & M-W & ~ & 0 & -\frac{q_y}{2} & 0 \cr
~ & ~ & ~ & 0 & 0 & -\frac{q_y}{2} \cr\-
 -\frac{q_x}{2} & 0 & 0  & \sigma_{x} & 0 & 0 \cr
0 & -\frac{q_y}{2} & 0  & 0 & \sigma_{y}  & 0   \cr
0 & 0 & -\frac{q_y}{2}  & 0 & 0 & \sigma_{z} \cr \end{pmat} \succcurlyeq 0 \nonumber
\end{equation}
with $M$ given in~\eqref{sddfsdf} and $W=\left[\begin{smallmatrix} q_x \psi_x & 0 & 0\\ 0 & q_y \psi_y & 0\\0& 0 & q_y \psi_z \end{smallmatrix}\right]$.  We fix $\psi_i = 10^{-4},~i=x,y,z$ and $L=2\pi$. 
In this case, we obtain $Re_{ISS} = 316$. {The quantity $Re_{ISS}=316$ is the closest estimate to the empirical Reynolds number $Re \approx 350$ obtained by~\cite{TA92} above which transition to turbulence is observed. In this sense, it turns out that the $Re_{ISS}$ gives lower bounds on the Reynolds number above which transition occurs.}

\begin{figure}

\centering{
                \includegraphics[scale=.35]{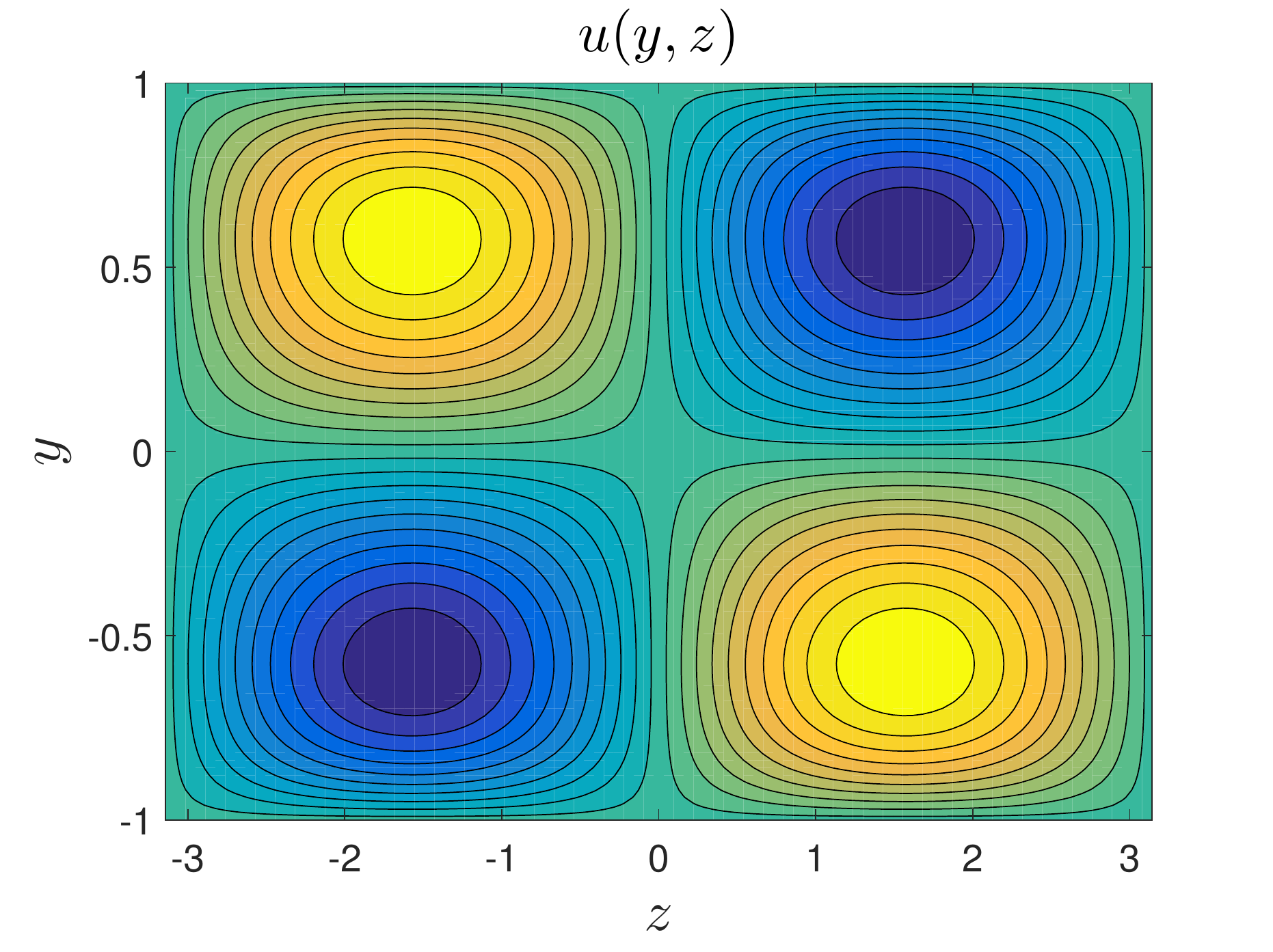}
                \includegraphics[scale=.35]{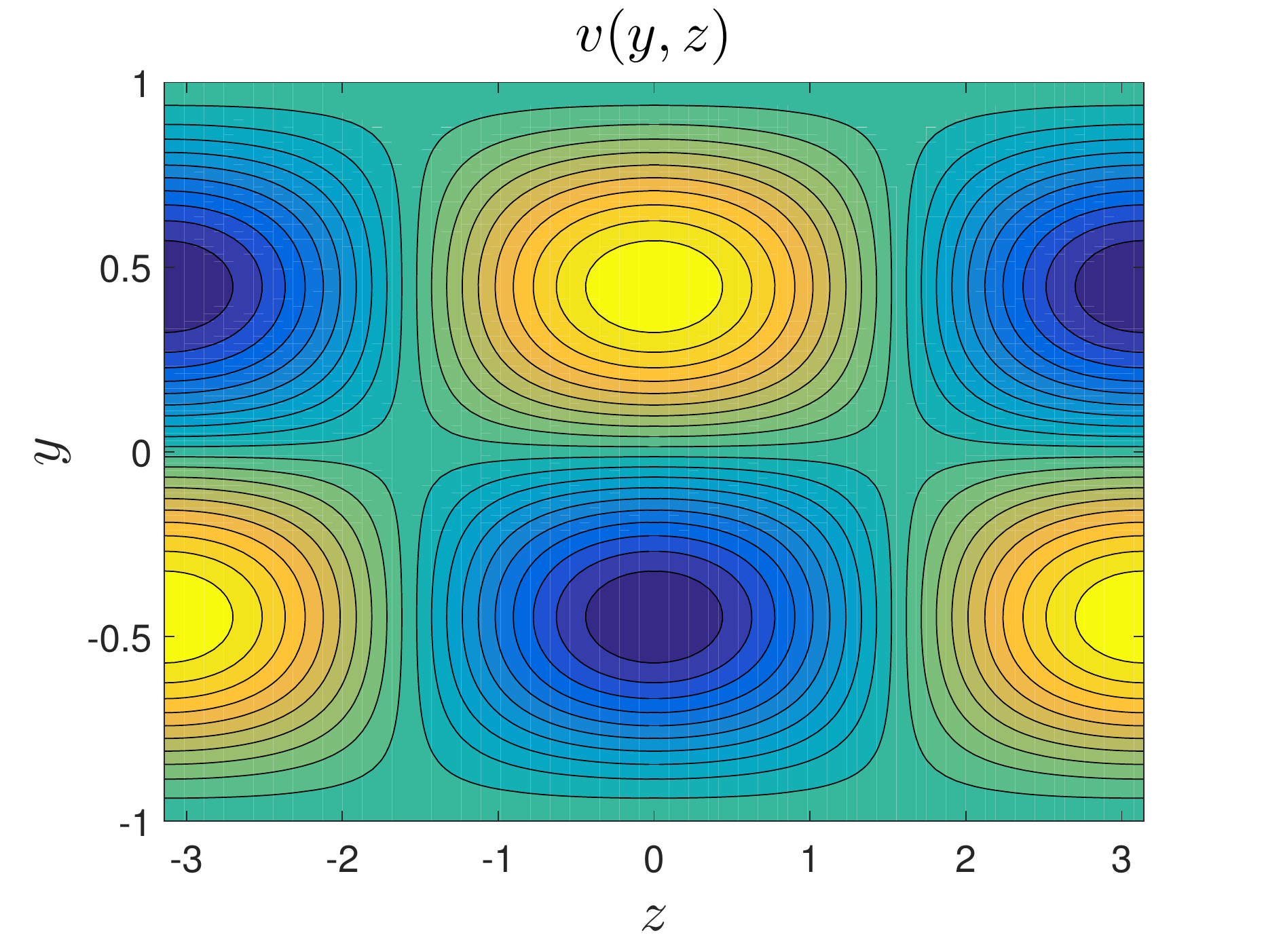}\\                
                \includegraphics[scale=.35]{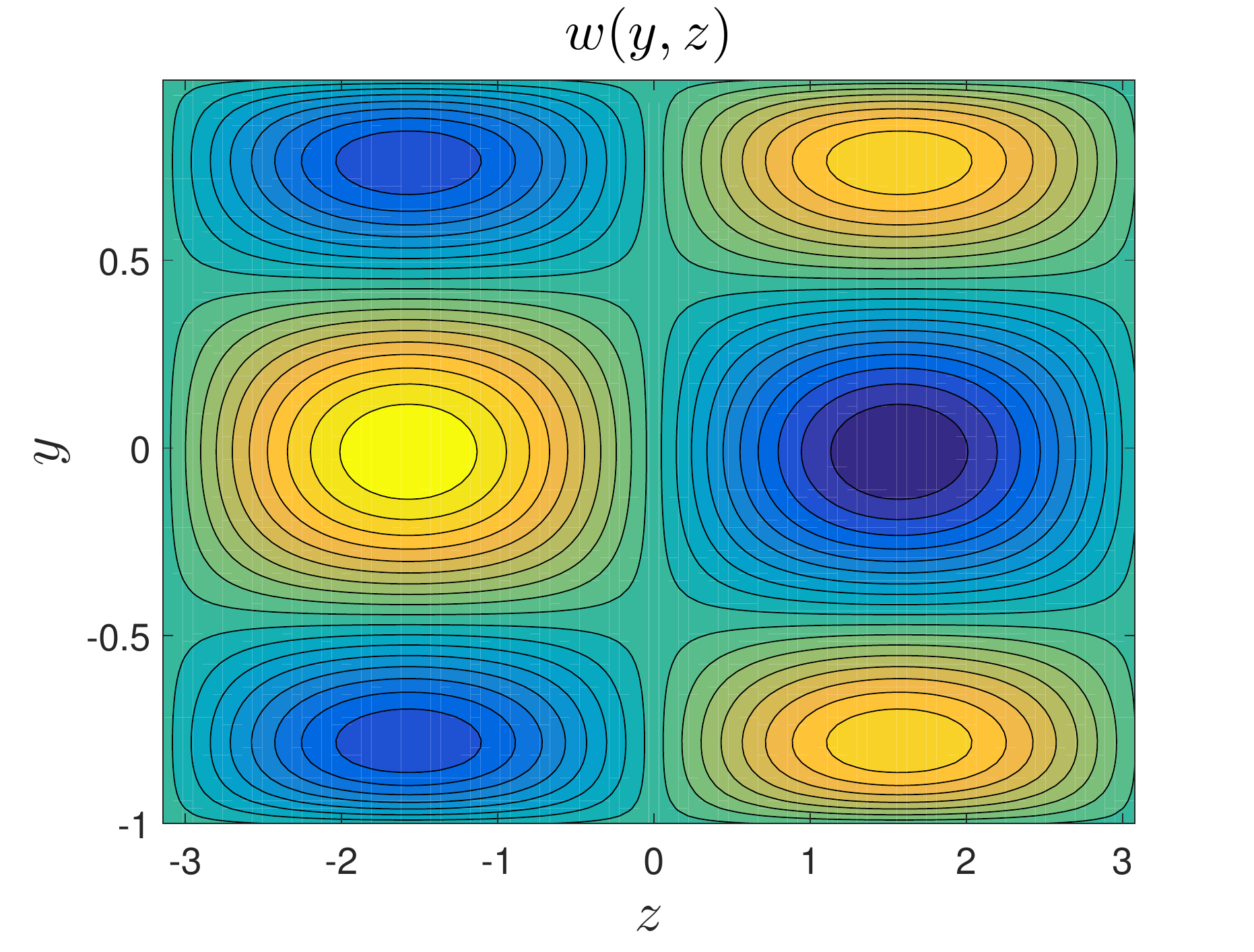}
}
        \caption{The perturbation flow structures with maximum amplification from persistent forcings at $Re=316$ for plane Couette flow.}\label{figISSflowstructCou}
\end{figure}

In order to understand the above result on stability to persistent disturbances, we carried out numerical experiments to obtain the flow structures that receive maximum amplification from persistent disturbances. The experiments were undertaken for the linearized Navier-Stokes equation through the Orr-Somerfield equations. Appendix~\ref{app:NEFS} discusses the details of these numerical experiments. Notice that these results are based on solving linear matrix inequalities that ensure stability to persistent forcings for the ODE space-discretizations of the Orr-Somerfield equations. This is carried out by making a $50 \times 50$ grid on the wave number space $k_x-k_z$ ($k_x,k_z \in [0,150]$) and running the linear matrix inequalities for each point in the grid. Then, the wave numbers corresponding to the maximum  amplification are selected (especially, we are interested to find $k_x$ corresponding to maximum amplification, as this is the streamwise direction) and the corresponding flow structure is simulated. It turns out that the maximum  amplification corresponds to the streamwise constant case $k_x=0$. Figure~\ref{figISSflowstructCou} illustrates the flow structures that receive maximum  amplification at $Re=316$. 

It is also worth mentioning that  certificates for stability to persistent disturbances of the linearized Navier-Stokes equation, as discussed in Appendix~\ref{app:NEFS}, could be constructed for all Reynolds numbers, which is in contrast to the nonlinear case. This illustrates that stability to persistent disturbances is a fundamentally nonlinear phenomenon. 


\begin{figure}

\centering{
                \includegraphics[width=15cm]{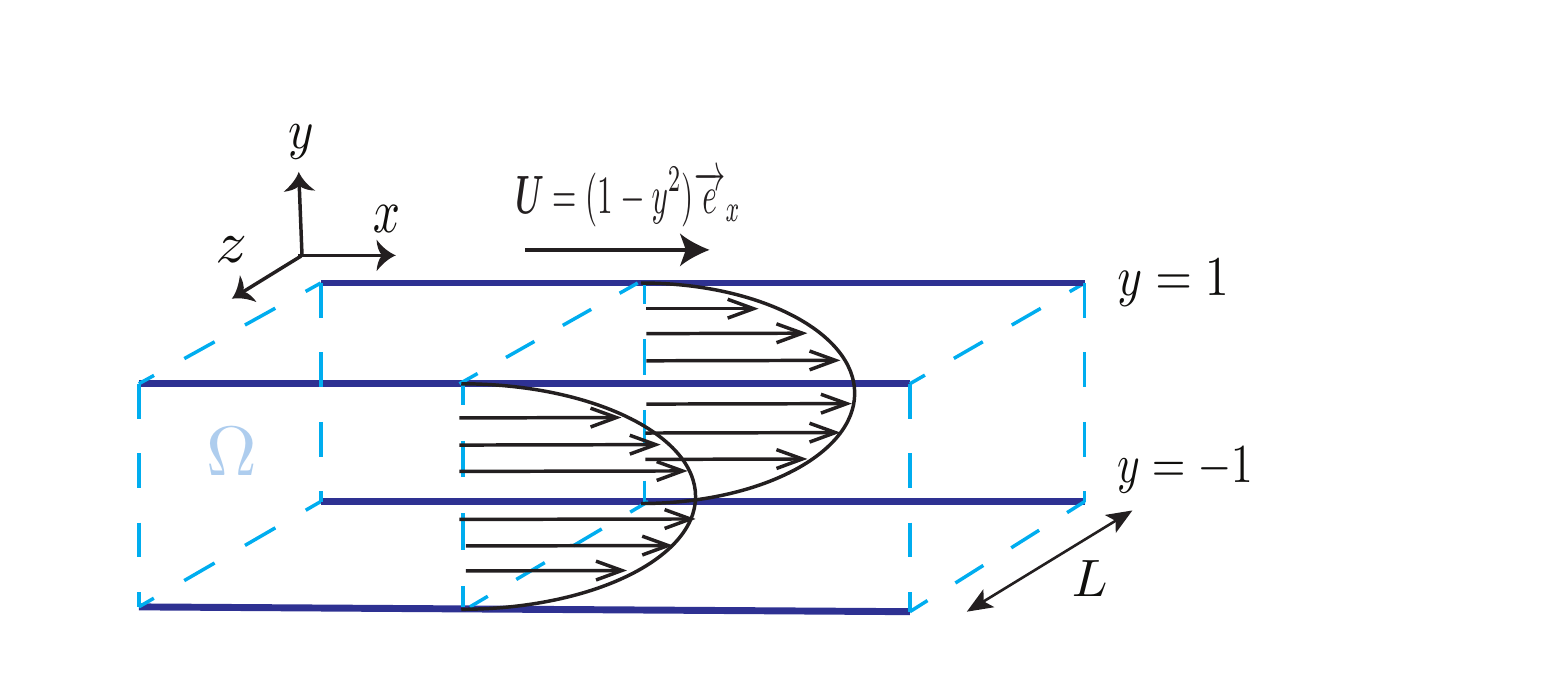}}

        \caption{Schematic of  the  plane Poiseuille flow geometry.}\label{figccjldfdsds2}
\end{figure}

\subsection{Plane Poiseuille Flow}

Similar to the plane Couette flow, we consider the  flow of viscous fluid between two parallel plates, where the gap between the plates is much smaller than the length of the plates. Unlike the plane Couette flow, the plates are stationary and the flow is induced by a pressure gradient in the flow direction, flowing from the region of higher pressure to one of lower pressure. The flow geometry is depicted in Figure~\ref{figccjldfdsds2}. 

The domain $\Omega$ is defined as $\Omega=\{ (y,z) \mid -1<y<1,~0< z <L\}$. The flow perturbations are assumed  invariant in the streamwise direction $x$.  The base flow is given by $\boldsymbol{U}= U_m(y)  \overrightarrow{e}_x= (1-y^2) \overrightarrow{e}_x$ and $P = 1 - \frac{4x}{Re}$.  We consider  no-slip boundary conditions \mbox{$\boldsymbol{u}|_{y=-1}^1 = 0$} and  $\boldsymbol{u}(t,y,z)=\boldsymbol{u}(t,y,z+L)$. The Poincar\'e constant is then given by $C=\frac{\pi^2}{\sqrt{L^2+2^2}}$. We study the  the input-output properties of the flow using the storage functional~\eqref{ExampleCouette}.

For this flow ($m=x,j=y,i=z$), we have
\begin{equation} \label{sddfsdf2}
M(y) = \begin{bmatrix} \frac{q_xC}{Re} &    {  -yq_x} & 0 \\ { -yq_x} & \frac{q_yC}{Re}  & 0 \\
0 & 0 & \frac{q_yC}{Re}
\end{bmatrix}.
\end{equation}
 
  \begin{figure}

\centerline{
                \includegraphics[scale=.37]{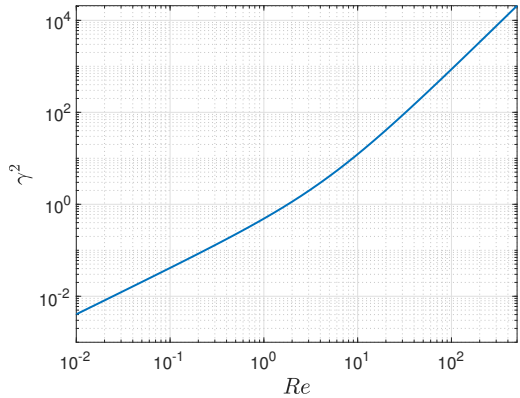}                           
}
        \caption{Upper bounds on the maximum energy growth for  plane Poiseuille  flow in terms of Reynolds numbers.}\label{figfjdfjkdoo1}
\end{figure}
 
 To find upper bounds on maximum energy growth for the plane Poiseuille flow, we solve the optimization problem~\eqref{opprobengr} with $M$ as given in~\eqref{sddfsdf2}. The results are illustrated in Figure~\ref{figfjdfjkdoo1}. This implies that the maximum energy amplification is described by $\gamma^2 = b_0 Re + b_1 Re^2$, with $b_0,b_1>0$.  This result tallies with transient growth calculations of (\cite{reddy_henningson_1993}), in which the authors showed that the transient growth of the linearized plane Poiseuille flow model behaves like $O(Re^2)$ for large Reynolds numbers.
 

For  worst-case amplification analysis, we use inequality~\eqref{eq:Nmat} which for this  flow is given by the following  matrix inequality

\begin{figure}

\centerline{
                \includegraphics[scale=.4]{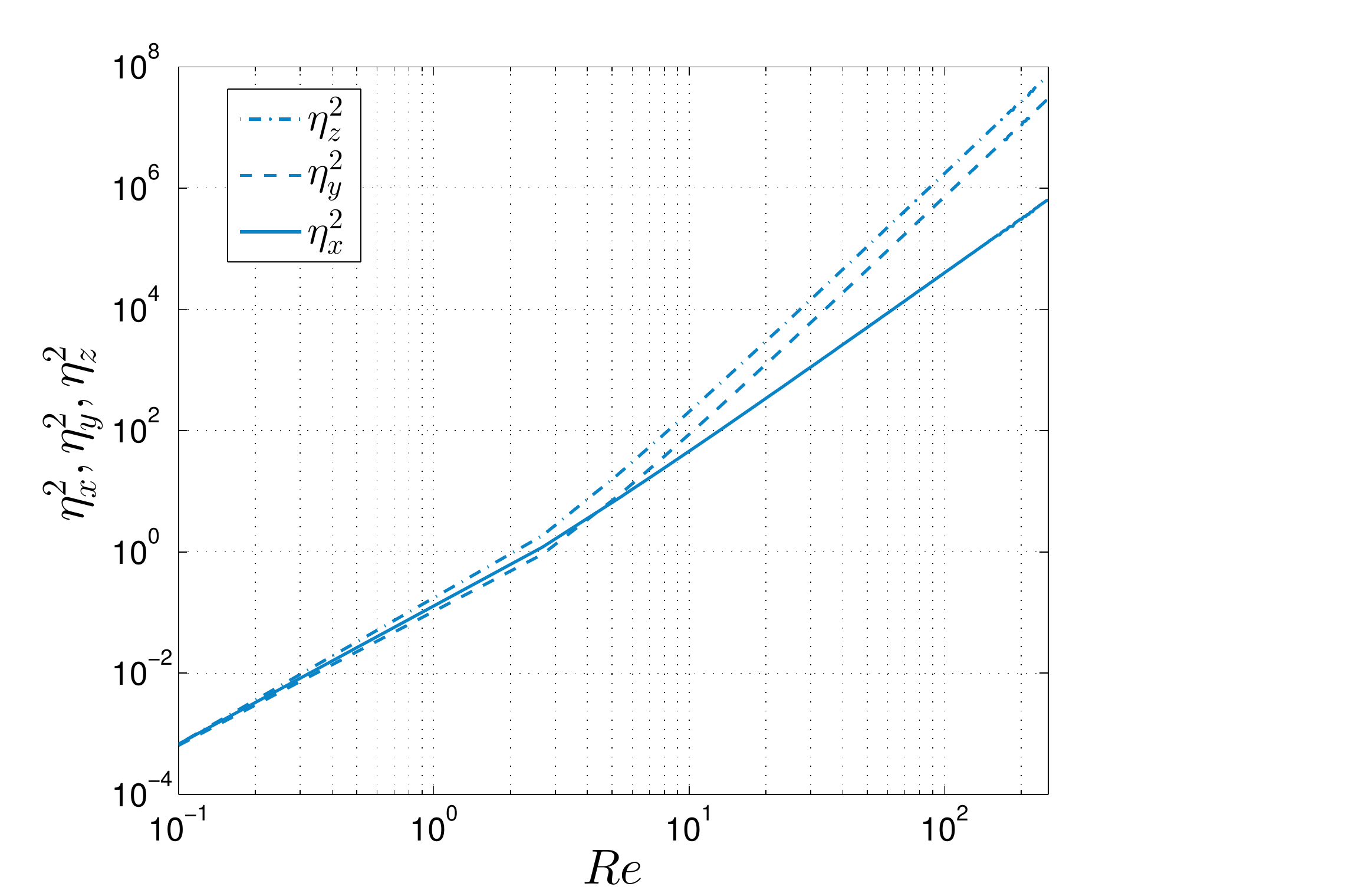}                           
}
        \caption{Upper bounds on the worst-case amplification  of plane Poiseuille flow for different Reynolds numbers.}\label{figx1}
\end{figure}

\begin{eqnarray}
N = \begin{pmat}[{..|}] 
~ & ~  & ~ & -\frac{q_x}{2} & 0 & 0\cr 
~ & M(y)-\mathrm{I}_{3\times 3} & ~ & 0 & -\frac{q_y}{2} & 0 \cr 
~ & ~& ~ & 0 & 0 & -\frac{q_y}{2} \cr\-
-\frac{q_x}{2} & 0 & 0  & \eta_{x}^2 & 0 & 0 \cr
0 & -\frac{q_y}{2} & 0  & 0 & \eta_{y}^2  & 0   \cr
0 & 0 & -\frac{q_y}{2}  & 0 & 0 & \eta_{z}^2  \cr
\end{pmat} \succcurlyeq 0,\quad y \in (-1,1), \nonumber
\end{eqnarray}
with $M$ as in~\eqref{sddfsdf2}. The obtained upper-bounds on the worst-case amplification for the plane Poiseuille  flow are also given in Figure~\ref{figx1}.
 Form Figure~\ref{figx1}, it can be inferred that  $\eta_x^2 = a_0 Re^{2}+a_1Re^{3}$, $\eta_y^2 = b_0Re^{2.2}+b_1 Re^4$ and $\eta_z^2 = c_0 Re^2 +c_1 Re^4$ with $a_0,a_1,b_0,b_1,c_0,c_1>0$. From this result, we can infer that the worst-case amplification in all three components of disturbances grow with a $Re^2$ ratio for low Reynold numbers. For Reynolds numbers approximately greater than $\approx 5$, the streamwise disturbances are amplified proportional to $Re^3$; whereas, the wall-normal and spanwise disturbance components are amplified relative to $Re^4$. Therefore, for high Reynolds numbers, worst-case amplification from wall-normal and spanwise forcings are approximately $Re$ times larger than from the worst-case amplification from streamwise forcings.


For stability to persistent disturbances, we check inequality~\eqref{eq:Pmat} from Corollary~\ref{LMIcor}  for plane Poiseuille flow,~\textit{i.e.,}
\begin{equation}
Z =\begin{pmat} [{..|}]  ~ & ~  & ~ & -\frac{q_x}{2} & 0 & 0 \cr
 ~ & M(y)-W & ~ & 0 & -\frac{q_y}{2} & 0 \cr
~ & ~ & ~ & 0 & 0 & -\frac{q_y}{2} \cr\-
 -\frac{q_x}{2} & 0 & 0  & \sigma_{x}(y) & 0 & 0 \cr
0 & -\frac{q_y}{2} & 0  & 0 & \sigma_{y}(y)  & 0   \cr
0 & 0 & -\frac{q_y}{2}  & 0 & 0 & \sigma_{z}(y) \cr \end{pmat} \succcurlyeq 0, \quad y \in (-1,1), \nonumber
\end{equation}
with $M$ given in~\eqref{sddfsdf2} and $W=\left[\begin{smallmatrix} q_x \psi_x & 0 & 0\\ 0 & q_y \psi_y & 0\\0& 0 & q_y \psi_z \end{smallmatrix}\right]$.  We fix $\psi_i = 10^{-4},~i=x,y,z$ and $L=2\pi$. 
In this case, we obtain $Re_{ISS} = 1855$. {The quantity $Re_{ISS}=1855$ can be compared with the empirical Reynolds number at the onset of turbulence $Re \approx 2000$ as discussed by \cite{RevModPhys.72.603}. Once again, we infer that $Re_{ISS}$ provides a lower bound for the Reynolds number for which transition to turbulence occurs.}

\begin{figure}

\centering{
                \includegraphics[scale=.35]{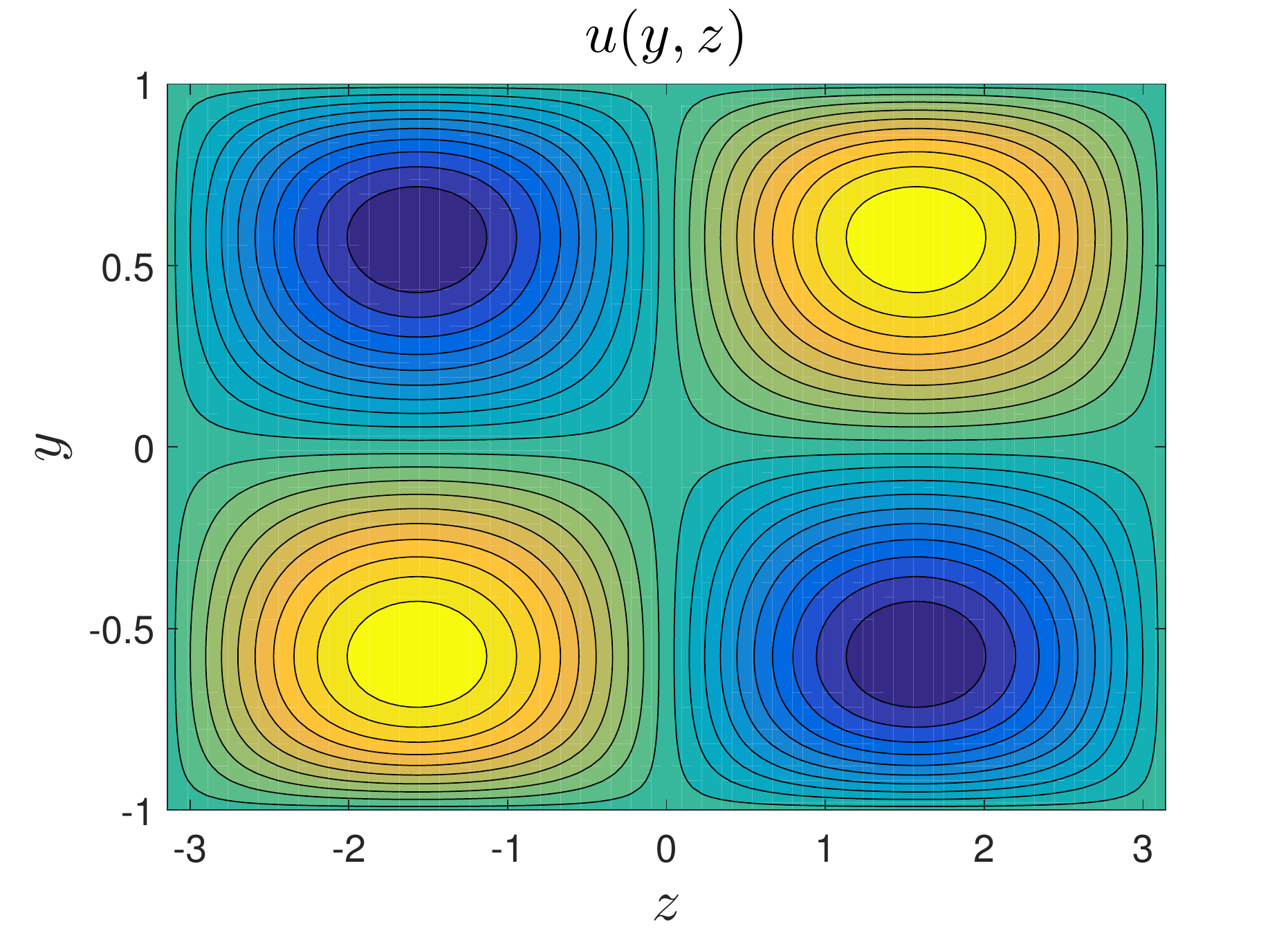}
                \includegraphics[scale=.35]{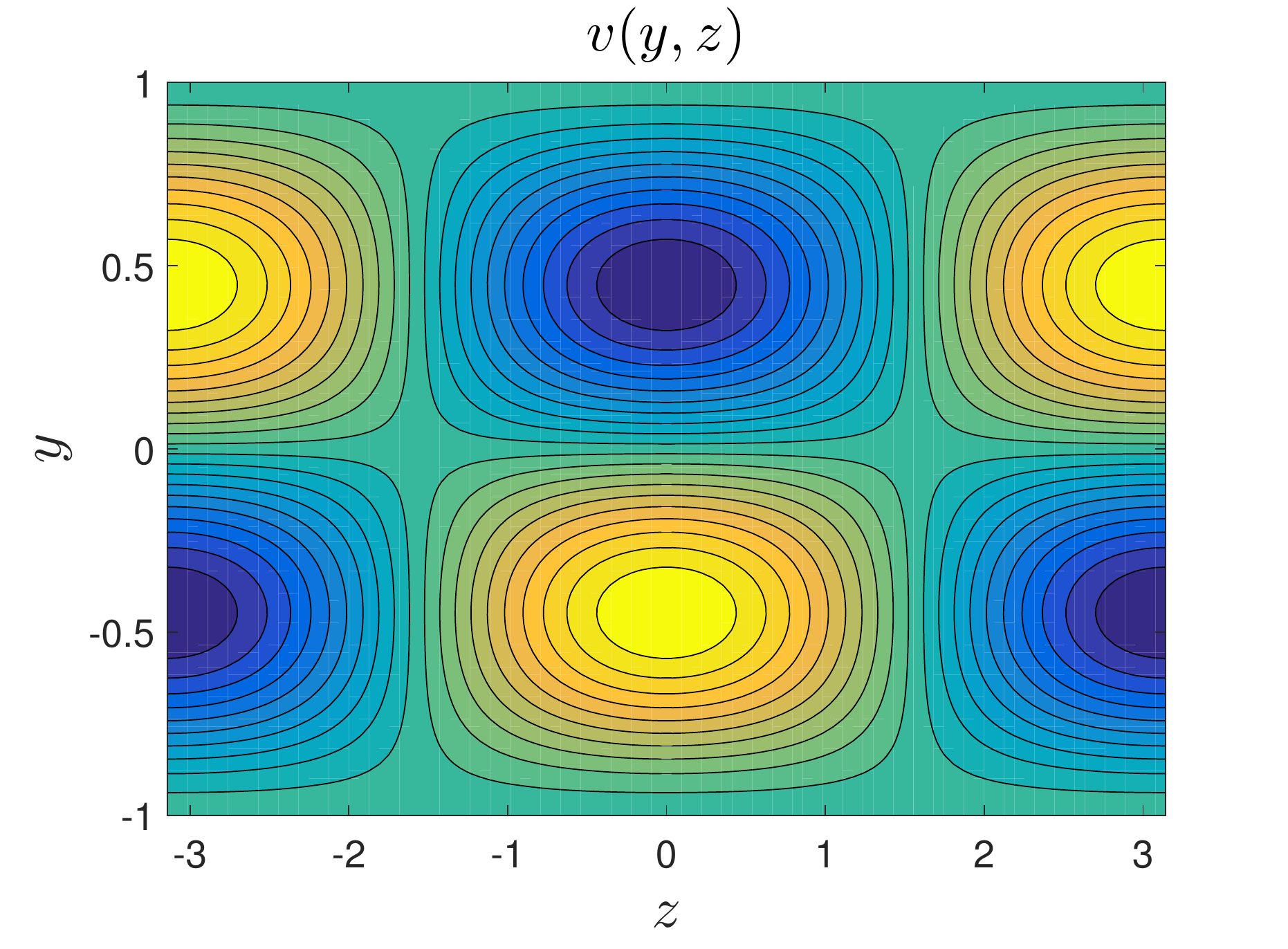} \\
                \includegraphics[scale=.35]{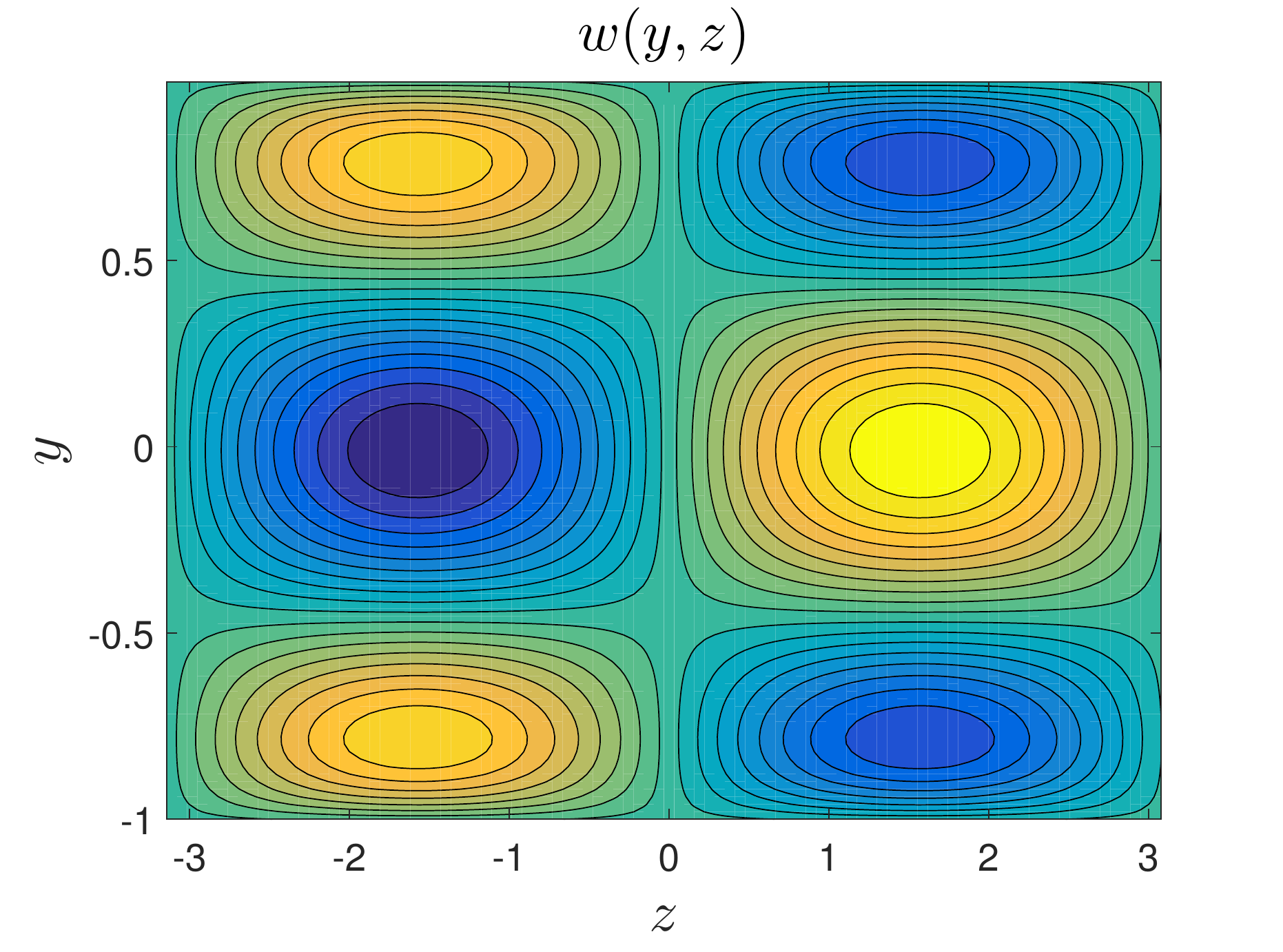}
}
        \caption{The perturbation flow structures with maximum amplification to persistent disturbances  $Re=1855$ for plane Poiseuille flow.}\label{figISSflowstructPois}
\end{figure}

Analogous to the plane Couette flow, we undertook numerical experiments to find the flow structures subject to maximum amplification from persistent forcings.  Again, we found that the maximum amplification corresponds to the streamwise constant case $k_x=0$. Figure~\ref{figISSflowstructPois} illustrates the flow structures that receive maximum  amplification from persistent forcings at $Re=1855$.

\begin{figure}

\centering{
                \includegraphics[width=8cm]{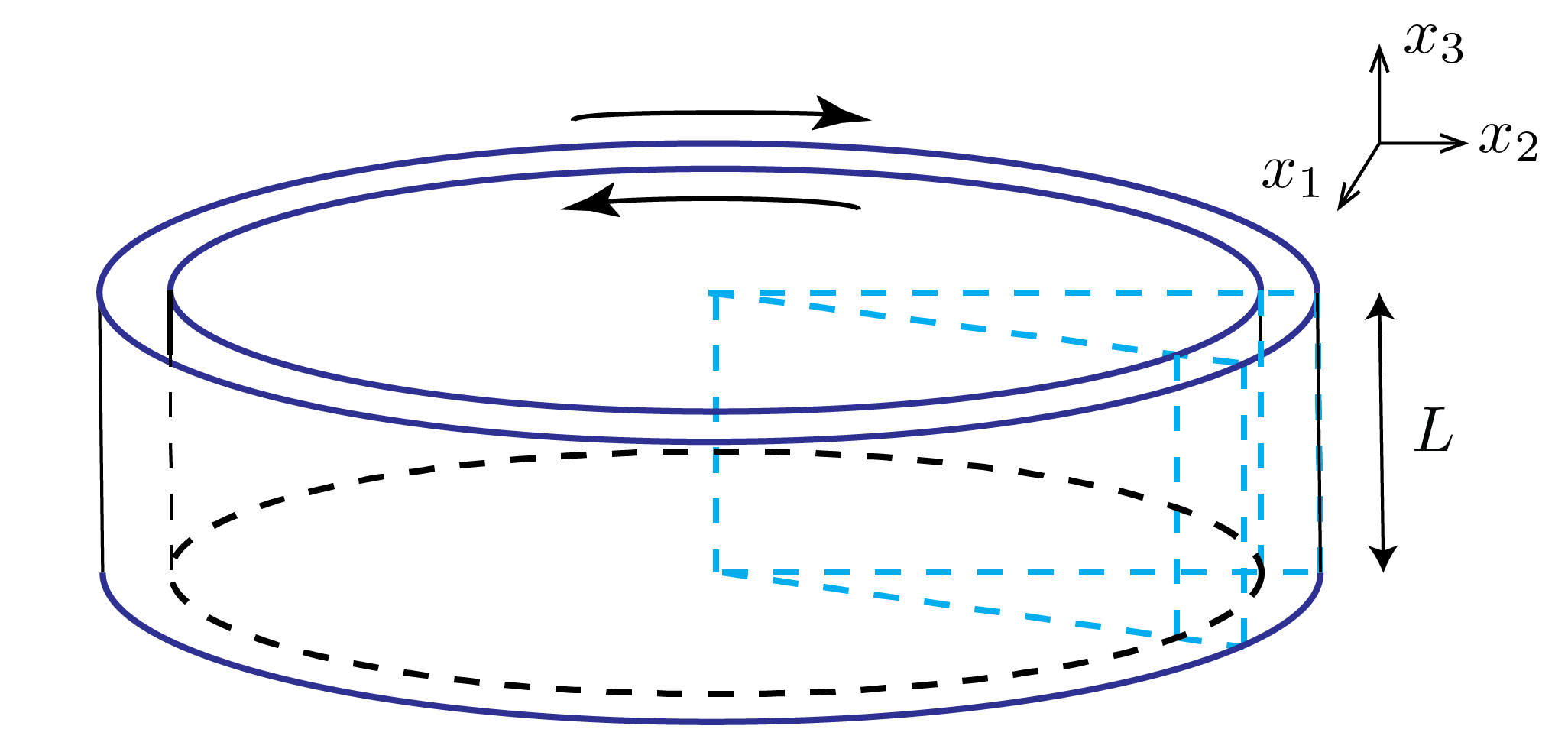}\\
                \includegraphics[width=8cm]{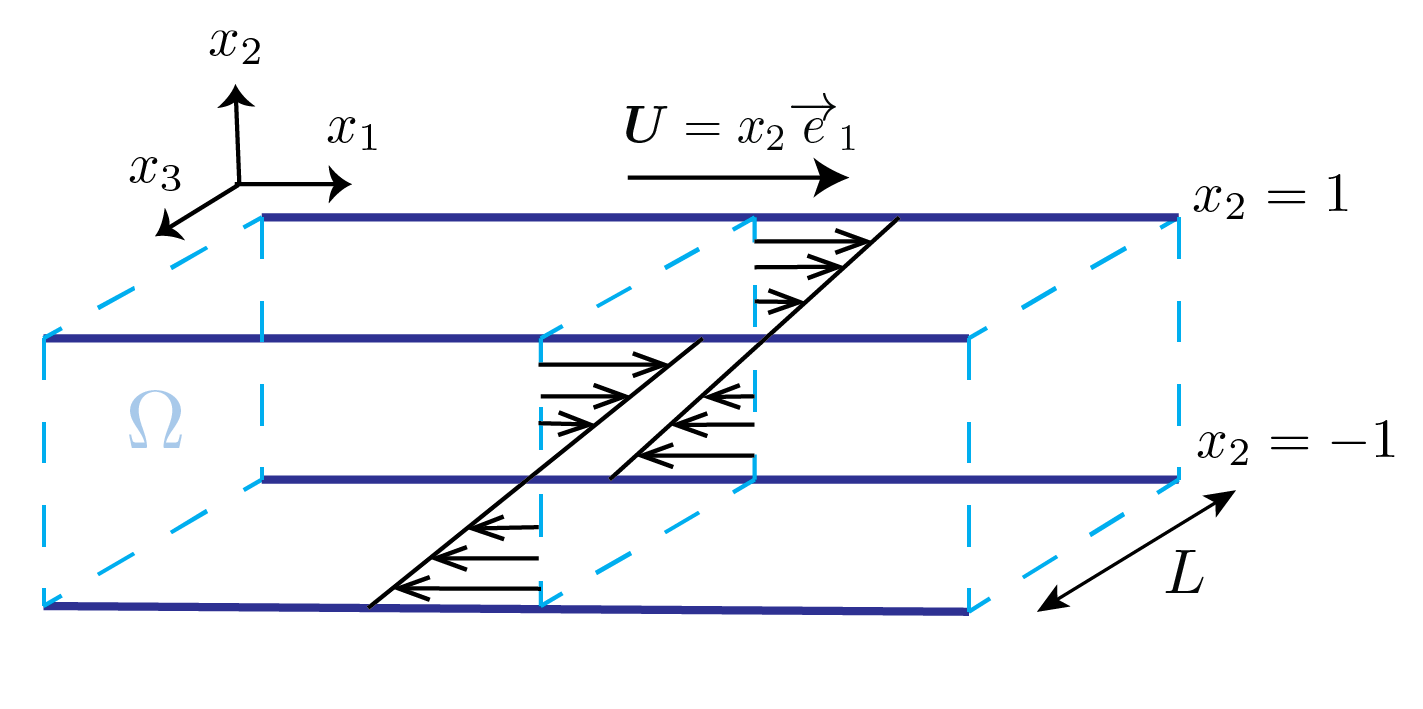}}

        \caption{ The Taylor-Couette flow, where the gap between the cylinders is much smaller than  their radii (top).  Schematic of the rotating Couette flow geometry with rotation about the $x_3$-axis (bottom).}\label{fig1xxx}
\end{figure}

\subsection{Rotating Couette Flow}

We consider the  flow  between two co-axial cylinders, where the gap between the cylinders is much smaller than  their radii. In this setting, the flow can be represented by the Couette flow subject to rotation (\cite{Lasagna2016176}) as illustrated in Figure~\ref{fig1xxx}. The axis of rotation is parallel to the $x_3$-axis and the circumferential direction corresponds to $x_1$-axis.  Then, the dynamics of the perturbation velocities is  described by~\eqref{eq:mainNS}. The perturbations are assumed to be invariant with respect to $x_1$ \mbox{($\partial_{x_1}=0$)} and periodic   in $x_3$ with period $L$. The domain is, therefore, defined as $$\Omega = \left\{ (x_2,x_3) \mid (x_2,x_3) \in (-1,1)\times (0,L) \right\}.$$ Note that $\Omega$ is indeed a semialgebraic set as given by
$$
\Omega = \left\{ (x_2,x_3) \mid ~~ (1-x_2)(1+x_2)>0 ~~ \text{and} ~~ x_3(x_3-L)>0\right\}.
$$
The base flow is given by $\boldsymbol{U}=(x_2,0,0)'=x_2\overrightarrow{e}_1$ and $P=P_0$. In~addition, 
~$
F =\left[ \begin{smallmatrix} 0 & Ro & 0 \\ -Ro & 0 & 0 \\ 0 & 0 & 0  \end{smallmatrix} \right],
$~
where $Ro \in [0,1]$ is a parameter representing the Coriolis force. That is,  $Ro=0$  corresponds to the case where the outer and inner cylinders are rotating with the same speed but in opposite directions and $Ro=1$ is the case where both cylinders are rotating with the same velocity in the same direction. Notice that the cases that correspond to plane Couette flow  was  discussed  in detail in Section~\ref{example:planecouette}. The case $Ro=1$ is globally stable for all Reynolds numbers due to  Rayleigh criterion (\cite{PhysRevE.95.021102}).  In this example, we focus on $Ro \in (0,1)$.


For comparison purposes, we consider  periodic boundary conditions \mbox{$\boldsymbol{u}(t,-1,x_3)=\boldsymbol{u}(t,1,x_3)$} and  $\boldsymbol{u}(t,x_2,x_3)=\boldsymbol{u}(t,x_2,x_3+L)$. The Poincar\'e constant is then given by $C=\frac{\pi^2}{\sqrt{L^2+2^2}}$. The linear stability limit of the flow can be computed by studying the spectrum of the linearized model~(\cite{Lasagna2016176}). That is,
$$
Re_L = \frac{2\sqrt{2}}{\sqrt{1-Ro}\sqrt{Ro}},
$$
with a minima at $Ro= 0.5$ corresponding to $Re = 4 \sqrt{2}$. Linear stability analysis suggests that the flow is stable for all Reynolds numbers for $Ro=0,1$. Moreover, the energy stability limit of the flow is found as $Re_E = 4 \sqrt{2}$~(\cite{huang2015sum}). 


   We consider the following storage functional 
$$
V(u) = \int_0^{L} \int_{-1}^1 \left[\begin{smallmatrix} u_1 \\ u_2 \\ u_3 \end{smallmatrix}\right]^\prime \left[\begin{smallmatrix} q_1 & 0 & 0 \\ 0 & q_2 & 0 \\ 0 & 0 & q_2 \end{smallmatrix} \right]\left[\begin{smallmatrix} u_1 \\ u_2 \\ u_3 \end{smallmatrix} \right] \,\, dx_2dx_3,
$$
which is the same as storage functional~\eqref{eq:Lyap} assuming invariance with respect to $x_1$. 

  Although our main focus is on input-output analysis, for this particular flow, we also study global stability for the sake of comparison with the nonlinear stability analysis method in~(\cite{huang2015sum}). Note that for the  rotating Couette flow the global stability bound and the linear stability bounds should coincide (\cite{Taylor289,huang2015sum}).  To study stability, we simply check the following inequality
$$
\frac{dV(u)}{dt} \le -\psi V(u),
$$ 
for some positive constant $\psi$. Setting $\psi = 10^{-2}$, we check the following matrix inequality
$$
M - \psi \begin{bmatrix} q_1  & 0 & 0\\ 0 & q_2  & 0\\0& 0 & q_2  \end{bmatrix} \succcurlyeq 0.
$$

Note that for this flow ($m=1,j=2,i=3$), we have
\begin{equation} \label{sddfsddsdsdsszzxxf}
M = \begin{bmatrix} \frac{q_1C}{Re} &    \frac{q_2Ro - q_1(Ro-1)}{2} & 0 \\ \frac{q_2Ro - q_1(Ro-1)}{2} & \frac{q_2C}{Re}  & 0 \\
0 & 0 & \frac{q_2C}{Re}
\end{bmatrix} .
\end{equation}

\begin{figure}

\centerline{
                \includegraphics[scale=.45]{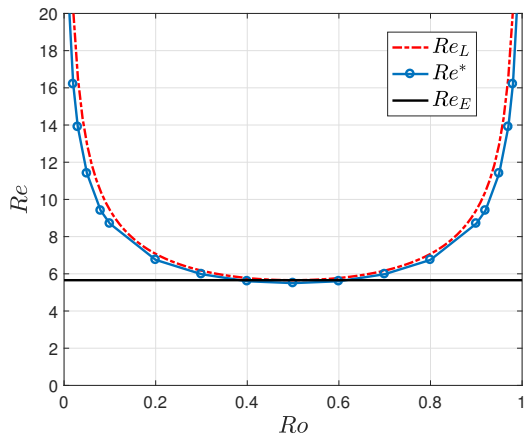}
}
        \caption{Stability bounds $Re_{E}$ (using energy method), $Re_L$ (linear stability limit), and $Re^*$ (using the proposed method) in terms of $Ro$ for rotating Couette flow.}\label{fig23e442342}
\end{figure}

The stability results are depicted in Figure~\ref{fig23e442342}. Interestingly, the stability bounds obtained using the proposed method can  effectively approximate the linear stability limit for all $Ro \in (0,1)$, which is indeed the case for this flow. This result can be compared with the stability method in (\cite{huang2015sum,GC12}) where the global stability bounds only converge to the linear stability bound for $Ro \in  [0.2529, 0.7471]$. This improved accuracy illustrates the significance of considering the full nonlinear PDE model of the flow rather than finite-dimensional truncations of the flow dynamics. 

\begin{figure}

\centerline{
                \includegraphics[scale=.45]{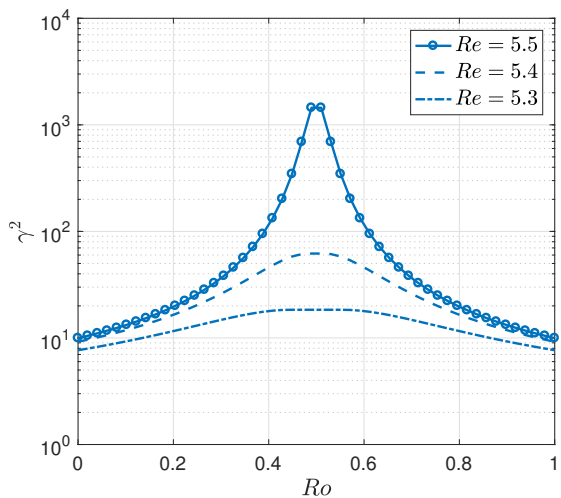}
}
        \caption{Energy growth  for rotating Couette flow with respect to the parameter~$Ro$.}\label{fig23e44234asdsdas2}
\end{figure}

\begin{figure}

\centerline{
                \includegraphics[scale=.45]{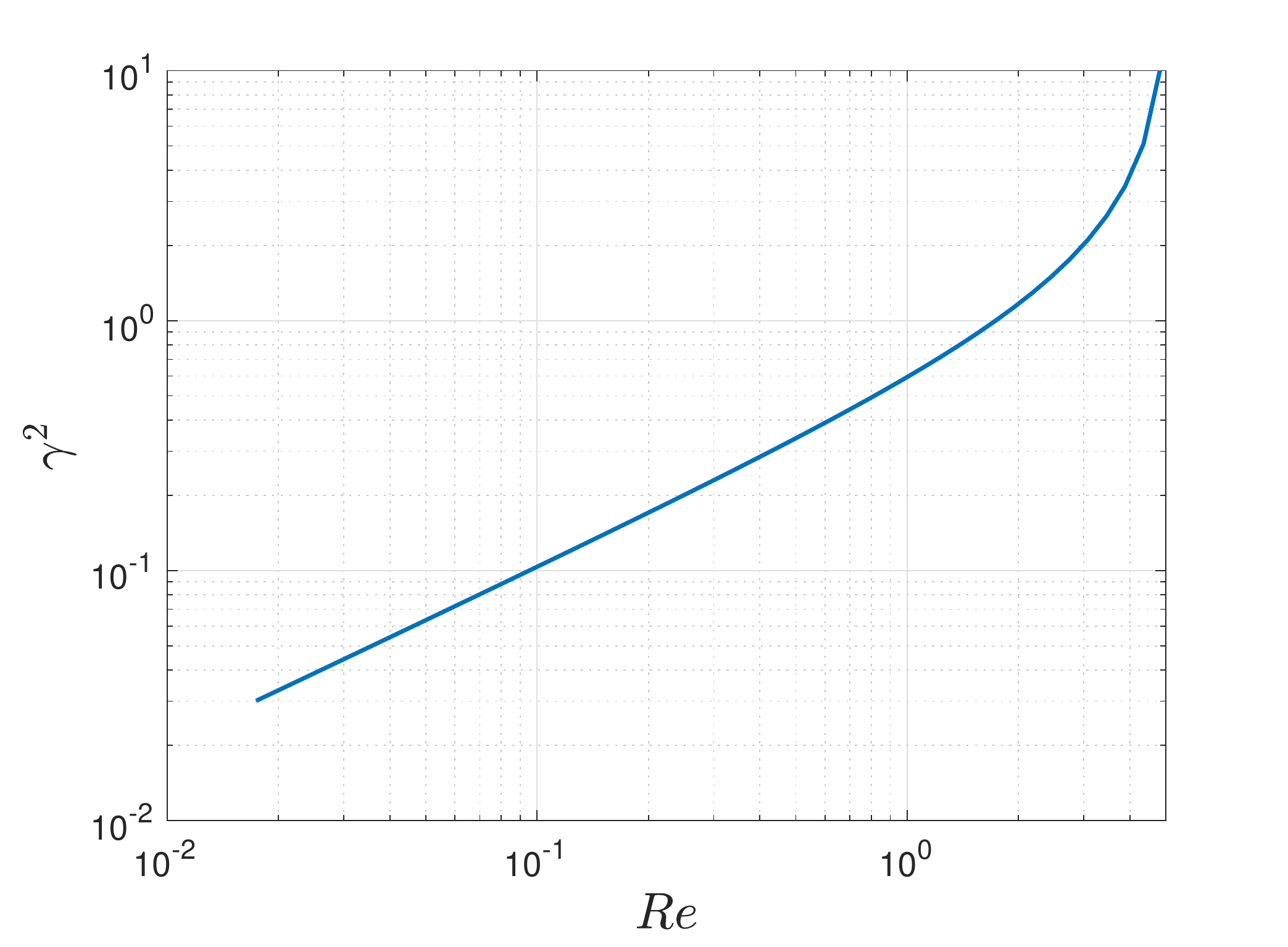}
}
        \caption{Energy growth  for rotating Couette flow with respect to the Reynolds number $Re$ for fixed~$Ro=0.5$.}\label{fig23e442dsdsdssd342}
\end{figure}

We next demonstrate how the proposed framework can be used to determine energy growth.  We solve optimization problem~\eqref{opprobengr} with matrix $M$ given in~\eqref{sddfsddsdsdsszzxxf}. Figure~\ref{fig23e44234asdsdas2} illustrates the maximum energy growth curves of the flow with respect to $Ro$. 
The figure demonstrates that as the Reynolds number approaches the global stability bound $Re_G = 4 \sqrt{2}$, the energy growth from initial perturbation velocities increases. Furthermore, this growth is more significant for $Ro=0.5$, i.e., the least stable rotation configuration. To compare the energy growth results here with the ones available in the literature, we fix $Ro=0.5$ and observe  how the energy growth evolves as the Reynolds number approaches the global stability bound. These results are  depicted in Figure~\ref{fig23e442dsdsdssd342}, which shows for  stable Reynolds numbers the energy growth scales with $O(Re^{\frac{2}{3}})$. This is consistent with analytical transient growth computations in~(\cite{maretzke_hof_avila_2014}) based on Wentzel-Kramers-Brillouin theory and the calculations and empirical results of ~(\cite{2004AA425385Y}) that furthermore showed that the maximum transient growth correspond to perturbations that are ``uniform along the direction of the rotation axis" (streamwise constant perturbations in our model). Note that both of these aforementioned studies were carried out based on the linearized (linearly stable) model of the flow. Figure~\ref{fig23e442dsdsdssd342} also shows that for Reynolds numbers closer to the global stability bound $Re_G = 4 \sqrt{2}$, the relationship between the energy growth and  the Reynolds number becomes significantly nonlinear, as the flow is becoming unstable.



\begin{figure*}

\centering{
                \includegraphics[scale=.25]{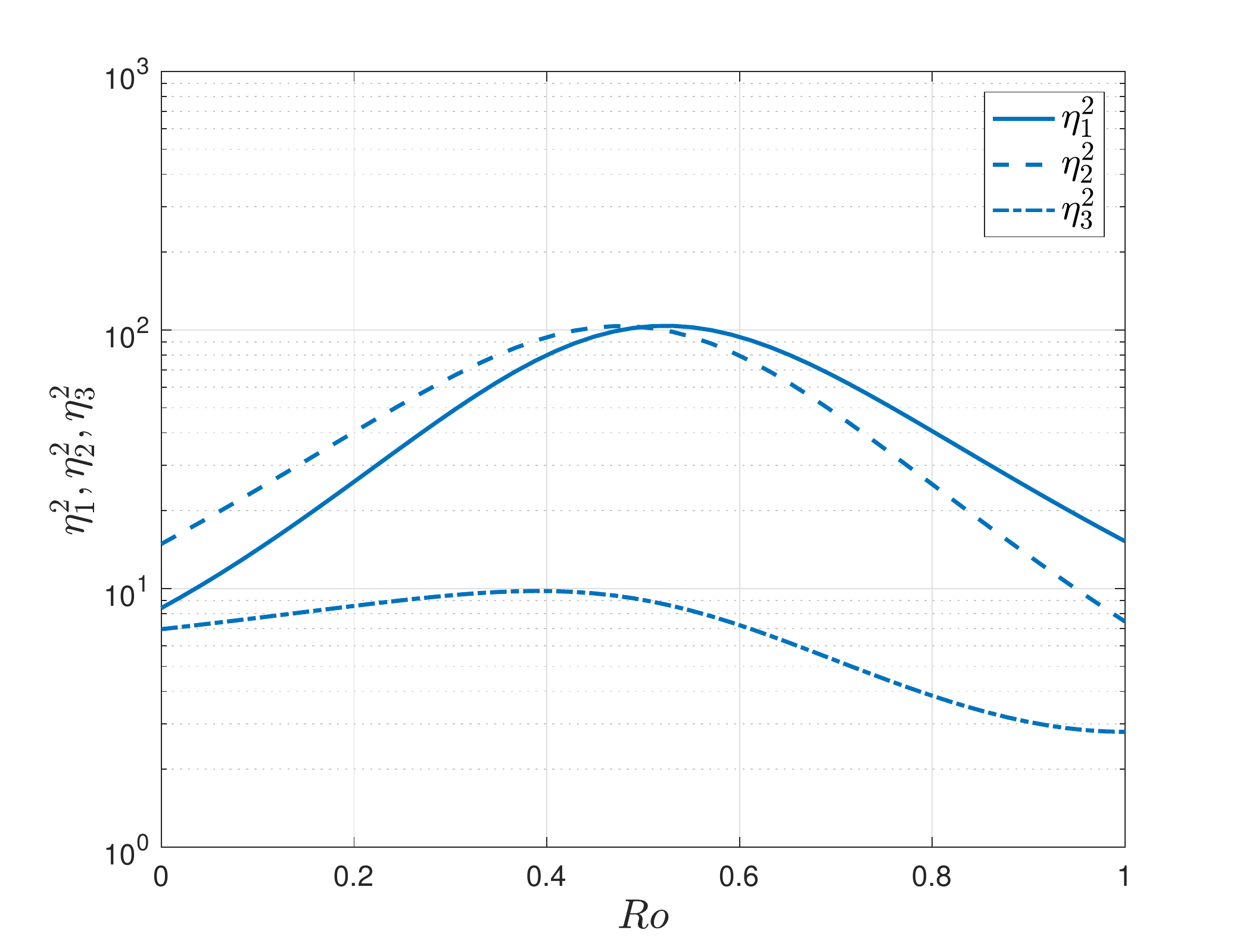}
                \includegraphics[scale=.25]{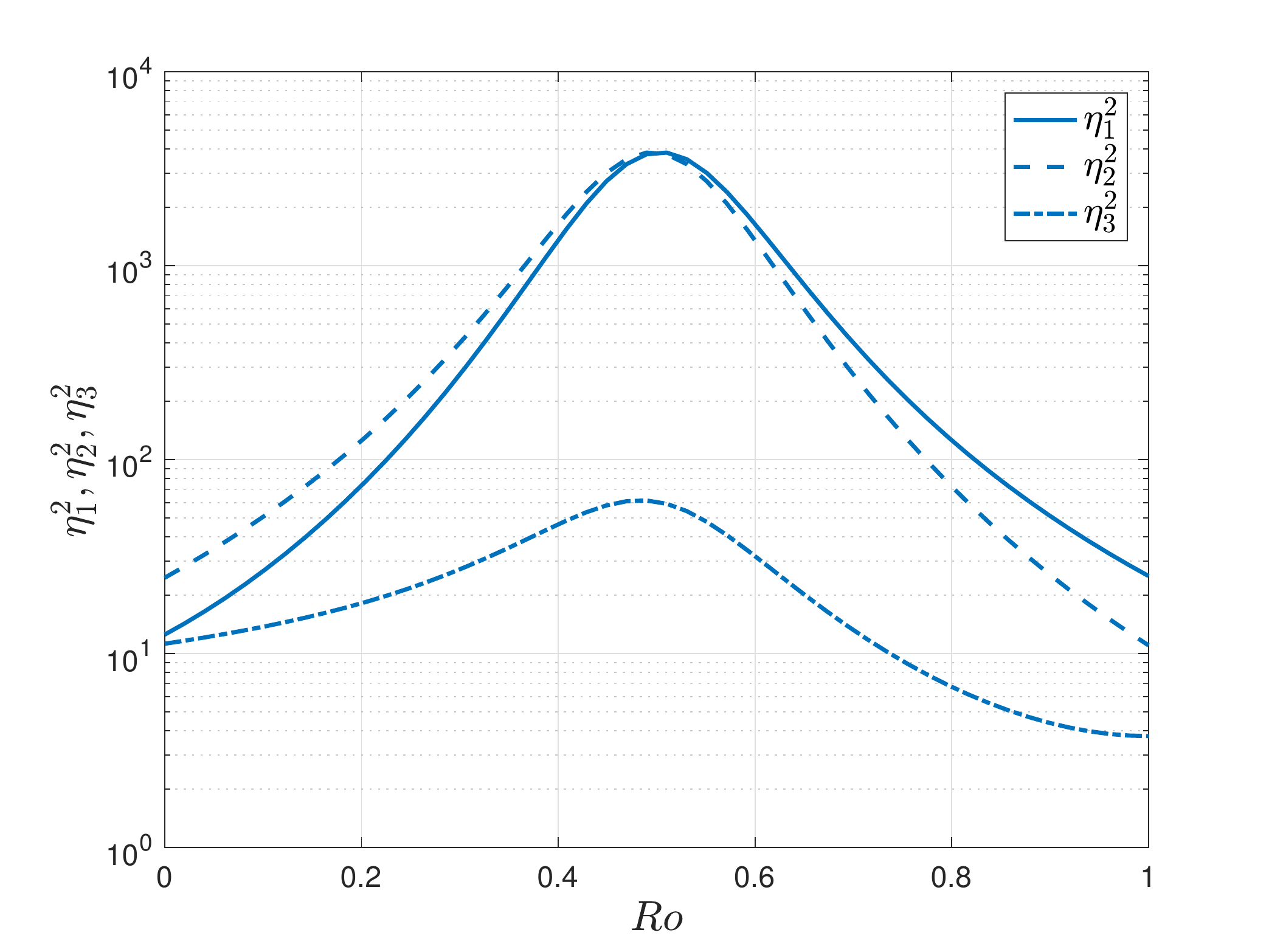}\\
                \includegraphics[scale=.3]{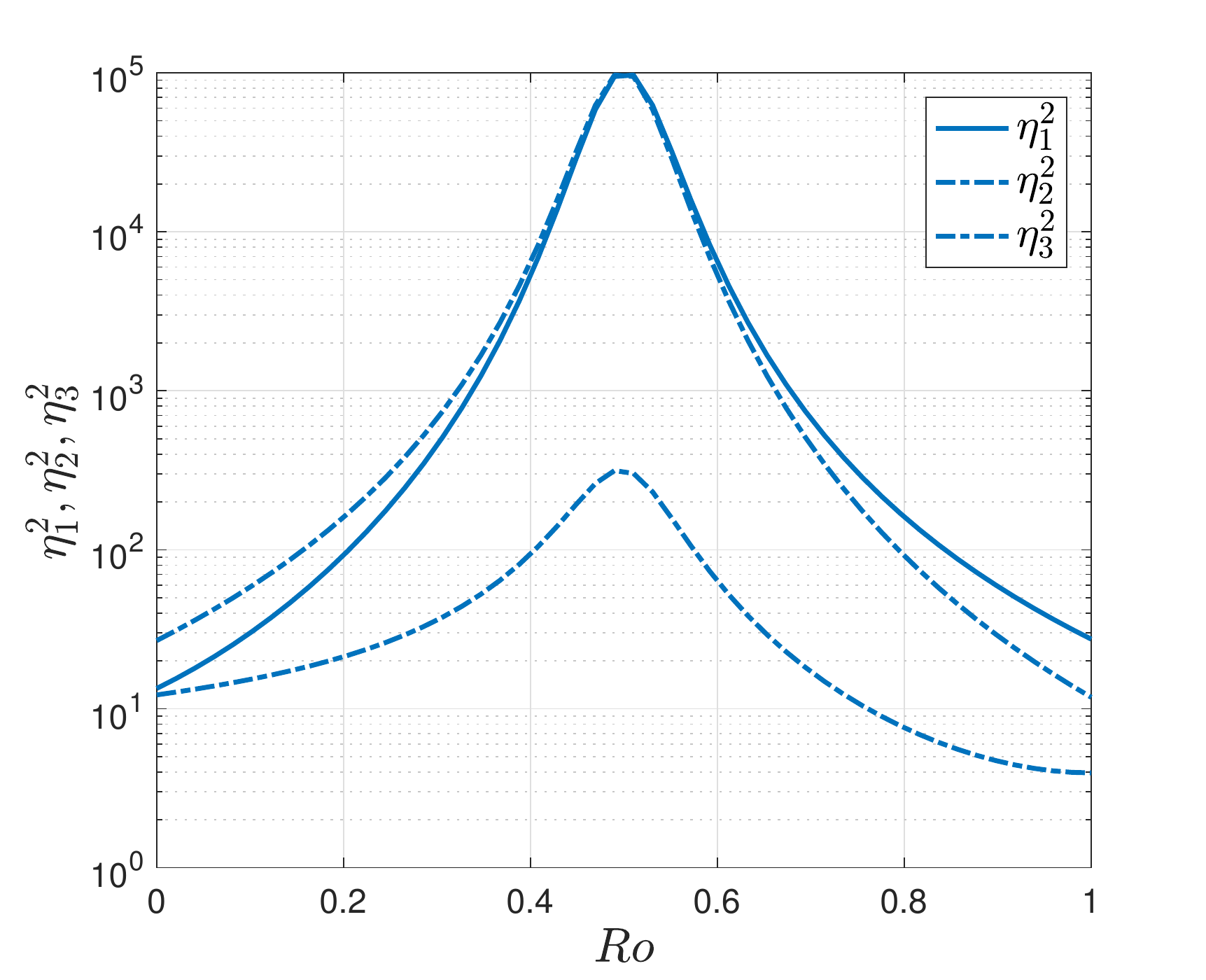}                              
}
        \caption{Upper bounds on  worst-case amplification from $\boldsymbol{d}$ to perturbation velocities $\boldsymbol{u}$ of rotating Couette flow for different Reynolds numbers: $Re=5$ (top left), $Re=5.3$ (top right), and $Re=5.6$ (bottom).}\label{fig3assaasdxx}
\end{figure*}


Finally, we use inequality~\eqref{eq:Nmat} to evaluate worst-case disturbance amplification (induced $\mathcal{L}^2$-norm),  which for this particular flow is given by the following linear matrix inequality

\begin{eqnarray}
N = \begin{pmat}[{..|}] 
~ & ~  & ~ & -\frac{q_1}{2} & 0 & 0\cr 
~ & M-\mathrm{I}_{3\times 3} & ~ & 0 & -\frac{q_2}{2} & 0 \cr 
~ & ~& ~ & 0 & 0 & -\frac{q_2}{2} \cr\-
-\frac{q_1}{2} & 0 & 0  & \eta_{1}^2 & 0 & 0 \cr
0 & -\frac{q_2}{2} & 0  & 0 & \eta_{2}^2  & 0   \cr
0 & 0 & -\frac{q_2}{2}  & 0 & 0 & \eta_{3}^2  \cr
\end{pmat} \succcurlyeq 0, \nonumber
\end{eqnarray}
with $M$ as in~\eqref{sddfsddsdsdsszzxxf}. 

Figure~\ref{fig3assaasdxx} depicts the obtained results for three different Reynolds numbers. As the Reynolds number approaches $Re_G=4\sqrt{2}$ for $Ro=0.5$, the upper-bounds on the worst-case disturbance amplification from the body forces $\boldsymbol{d}$ to perturbation velocities $\boldsymbol{u}$ increase dramatically. Furthermore, worst-case amplification from streamwise and wall-normal disturbances is significantly larger than the amplification from spanwise disturbances. For example, for  $Re=5.6$, worst-case amplification from streamwise and wall-normal disturbances is $10000$-times larger than the amplification from spanwise disturbances.

\subsection{Hagen-Poiseuille Flow}

In Appendix~\ref{app:cylindr}, we extended the proposed input-output analysis framework to pipe flows. In this example, we show the applicability of the proposed method for pipe flows through studying input-output properties of the Hagen-Poiseuille flow. 

\begin{figure}

\centering{
                \includegraphics[width=8cm]{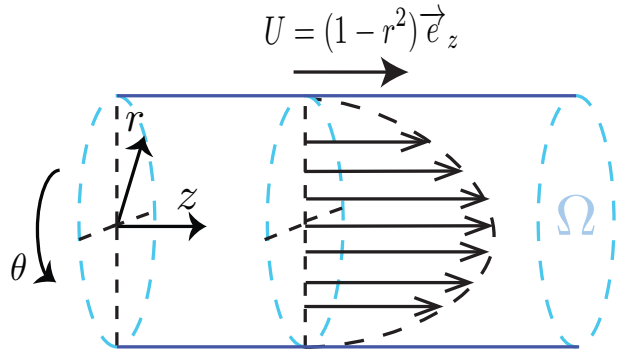}}

        \caption{Schematic of  the Hagen-Poiseuille flow geometry.}\label{figccjldf}
\end{figure}

We consider the flow of viscous fluid driven by the pressure gradient in a pipe as illustrated in Figure~\ref{figccjldf}. The domain $\Omega$ is defined as $\Omega=\{ (r,\theta) \mid 0<r<1,~0< \theta <2\pi\}$. The flow is invariant in the streamwise direction $z$. It was shown by \cite{SH94} that axial constant perturbations are subject to maximum background energy amplification in pipe flow.  The base flow is given by $\boldsymbol{U}= U_m(r)  \overrightarrow{e}_z = (1-r^2) \overrightarrow{e}_z$ and $P = 1 - \frac{4z}{Re}$. Then, the perturbation dynamics is given by~\eqref{eq:NScyl} in Appendix~\ref{app:cylindr} with $F\equiv 0$ and $U_m(r)=1-r^2$. Moreover, we assume no-slip boundary conditions $\boldsymbol{u}|_{r=1}=0$.

\begin{figure}

\centerline{
                \includegraphics[scale=.35]{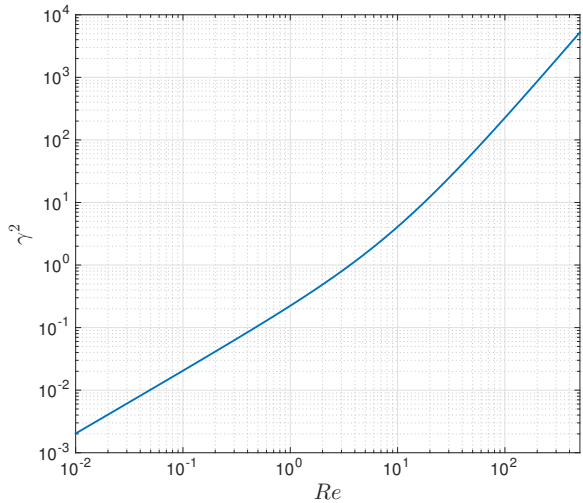}                           
}
        \caption{Upper bounds on the maximum energy growth for Hagen-Poiseuille  flow in terms of Reynolds numbers.}\label{figfjdfjkdoo}
\end{figure}

We consider the storage functional given in~\eqref{LyapCyr}. Then, substituting $U_m$ and $F$, we have
 \begin{equation} \label{rwerw}
M_c({r}) = \begin{bmatrix}  
 \frac{  q_zC}{Re}  & -rq_z   & 0  \\ -rq_z    &   \frac{  q_rC}{Re}  & 0 \\ 0 & 0   &   \frac{  q_\theta C}{Re} 
 \end{bmatrix}.
 \end{equation}
%
 In order to find upper bounds on maximum energy growth for Hagen-Poiseuille flow, we solve optimization problem~\eqref{opprobengr} with $M=M_c(r)$ as~\eqref{rwerw}. The results are illustrated in Figure~\ref{figfjdfjkdoo}. The results imply that the maximum energy growth is described by $\gamma^2 = b_0 Re + b_1 Re^2$, with $b_0,b_1>0$. This is consistent with the calculations and numerical experiments of (\cite{SH94}) on the transient growth based on the linearized Navier-Stokes equations for the pipe flow. 
 
 
 
\begin{figure}

\centerline{
                \includegraphics[scale=.4]{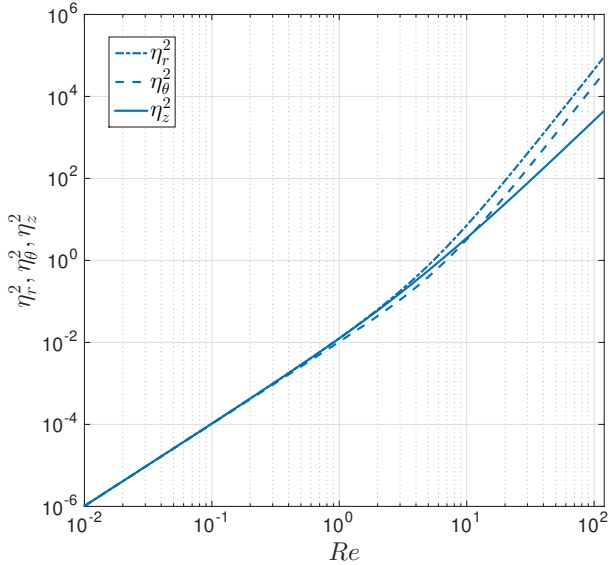}                           
}
        \caption{Upper bounds on the worst-case amplification of the Hagen-Poiseuille flow in terms of different Reynolds numbers.}\label{fig6}
\end{figure}

 Considering $M_c(r)$ as in \eqref{rwerw}, inequality~\eqref{eq:Nmat2} becomes
 \begin{equation} \label{eq:Nmat22}
N_c(r) = \begin{pmat}[{..|}] 
~ & ~  & ~ & -\frac{q_z}{2} & 0 & 0\cr 
~ & M_c(r)-\mathrm{I}_{3\times 3} & ~ & 0 & -\frac{q_r}{2} & 0 \cr 
~ & ~& ~ & 0 & 0 & -\frac{q_\theta}{2} \cr\-
-\frac{q_z}{2} & 0 & 0  & \eta_{z}^2 & 0 & 0 \cr
0 & -\frac{q_r}{2} & 0  & 0 & \eta_{r}^2  & 0   \cr
0 & 0 & -\frac{q_\theta}{2}  & 0 & 0 & \eta_{\theta}^2  \cr
\end{pmat} \succcurlyeq 0,\quad r \in (0,1).
\end{equation}
Minimizing $\eta_{z}^2$, $\eta_{r}^2$ and $\eta_{\theta}^2$ subject to the above inequality provides upper-bounds on the worst-case disturbance amplification for the Hagen-Poiseuille flow. The results are depicted in Figure~\ref{fig6}. The interesting conclusion from the figure is that the perturbations are amplified as $\eta_z^2=a_0Re^2+a_1Re^3$, $\eta_\theta^2=b_0Re^2+b_1Re^4$, and $\eta_r^2=c_0Re^2+c_1Re^4$ with $a_0,a_1,b_0,b_1,c_0,c_1>0$. Thus, similar to channel flows, for low Reynolds numbers, worst-case amplification from all three disturbance components scale to $Re^2$. For Reynolds numbers greater than $\approx 8$, the amplification from axial (which is the direction of the base flow) disturbances grow proportional to $Re^3$; whereas, the worst-case amplification from azimuthal and radial disturbances increase with respect to $Re^4$. This implies that for sufficiently large Reynolds numbers, the worst-case amplification growth from azimuthal and radial external forcings are $Re$-times larger than the amplification from axial forcings.

 Note that~(\cite{JB05}) just considered channel flows which does not include the Hagen-Poiseuille flow. 

In order to check stability to persistent forcings, the following polynomial matrix inequality
\begin{equation}
Z_c(r) =\begin{pmat} [{..|}]  ~ & ~  & ~ & -\frac{q_z}{2} & 0 & 0 \cr
 ~ & M_c(r)-W_c & ~ & 0 & -\frac{q_r}{2} & 0 \cr
~ & ~ & ~ & 0 & 0 & -\frac{q_\theta}{2} \cr\-
 -\frac{q_z}{2} & 0 & 0  & \sigma_{z}(r) & 0 & 0 \cr
0 & -\frac{q_r}{2} & 0  & 0 & \sigma_{r}(r)  & 0   \cr
0 & 0 & -\frac{q_\theta}{2}  & 0 & 0 & \sigma_{\theta}(r) \cr \end{pmat} \succcurlyeq 0,\quad r \in (0,1),
\end{equation}
where $W_c=\left[\begin{smallmatrix}\psi_zq_z & 0 & 0\\0&\psi_r q_r &0\\0&0&\psi_\theta q_\theta\end{smallmatrix} \right]$ was checked. The maximum Reynolds number for which certificates of ISS could be found was $Re_{ISS} = 1614$ using degree 10 polynomials in $\sigma_{z}(r),\sigma_{\theta}(r)$ and $ \sigma_{r}(r)$. {Remarkably, this is a lower bound to the  Reynolds  number for which transition to turbulence was observed empirically by~\cite{PM06}, \textit{i.e.}, $Re \approx 1800$.} Therefore, even in the case of the Hagen-Poiseuille flow, stability to persistent disturbances analysis can be used to predict transition.

 \begin{table}
\begin{center}
\begin{minipage}{12cm}
\begin{tabular}{@{}c|c|c|cl@{}}
{Flow} & {Energy Growth} & {Worst-Case Amplification}
& \multicolumn{1}{c@{}}{Transition}\\[3pt]
\hline
Plane Couette  & $\boldsymbol{O({Re}^3)}$, $O(Re^3)$\footnote{\cite{BBD02}}
& $\left(\begin{matrix} \boldsymbol{O({Re}^3)} \\  \boldsymbol{O({Re}^4)} \\  \boldsymbol{O({Re}^4)} \end{matrix} \right)$, $\left(\begin{matrix} O({Re}^3) \\  O({Re}^4) \\  O({Re}^4) \end{matrix} \right)$\footnote{\cite{JB05}} & $\boldsymbol{316}$, $350$\footnote{\cite{TA92}}\\
\hline
Plane Poiseuille  & $\boldsymbol{O({Re}^2)}$, $O(Re^2)$\footnote{\cite{reddy_henningson_1993}} & $\left(\begin{matrix} \boldsymbol{O({Re^3})} \\ \boldsymbol{O({Re^4})} \\  \boldsymbol{O({Re^4})} \end{matrix} \right)$, $\left(\begin{matrix} O({Re^3}) \\  O({Re^4}) \\  O({Re^4}) \end{matrix} \right)$\footnote{\cite{JB05}} & $\boldsymbol{1855}$, $2000$\footnote{\cite{RevModPhys.72.603}}\\
\hline
Hagen-Poiseuille & $\boldsymbol{O(Re^2)}$, $O(Re^2)$\footnote{\cite{SH94}} & $\left(\begin{matrix} \boldsymbol{O({Re^3})} \\ \boldsymbol{O({Re^4})} \\  \boldsymbol{O({Re^4})} \end{matrix} \right)$, -- & $\boldsymbol{1614}$, $1800$\footnote{\cite{PM06}}
\end{tabular}
\end{minipage}
\end{center}
\caption{Summary of the numerical results using the proposed framework (boldfaced), and results obtained in the literature.} \label{table1}
\end{table}

\section{Discussions}\label{sec:conclusions}
We studied stability and input-output properties of fluid flows with spatially invariant perturbations in one of the directions using dissipation inequalities.  Our framework generalizes certain types of input-output analysis techniques to the nonlinear Navier-Stokes equations, thereby matching  more closely with experimental results.  The proposed input-output analysis method introduces a unified framework for addressing  a broad range of questions related to transition (transient growth and input-output analysis) that can be adapted to a large class of flow conditions. Whenever the base flow is given by a polynomial of spatial coordinates and the flow geometry is described by a semi-algebraic set, we showed how the input-output framework can be computationally implemented based on convex optimization. For illustration purposes, we applied the proposed method to study  several examples of flows between parallel plates and a pipe flow. A toolbox is under development which can be used to apply the proposed framework to investigate more flows and input-output properties.

Table~\ref{table1} lists the numerical results based on the proposed framework for plane Couette flow, plane Poiseuille flow, and the Hagen-Poiseuille flow. For energy growth and worst-case amplification,  the table outlines the amplification scalings at high Reynolds numbers.   Energy growth results for all three flows tally with the theoretical and experimental amplification scalings in the literature. Our worst-case amplification scalings  for  plane Couette flow and plane Poiseuille flow were consistent with the scalings calculated by~\cite{JB05}. In addition to comparing the scalings we obtained using our framework for channel flows, we carried out numerical experiments to study the worst-case amplification scalings in Hagen-Poiseuille flow. This indicates that, similar to channel flows, perturbations in the direction of the base flow are least amplified in Hagen-Poiseuille flow. For transition analysis, we compare the maximum Reynolds numbers for which stability to persistent disturbances could be certified to the Reynolds numbers for which transition to turbulence was observed experimentally. We inferred from the results that $Re_{ISS}$ can be used as an acceptable theoretical estimate to predict transition to turbulence.

In addition to the aforementioned three flows, we undertook global stability analysis, energy growth analysis, and worst-case amplification analysis for  the rotating Couette flow. Global stability analysis results could replicate the actual global stability bounds calculated by \cite{Taylor289}. Our results for energy growth implied a scaling of $O(Re^{\frac{2}{3}})$, which is consistent with the  transient growth calculations in~(\cite{maretzke_hof_avila_2014}) and the calculations and empirical results of ~(\cite{2004AA425385Y}).

Future research will focus on applying the framework obtained here to turbulent flows~(\cite{annurev-fluid-010814-014637}). In particular, we study \textit{time-averaged mechanical energy dissipation}. For a channel flow of channel length $h$, the mechanical energy dissipation per unit mass is given by 
$$
\varepsilon := \frac{\nu^3}{h^4} \| \nabla \boldsymbol{u} \|^2_{\mathcal{L}^2_\Omega},
$$
where $\nu$ is the kinematic viscosity. \cite{PhysRevE.49.4087} proposed a variational method for bounding  this quantity based on the \textit{background flow} decomposition. The method has been significantly successful in finding the time-averaged mechanical energy dissipation scaling with respect to the root-mean-square velocity $U$ and $\ell$ the longest length scale, i.e., it was shown  that
$$
\varepsilon \le c_1 \nu \frac{U^2}{\ell^2}+c_2 \frac{U^3}{\ell}.
$$
and bounds on $c_1$ and $c_2$ were obtained for different flows~(\cite{doering_foias_2002,CHILDRESS2001105,ALEXAKIS2006652,rollin_dubief_doering_2011,tang_caulfield_young_2004}). In order to find bounds on the time-averaged mechanical energy dissipation, we can consider the following dissipation inequality
\begin{equation} \label{dsse3dss}
\frac{d V(\boldsymbol{u})}{dt} \le \frac{\nu^3}{h^4} \| \nabla \boldsymbol{u} \|^2_{\mathcal{L}^2_\Omega} - C,
\end{equation}
where $C>0$ is a constant. Minimizing $C$ while searching over the storage functional $V(\boldsymbol{u})$ gives upper bounds on the time-averaged mechanical energy dissipation.


Another interesting problem for future research is identifying the regions of attraction for different flow configurations. For example, in the case of  Taylor-Couette flow, after decomposing the Navier-Stokes equation about different flow regimes, one can search for estimates of the region of attraction inside which each flow regime is stable. 

In addition, input-output amplification mechanisms of turbulent flows is also an intriguing prospective research direction. In this regard,~\cite{FLM454650,PGCD09}, consider a non-polynomial model for  turbulent mean velocity profiles and turbulent eddy viscosities. Polynomial approximations (of high degrees) of such nonlinear models fit the formulation given in this paper.

Lastly, more general  storage functional structures can be considered. More specifically, given the nonlinear dynamics of the Navier-Stokes equations, one can consider the following class of storage functionals
$$
V(\boldsymbol{u}) = \int_\Omega \begin{bmatrix} \boldsymbol{u} \\ \boldsymbol{u}^2  \end{bmatrix}^\prime Q \begin{bmatrix} \boldsymbol{u} \\ \boldsymbol{u}^2  \end{bmatrix} \,\, {d}\Omega.
$$
However, a convex formulation using the above structure is not clear at the moment.

\appendix

\section{Proof of Proposition~\ref{fluidsprop1}} \label{appfdsfdfdsfcccvr}

The time derivative of  storage functional~\eqref{eq:Lyap} along the solutions of~\eqref{eq:mainNSEin} can be computed as
\begin{multline} \label{eq:Lyapdt}
\partial_t V(\boldsymbol{u}) = \sum_{i\in I}\int_\Omega q_i \bigg( \frac{1}{Re} u_i\nabla^2 u_i - u_j  u_i \partial_{x_j} u_i \\- U_j  u_i \partial_{x_j} u_i - u_j u_i  \partial_{x_j} U_i - u_i \partial_{x_i} p + u_i F_{ij} u_j +u_id_i\bigg) \,\, d\Omega.
\end{multline}
Consider  $\int_\Omega  q_i u_j  u_i \partial_{x_j} u_i\,\, d\Omega$. Using the  boundary conditions, integration by parts and the incompressibility condition $ \partial_{x_j} u_j = 0$, we obtain
\begin{equation*}
\int_\Omega  q_i u_j  u_i \partial_{x_j} u_i\,\, d\Omega = \frac{1}{2}\int_{ \Omega_i}q_i u_j u_i^2|_{\partial \Omega_j} \,\, dx_i  - \frac{1}{2}\int_\Omega  q_i  u_i^2 \left( \partial_{x_j}  u_j  \right) \,\, d\Omega = 0.
\end{equation*}
Consider the pressure terms $\int_\Omega q_i u_i \partial_{x_i} p \,\, d\Omega$. Since the perturbations are assumed invariant in $x_1$, we have 
\begin{multline} \label{fsfstttd}
\int_\Omega \left(q_2 u_2 \partial_{x_2} p + q_3 u_3 \partial_{x_3} p\right) \,\, d\Omega \\ = \int_{\Omega_3} (q_2 u_2 p)|_{\partial\Omega_2} \,\, dx_3 + \int_{\Omega_2}(q_3 u_3 p)|_{\partial\Omega_3} \,\, dx_2-\int_\Omega \left(q_2 \partial_{x_2} u_2  p + q_3 \partial_{x_3}  u_3 p\right) \,\, d\Omega \\ = -\int_\Omega \left(q_2 \partial_{x_2} u_2 + q_3 \partial_{x_3}  u_3 \right) p \,\, d\Omega,
\end{multline}
where, in the first equality above, we use integration by parts and, in the second inequality, we use     the boundary conditions. Then, if $q_2=q_3$, using the incompressibility condition $\partial_{x_2} u_2 +  \partial_{x_3}  u_3=0$,  \eqref{fsfstttd} equals zero. Therefore, the time derivative of the storage functional~\eqref{eq:Lyapdt} is modified to 
\begin{multline} \label{eq:Lyapmaindt1}
\partial_t V(\boldsymbol{u}) = \sum_{i\in I}\int_\Omega q_i \bigg( \frac{1}{Re} u_i\nabla^2 u_i  - U_j  u_i \partial_{x_j} u_i - u_j u_i  \partial_{x_j} U_i  + u_i F_{ij} u_j +u_id_i\bigg) \,\, d\Omega.
\end{multline} 
Integrating by parts the $u_i\nabla^2 u_i$ term and using the   boundary conditions, we get
\begin{multline} \label{eq:Lyapmaindt2sdsds}
\partial_t V(\boldsymbol{u}) = \sum_{i\in I}\int_\Omega q_i \bigg( \frac{-1}{Re} (\partial_{x_i}u_i)^2 - U_j  u_i \partial_{x_j} u_i - u_j u_i  \partial_{x_j} U_i  + u_i F_{ij} u_j + u_id_i \bigg) \,\, d\Omega.
\end{multline} 
 Applying the Poincar\'e inequality  to~\eqref{eq:Lyapmaindt2sdsds}, we obtain \eqref{eq:Lyapmaindt}.

\section{Derivation of the Convex Programs for Channel Flows}

The next corollary  proposes integral inequalities under which  properties such as  energy growth bounds,  worst-case amplification and stability to persistent forcings can be inferred for the flow described by~\eqref{eq:mainNSEin}.

\begin{corollary} \label{cor1}
Consider the perturbation dynamics described by~\eqref{eq:mainNSEin} subject to periodic or     no-slip boundary conditions $\boldsymbol{u}|_{\partial\Omega} = 0$. Assume the velocity perturbations are constant with respect to $x_1$. Let $I_0=\{2,3\}$. If there exist positive constants $q_i$, $i\in I$, with $q_i=q_j$, $i,j \in I_0$, positive scalars $\{\psi_i\}_{i\in I}$, $\{\eta_i\}_{i\in I}$, and $\sigma \in \mathcal{K}$ such that\\
\mbox{I)}  when $\boldsymbol{d}\equiv 0$, 
\begin{multline}\label{eq:conengr2}
\sum_{i\in I}\int_\Omega  \bigg( \left(\frac{ q_i C(\Omega)}{Re}-1\right) u_i^2   +  q_iU_j  u_i \partial_{x_j} u_i    +q_iu_j u_i  \partial_{x_j} U_i  -  q_iu_i F_{ij} u_j  \bigg) \,\, d\Omega \ge 0,
\end{multline}
\mbox{II)}
\begin{multline}\label{eq:conL22}
\sum_{i\in I}\int_\Omega  \bigg( \left(\frac{ q_i C(\Omega)}{Re}-1\right) u_i^2   +  q_iU_j  u_i \partial_{x_j} u_i    \\+q_iu_j u_i  \partial_{x_j} U_i  -  q_iu_i F_{ij} u_j -q_iu_id_i + \eta_i^2 d_i^2\bigg) \,\, d\Omega \ge 0
\end{multline}
\mbox{III)}
\begin{multline}\label{eq:conISS}
\sum_{i\in I} \int_\Omega  \bigg( \left(\frac{ q_i C(\Omega)}{Re}-\psi_i q_i\right) u_i^2   +  q_iU_j  u_i \partial_{x_j} u_i    +q_iu_j u_i  \partial_{x_j} U_i  -  q_iu_i F_{ij} u_j \\-q_iu_id_i +\sigma(|\boldsymbol{d}|)\bigg) \,\, d\Omega \ge 0
\end{multline}
Then, \\
\mbox{I)} system~\eqref{eq:mainNSEin} has bounded energy growth as described by~\eqref{eq:engr} with $\gamma^2 = \max_{i \in I} q_i$;\\
\mbox{II)} under zero initial perturbations $\boldsymbol{u}(0,\mathrm{x})\equiv 0$, the worst-case amplification from disturbances to perturbation velocities is bounded by $\eta_i$, $i\in I$ as in~\eqref{eq:L2};\\
\mbox{III)} the perturbation velocities described by~\eqref{eq:mainNSEin} are stable to persistent forcings in the sense of~\eqref{eq:iss}.
\end{corollary}

\begin{pf}
Each item is proven as follows. \\
\mbox{I)} Given storage functional structure~\eqref{eq:Lyap}, we have 
$$
V(\boldsymbol{u}(t,\mathrm{x})) \le \lambda_M(Q)  \int_\Omega \boldsymbol{u}^\prime \boldsymbol{u} \,\, d\Omega,
$$
where $\lambda_M(Q)$ denotes the maximum eigenvalue of $Q$. Since $Q$ is diagonal, we have $\lambda_M(Q) = \max_{i \in I} q_i$. Therefore, \eqref{fjffjhfjghfjfh} is satisfied with $\gamma^2 = \max_{i \in I} q_i$. Re-arranging terms in~\eqref{eq:conengr2} yields
$$
-\sum_{i\in I}q_i\int_\Omega  \bigg( \frac{ C(\Omega)}{Re} u_i^2   +  U_j  u_i \partial_{x_j} u_i  \nonumber +u_j u_i  \partial_{x_j} U_i   -  u_i F_{ij} u_j \bigg) \,\, d\Omega  \le \sum_{i\in I} \int_\Omega u_i^2 \,\, d\Omega.
$$
Applying Proposition~\ref{fluidsprop1} with $d\equiv 0$, we obtain
$$
\frac{ dV(\boldsymbol{u})}{dt} \le \sum_{i\in I} \int_\Omega u_i^2 \,\, d\Omega.
$$
Thus, inequality~\eqref{con:engr} is also satisfied. Applying Item I from Theorem~\ref{bigthmfluidfluid}, we infer that the system has bounded energy growth. \\
\mbox{II)} Re-arranging terms in \eqref{eq:conL22} yields 
\begin{multline}\label{eq:conL2}
- \sum_{i\in I}q_i\int_\Omega  \bigg( \frac{  C(\Omega)}{Re} u_i^2   +  U_j  u_i \partial_{x_j} u_i    +u_j u_i  \partial_{x_j} U_i  -  u_i F_{ij} u_j -u_id_i \bigg) \,\, d\Omega \\\le -\sum_{i\in I}\int_\Omega u_i^2 \,\, d\Omega + \sum_{i\in I}\int_\Omega \eta_i^2 d_i^2\,\, d\Omega
\end{multline}
Then, from \eqref{eq:Lyapmaindt} in Proposition~\ref{fluidsprop1}, we deduce that
$$
\frac{ dV(\boldsymbol{u})}{dt} \le -\sum_{i\in I}\int_\Omega u_i^2 \,\, d\Omega + \sum_{i\in I}\int_\Omega \eta_i^2 d_i^2\,\, d\Omega.
$$
From Item II in Theorem~\ref{bigthmfluidfluid}, we infer that, under zero initial conditions, the perturbation velocities satisfy \eqref{eq:L2}.\\
\mbox{III)} Adopting~\eqref{eq:Lyap} as a storage functional,~\eqref{fluidse11} is satisfied with $\beta_1(\cdot)=\min_{i\in I}q_i (\cdot)^2$ and $\beta_2(\cdot)=\max_{i\in I}q_i (\cdot)^2$. 
Re-arranging the terms in \eqref{eq:conISS}, we obtain
\begin{multline}
-\sum_{i\in I}\int_\Omega  \bigg( \frac{ q_i C(\Omega)}{Re}u_i^2   +  q_iU_j  u_i \partial_{x_j} u_i    +q_iu_j u_i  \partial_{x_j} U_i  -  q_iu_i F_{ij} u_j -q_iu_id_i \bigg) \,\, d\Omega \\ \le -\sum_{i\in I}\psi_i \int_\Omega q_i u_i^2 \,\, d\Omega + \int_\Omega \sigma(|\boldsymbol{d}|)\,\, d\Omega
\end{multline}
From \eqref{eq:Lyapmaindt} in Proposition~\ref{fluidsprop1}, it follows that 
\begin{equation}
\frac{ dV(\boldsymbol{u})}{dt}  \le - \psi V(\boldsymbol{u}) + \int_\Omega \sigma(|\boldsymbol{d}|)\,\, d\Omega,
\end{equation}
with $\psi= \min_{i\in I} \psi_i$. Then, from Item III in Theorem~\ref{bigthmfluidfluid}, we infer that the perturbation velocities are stable to persistent focings~\eqref{eq:iss}.
\end{pf}

\subsection{Proof of Corollary~\ref{LMIcor}}

The proof is straightforward and  follows from computing conditions~\eqref{eq:conL22}-\eqref{eq:conISS} considering perturbations that are constant in $x_1$,  the base flow $\boldsymbol{U} = U_m \overrightarrow{e}_1$, and $\sigma(|\boldsymbol{d}|) =  \sum_{i\in I} \sigma_i(\mathrm{x}) d_i^2$. Since the flow perturbations are constant in $x_1$ and  the base flow is given by $\boldsymbol{U} = U_m \overrightarrow{e}_1$, we have $U_j  u_i \partial_{x_j} u_i=0$, $i\in I$. The right hand side of~\eqref{eq:Lyapmaindt} hence changes to
  \begin{multline} 
\mathcal{A} = \int_\Omega  \bigg( \left(\frac{ C(\Omega)}{Re}-F_{ii}\right)q_i u_i^2  - u_i(q_iF_{ij})u_j -u_i(q_iF_{i1})u_1 \\
+\left(\frac{ C(\Omega)}{Re}-F_{jj}\right)q_j u_j^2  - u_j(q_jF_{ji})u_i -u_j(q_jF_{j1})u_1 \\
\left(\frac{ C(\Omega)}{Re}-F_{11}\right)q_1 u_1^2  + u_1(\partial_{x_i}U_m -F_{1i})u_i + u_1(\partial_{x_j}U_m - F_{1j})
 \bigg) \,\, d\Omega  
\end{multline}
for $i,j \in I_0$, $i\neq j$, which can be rewritten as
\begin{equation} \label{p2323}
\mathcal{A}  = \int_\Omega \left[ \begin{smallmatrix} u_1 \\ u_j \\ u_i  \end{smallmatrix} \right]^\prime M(\mathrm{x}) \left[ \begin{smallmatrix} u_1 \\ u_j \\ u_i  \end{smallmatrix} \right] \,\, d\Omega .
\end{equation}
with $M(\mathrm{x})$ given in~\eqref{eq:Mmat}. \\
\mbox{I)} Given storage functional structure~\eqref{eq:Lyap}, we have 
$$
V(\boldsymbol{u}(t,\mathrm{x})) \le \lambda_M(Q)  \int_\Omega \boldsymbol{u}^\prime \boldsymbol{u} \,\, d\Omega,
$$
where $\lambda_M(Q)$ denotes the maximum eigenvalue of $Q$. Since $Q$ is diagonal, we have $\lambda_M(Q) = \max_{i \in I} q_i$. Therefore, \eqref{fjffjhfjghfjfh} is satisfied with $\gamma^2 = \max_{i \in I} q_i$. Re-arranging terms in~\eqref{eq:conengr2} yields
$$
-\sum_{i\in I}q_i\int_\Omega  \bigg( \frac{ C(\Omega)}{Re} u_i^2   +  U_j  u_i \partial_{x_j} u_i  \nonumber +u_j u_i  \partial_{x_j} U_i   -  u_i F_{ij} u_j \bigg) \,\, d\Omega  \le \sum_{i\in I} \int_\Omega u_i^2 \,\, d\Omega.
$$
Applying Proposition~\ref{fluidsprop1} with $d\equiv 0$, we obtain
$$
\frac{ dV(\boldsymbol{u})}{dt} \le \sum_{i\in I} \int_\Omega u_i^2 \,\, d\Omega.
$$
Thus, inequality~\eqref{con:engr} is also satisfied. Applying Item II from Theorem~\ref{bigthmfluidfluid}, we infer that the system has bounded energy growth.
\mbox{II)} Inequality~\eqref{eq:conL22} is changed to 
\begin{multline}
\mathcal{A} +\int_\Omega (q_iu_id_i+q_ju_jd_j+q_1u_1d_1)\,\, d\Omega-\int_\Omega (u_i^2+u_j^2+u_1^2)\,\, d\Omega \\+ \int_\Omega (\eta_i^2d_i^2+\eta_j^2d_j^2+\eta_1^2d_1^2)\,\, d\Omega \ge 0,
\end{multline}
for $i,j \in I_0$, $i\neq j$, which can be rewritten as
\begin{equation} \label{jhjlhljll}
\int_\Omega \left[ \begin{smallmatrix} u_1 \\ u_j \\ u_i \\ d_1 \\ d_j \\ d_i  \end{smallmatrix} \right]^\prime N(\mathrm{x}) \left[ \begin{smallmatrix} u_1 \\ u_j \\ u_i  \\ d_1 \\ d_j \\ d_i\end{smallmatrix} \right] \,\, d\Omega \ge 0,
\end{equation}
where $N$ is defined in~\eqref{eq:Nmat}. Consequently, if \eqref{eq:Nmat}  is satisfied for all $\mathrm{x}\in \Omega$, \eqref{jhjlhljll}  holds and from Item II in Corollary~\ref{cor1} we infer that, subject to zero initial conditions, the worst-case amplification from disturbances to perturbation velocities is bounded by $\eta_i$, \mbox{$i \in I$} as in~\eqref{eq:L2}.\\\
\mbox{III)} The proof follows the same lines as the proof of Item II above.

 \section{Pipe Flows: Cylindrical Coordinates} \label{app:cylindr}
 
In this appendix, we extend the proposed method to flows in cylindrical coordinates $(r,\theta,z)$. In cylindrical coordinates, the gradient and Laplacian operators are, respectively, defined as $\nabla_c(\cdot) = \partial_r(\cdot) \overrightarrow{e}_r+\frac{1}{r}\partial_\theta(\cdot)\overrightarrow{e}_\theta+\partial_z(\cdot)\overrightarrow{e}_z$ and $\nabla^2_c(\cdot) = \frac{1}{r} \partial_r\left( r \partial_r (\cdot)   \right) + \frac{1}{r^2} \partial_\theta^2 (\cdot)+ \partial_z^2(\cdot)$.  The Navier-Stokes equations in cylindrical coordinates are then given by
\begin{eqnarray} \label{NSCyrMain}
\partial_t \bar{u}_r &=& \frac{1}{Re} \left( \nabla_c^2 \bar{u}_r -\frac{\bar{u}_r}{r^2} -\frac{2}{r^2}\partial_\theta \bar{u}_\theta  \right) - \boldsymbol{\bar{u}} \cdot \nabla_c \bar{u}_r +\frac{\bar{u}_\theta^2}{r} - \partial_r \bar{p} + F_r^\prime \boldsymbol{\bar{u}} +d_r \nonumber \\
\partial_t \bar{u}_\theta &=& \frac{1}{Re} \left( \nabla_c^2 \bar{u}_\theta -\frac{\bar{u}_\theta}{r^2} +\frac{2}{r^2} \partial_\theta \bar{u}_r \right) - \boldsymbol{\bar{u}} \cdot \nabla_c \bar{u}_\theta -\frac{\bar{u}_\theta \bar{u}_r}{r} -\frac{1}{r} \partial_\theta\bar{p} + F_\theta^\prime \boldsymbol{\bar{u}} +d_\theta  \nonumber \\
\partial_t \bar{u}_z &=& \frac{1}{Re} \nabla_c^2 \bar{u}_z  - \boldsymbol{\bar{u}} \cdot \nabla_c \bar{u}_z - \partial_z\bar{p} + F_z^\prime \boldsymbol{\bar{u}} +d_z  \nonumber \\
0 &=& \frac{1}{r} \partial_r \left( r \bar{u}_r  \right) + \frac{1}{r} \partial_\theta \bar{u}_\theta +\partial_z \bar{u}_z,
\end{eqnarray}
where $\boldsymbol{\bar{u}} = (\bar{u}_r, \bar{u}_\theta, \bar{u}_z)^\prime$ and $\left[\begin{smallmatrix} F_r^\prime & F_\theta^\prime & F_z^\prime \end{smallmatrix}\right]^\prime = F \in \mathbb{R}^{3\times 3}$. 

We consider the flow perturbations that are invariant in the axial direction  ($z$-direction).  The base flow is given by $\boldsymbol{U} = U_m(r,\theta) \overrightarrow{e}_z$ and $P$. For such flows, substituting $\boldsymbol{\bar{u}} = \boldsymbol{u} + \boldsymbol{U}$ and $\bar{p}=P+p$ in  \eqref{NSCyrMain}, the perturbation dynamics is obtained as
\begin{eqnarray} \label{eq:NScyl}
\partial_t u_r &=& \frac{1}{Re}\nabla^2_c u_r -u_r \partial_r u_r  -\frac{u_\theta \partial_\theta u_r}{r}  + \frac{u_\theta^2}{r}-\frac{u_r}{r^2Re} -\frac{2\partial_\theta u_\theta}{r^2Re}-\partial_r p + F_r^\prime \boldsymbol{u} +d_r, \nonumber \\
\partial_t u_\theta &=& \frac{1}{Re}\nabla^2_c u_\theta -u_r \partial_r u_\theta  -\frac{u_\theta \partial_\theta u_\theta}{r}  - \frac{u_r u_\theta}{r}-\frac{u_\theta}{r^2Re} -\frac{2 \partial_\theta u_\theta}{r^2Re}-\frac{1}{r}\partial_\theta p +F_\theta^\prime \boldsymbol{u}+ d_\theta, \nonumber \\
\partial_t u_z &=& \frac{1}{Re}\nabla^2_c u_z -u_r \partial_r u_z - u_r \partial_r U_m -  \frac{u_\theta \partial_\theta U_m}{r}-  \frac{u_\theta \partial_\theta u_z}{r} + F_z^\prime \boldsymbol{u} + d_z, \nonumber\\
0 &=& \partial_r (r u_r) + \partial_\theta u_\theta,
\end{eqnarray}
 wherein  $\boldsymbol{u}=(u_r,u_\theta,u_z)^\prime$. 
 
 \begin{prop} \label{Proposition2}
 Consider the perturbation dynamics in cylindrical coordinates~\eqref{eq:NScyl} with periodic or no-slip boundary conditions $u\mid_{\partial \Omega}=0$. The time derivative of storage functional 
 \begin{equation} \label{LyapCyr}
 V(\boldsymbol{u}) = \frac{1}{2} \int_\Omega \left[\begin{smallmatrix} u_r \\ u_\theta \\ u_z \end{smallmatrix} \right]^\prime \left[ \begin{smallmatrix} q_r & 0 & 0 \\ 0 & q_\theta & 0 \\ 0 & 0 & q_z  \end{smallmatrix}\right]  \left[\begin{smallmatrix} u_r \\ u_\theta \\ u_z \end{smallmatrix} \right] \,\, r dr d\theta,
 \end{equation}
 with $q_r = q_\theta$, satisfies 
 \begin{multline} \label{LyapDerCyr}
\frac{ dV(\boldsymbol{u})}{dt}\le - \int_\Omega \bigg( \frac{q_r C}{Re} u_r^2 + q_z \partial_r U_m  u_r u_z + \frac{q_z C}{Re} u_z^2 +  \frac{q_z}{r}\partial_\theta U_m u_\theta u_z+ \frac{q_\theta C}{Re}u_\theta^2 \\-q_ru_rF_r^\prime \boldsymbol{u} -q_\theta u_\theta F_\theta^\prime \boldsymbol{u} -q_z u_zF_z^\prime \boldsymbol{u}  - q_r u_r d_r -q_\theta u_\theta d_\theta - q_z u_z d_z \bigg) \,\, rdrd\theta,
 \end{multline}
 where $C>0$.
 \end{prop}

\begin{pf}
The time derivative of the storage functional~\eqref{LyapCyr}  is given by 
 \begin{multline} \label{ddds}
\frac{ dV(\boldsymbol{u})}{dt}= \int_\Omega \bigg( -rq_r u_r^2 \partial_r u_r -q_r u_r u_\theta \partial_\theta u_r +q_r u_r u_\theta^2 -rq_r \partial_r p u_r \\+\frac{q_r}{Re} ru_r\nabla^2_c u_r -\frac{q_ru_r^2}{Re r} -\frac{2q_r u_r \partial_\theta u_\theta}{rRe} + q_r ru_rF_r^\prime \boldsymbol{u}  +q_r r u_r d_r \bigg) \,\, d\theta dr \\
 + \int_\Omega \bigg( -rq_\theta u_ru_\theta \partial_r u_\theta -q_\theta u_r u_\theta \partial_\theta u_r -q_\theta u_r u_\theta^2 - q_\theta \partial_\theta p u_\theta \\
 +\frac{q_\theta}{Re} ru_\theta\nabla^2_c u_\theta -\frac{q_\theta u_\theta^2}{rRe} + \frac{2q_\theta \partial_\theta u_r u_\theta}{rRe} +q_\theta ru_\theta F_\theta^\prime \boldsymbol{u}   +q_\theta r u_\theta d_\theta   \bigg) \,\, dr d\theta \\
 + \int_\Omega \bigg( -rq_z u_r u_z \partial_r u_z - rq_z u_r u_z \partial_r U_m - {q_z}\partial_\theta U_m u_\theta u_z - q_z u_\theta u_z \partial_r u_z \\+ \frac{q_z}{Re} ru_z \nabla^2_c u_z +q_zru_zF_z^\prime \boldsymbol{u}   + q_z r u_z d_z    \bigg) \,\, drd\theta.
 \end{multline}
 From the incompressibility condition $\partial_r(r u_r) + \partial_\theta u_\theta=0$ and the fact that $q_r = q_\theta$, we obtain
\begin{multline}
 \int_\Omega \left( -rq_r \partial_r p u_r - q_\theta \partial_\theta p u_\theta \right) \,\, dr d\theta = \int_\Omega \left( q_r \partial_r (ru_r)p + q_\theta \partial_\theta u p   \right) \,\, dr d\theta \\ = \int_\Omega q_r p \left(  \partial_r (ru_r) +  \partial_\theta u    \right) \,\, dr d\theta = 0.
\end{multline}
where, in the first equality above, we used integration by parts and the boundary conditions. Furthermore, using integration by parts, boundary conditions and the incompressibility condition it can be shown that
$$
\int_\Omega \left( -rq_r u_r^2 \partial_r u_r -q_r u_r u_\theta \partial_\theta u_r \right) \,\, drd\theta = \int_\Omega \left( \frac{q_r u_r^2}{2} \partial_r(ru_r)  + \frac{q_r u_r^2}{2} \partial_\theta u_\theta \right) \,\, drd\theta = 0,
$$
$$
\int_\Omega \left( -rq_\theta u_r u_\theta \partial_r u_\theta -q_\theta u_\theta^2 \partial_\theta u_\theta \right) \,\, drd\theta = \int_\Omega \left( \frac{q_\theta u_\theta^2}{2} \partial_r(ru_r)  + \frac{q_\theta u_\theta^2}{2} \partial_\theta u_\theta \right) \,\, drd\theta = 0,
$$
$$
\int_\Omega \left( -rq_z u_r u_z \partial_r u_z -q_z u_\theta u_z \partial_\theta u_z \right) \,\, drd\theta = \int_\Omega \left( \frac{q_z u_z^2}{2} \partial_r(ru_r)  + \frac{q_z u_z^2}{2} \partial_\theta u_\theta \right) \,\, drd\theta = 0,
$$
and
$$
\int_\Omega \left( -\frac{2q_r u_r \partial_\theta u_\theta}{rRe} -\frac{2q_\theta \partial_\theta u_r u_\theta}{rRe}   \right) \,\, drd\theta = \int_\Omega \frac{2q_r}{rRe}  \left( -u_r \partial_\theta u_\theta +  u_r  \partial_\theta u_\theta   \right) \,\, drd\theta = 0.
$$
 Then, the time derivative expression~\eqref{ddds} simplifies to  
  \begin{multline}
\frac{ dV(\boldsymbol{u})}{dt}= \int_\Omega \bigg( \frac{q_r}{Re} ru_r\nabla^2_c u_r -\frac{q_ru_r^2}{Re r} +q_r ru_rF_r^\prime \boldsymbol{u}  +q_r r u_r d_r  \bigg) \,\, d\theta dr 
 \\+ \int_\Omega \bigg( \frac{q_\theta}{Re} ru_\theta\nabla^2_c u_\theta -\frac{q_\theta u_\theta^2}{rRe} +q_\theta ru_\theta F_\theta^\prime \boldsymbol{u}  + q_\theta r u_\theta d_\theta   \bigg) \,\, dr d\theta \\
 + \int_\Omega \bigg(  - rq_z u_r u_z \partial_r U_m -  {q_z}\partial_\theta U_m u_\theta u_z+ \frac{q_z}{Re} ru_z \nabla^2_c u_z +q_z ru_zF_z^\prime \boldsymbol{u}  +q_z r u_z d_z     \bigg) \,\, drd\theta.
 \end{multline}
 Factoring out $r$ yields
   \begin{multline}
\frac{ dV(\boldsymbol{u})}{dt}= \int_\Omega \bigg( \frac{q_r}{Re} u_r\nabla^2_c u_r -\frac{q_ru_r^2}{r^2Re}+q_r u_rF_r^\prime \boldsymbol{u}  + q_r  u_r d_r  \bigg) \,\, rd\theta dr 
 \\+ \int_\Omega \bigg( \frac{q_\theta}{Re} u_\theta\nabla^2_c u_\theta -\frac{q_\theta u_\theta^2}{r^2Re} +q_\theta u_\theta F_\theta^\prime \boldsymbol{u} +q_\theta  u_\theta d_\theta   \bigg) \,\, rdr d\theta \\
 + \int_\Omega \bigg(  - q_z u_r u_z \partial_r U_m-  \frac{q_z}{r}\partial_\theta U_m u_\theta u_z + \frac{q_z}{Re} u_z \nabla^2_c u_z+q_z u_z F_z^\prime \boldsymbol{u}   +q_z  u_z d_z    \bigg) \,\, rdrd\theta.
 \end{multline}
 Since the terms $\frac{q_ru_r^2}{r^2Re}$ and $\frac{q_\theta u_\theta^2}{r^2Re}$ are  non-negative, it follows that
    \begin{multline}
\frac{ dV(\boldsymbol{u})}{dt}\le \int_\Omega \bigg( \frac{q_r}{Re} u_r\nabla^2_c u_r  + \frac{q_\theta}{Re} u_\theta\nabla^2_c u_\theta  + \frac{q_z}{Re} u_z \nabla^2_c u_z    - q_z u_r u_z \partial_r U_m - \frac{q_z}{r}\partial_\theta U_m u_\theta u_z     \\- q_r u_rF_r^\prime \boldsymbol{u} - q_\theta u_\theta F_\theta^\prime \boldsymbol{u} -q_z u_zF_z^\prime \boldsymbol{u} - q_r u_r d_r -q_\theta u_\theta d_\theta - q_z u_z d_z \bigg) \,\, rdrd\theta \\ = -\int_\Omega \bigg( \frac{q_r}{Re} |\nabla_c u_r|^2  + \frac{q_\theta}{Re} |\nabla_c u_\theta|^2  + \frac{q_z}{Re}  |\nabla_c u_z|^2    + q_z u_r u_z \partial_r U_m + \frac{q_z}{r}\partial_\theta U_m u_\theta u_z \\ -q_r u_rF_r^\prime \boldsymbol{u} - q_\theta u_\theta F_\theta^\prime \boldsymbol{u} -q_z u_zF_z^\prime \boldsymbol{u} - q_r u_r d_r -q_\theta u_\theta d_\theta - q_z u_z d_z     \bigg) \,\, rdrd\theta,
 \end{multline}
 where  in the last equality above integration by parts and the boundary conditions were used. Applying the Poincar\'e inequality, we obtain~\eqref{LyapDerCyr}. 
\end{pf}

\subsection{Convex Formulation: Pipe Flows}

Similar to the case of channel flows, in the following, we propose a convex formulation for pipe flows. The method relies on inequality~\eqref{LyapDerCyr}. Note that for cylindrical coordinates $I=\{r,\theta,z\}$ and $I_0 = \{r,\theta\}$.
 
  \begin{corollary} \label{LMIcor2}
 Consider the perturbation dynamics given by~\eqref{eq:NScyl}, streamwise constant in the $z$-direction with base flow $\boldsymbol{U} = U_m(r,\theta) \overrightarrow{e}_z$.  Suppose that there exist positive constants $\{q_l\}_{l \in I}$ with $q_r=q_\theta$,  $\{\psi_l\}_{l \in I}$, and functions $\{\sigma_l\}_{l \in I}$ such that \\
 \begin{multline} \label{eq:Mmat2}
M_c({r,\theta}) =\\ \left[\begin{smallmatrix} 
 \left( \frac{  C}{Re} - F_{z,3} \right) q_z & \frac{1}{2}\left( q_z \partial_r U_m -q_r F_{r,3} - q_z F_{z,1}  \right) & \frac{q_z}{2} \left( \frac{\partial_\theta U_m}{r} - F_{z,2} \right) -  q_\theta F_{\theta,3}  \\ \frac{1}{2}\left( q_z \partial_r U_m -q_r F_{r,3} - q_z F_{z,1}  \right)  &   \left( \frac{  C}{Re} - F_{r,1} \right) q_r & -\frac{1}{2}\left( q_r F_{r,2} +q_\theta F_{\theta,2} \right) \\ \frac{q_z}{2} \left( \frac{\partial_\theta U_m}{r} - F_{z,2} \right) -  q_\theta F_{\theta,3}  & -\frac{1}{2}\left( q_r F_{r,2} +q_\theta F_{\theta,2} \right)   &  \left( \frac{  C}{Re} - F_{\theta,2} \right) q_\theta
 \end{smallmatrix} \right],
 \end{multline}\\
  \mbox{I)}  
 \begin{equation} \label{condengr2}
  M_c\left(r,\theta \right) - \mathrm{I}_{3 \times 3} \ge 0,\quad (r,\theta) \in \Omega,
  \end{equation} 
 \mbox{II)} 
 \begin{equation} \label{eq:Nmat2}
N_c(r,\theta) = \begin{pmat}[{..|}] 
~ & ~  & ~ & -\frac{q_z}{2} & 0 & 0\cr 
~ & M_c(r,\theta)-\mathrm{I}_{3\times 3} & ~ & 0 & -\frac{q_r}{2} & 0 \cr 
~ & ~& ~ & 0 & 0 & -\frac{q_\theta}{2} \cr\-
-\frac{q_z}{2} & 0 & 0  & \eta_{z}^2 & 0 & 0 \cr
0 & -\frac{q_r}{2} & 0  & 0 & \eta_{r}^2  & 0   \cr
0 & 0 & -\frac{q_\theta}{2}  & 0 & 0 & \eta_{\theta}^2  \cr
\end{pmat} \succcurlyeq 0,\quad (r,\theta) \in \Omega,
\end{equation}\\
\mbox{III)} $\sigma_l(r,\theta) \ge 0,~(r,\theta) \in \Omega$, $l \in I$ and 
\begin{equation}\label{eq:Pmat2}
Z_c(r,\theta) =\begin{pmat} [{..|}]  ~ & ~  & ~ & -\frac{q_z}{2} & 0 & 0 \cr
 ~ & M_c(r,\theta)-W_c & ~ & 0 & -\frac{q_r}{2} & 0 \cr
~ & ~ & ~ & 0 & 0 & -\frac{q_\theta}{2} \cr\-
 -\frac{q_z}{2} & 0 & 0  & \sigma_{z}(r,\theta) & 0 & 0 \cr
0 & -\frac{q_r}{2} & 0  & 0 & \sigma_{r}(r,\theta)  & 0   \cr
0 & 0 & -\frac{q_\theta}{2}  & 0 & 0 & \sigma_{\theta}(r,\theta) \cr \end{pmat} \succcurlyeq 0,\quad (r,\theta) \in \Omega,
\end{equation}
 where $W_c=\left[\begin{smallmatrix}\psi_zq_z & 0 & 0\\0&\psi_r q_r &0\\0&0&\psi_\theta q_\theta\end{smallmatrix} \right]$. Then, it follows that \\
 \mbox{I)} the flow has bounded energy growth $\gamma^2 = \max (q_r,q_\theta,q_z)$ as given by \eqref{eq:engr},\\
\mbox{II)} subject to zero initial conditions, the induced  $\mathcal{L}^2$ norm from inputs to perturbation velocities is bounded,\\
\mbox{III)} the perturbation velocities are ISS in the sense of~\eqref{eq:iss} with $\sigma(|\boldsymbol{d}|) =  \sum_{l \in I} \sigma_l(r,\theta) d_l^2$.
  \end{corollary}
  
Note that, depending on $\partial_\theta U_m$, $M_c$ and therefore $N_c$ and $Z_c$ can be functions of $\frac{1}{r}$. Then, inequalities \eqref{eq:Mmat2}-\eqref{eq:Pmat2} become intractable. To circumvent this problem, since $r$ is positive, we can multiply \eqref{eq:Mmat2}-\eqref{eq:Pmat2} by positive powers of $r$ making the resulting inequalities solvable by convex optimization methods.

\vspace{-.5cm}

\section{Polynomial Optimization and Sum-of-Squares Programming}  \label{app:sosp}

Let $\mathcal{R}[x]$ denote the set of polynomials in $x$ with real coefficients and $\Sigma[x] \subset \mathcal{R}[x]$ the set of such polynomials with a sum-of-squares decomposition. We employ sum-of-squares programming in our computational formulations. That is, we convert different analysis problems into a { sum-of-squares program (SOSP), \textit{i.e.,}} an optimization problem involving a linear objective function subject to a set of polynomial constraints as given below
\begin{align} 
&\underset{c \in \mathbb{R}^N} {\text{minimize}}\quad w^\prime c & \nonumber \\
&\text{subject to}&\nonumber \\
&a_{0,j}(x) + \sum_{i=1}^N p_i(x) a_{i,j}(x) = 0,~~j=1,2,\ldots, \bar{J},& \nonumber \\
& a_{0,j}(x)+\sum_{i=1}^N p_i(x) a_{i,j}(x) \in \Sigma[x],~~j=\bar{J}+1,\bar{J}+2,\ldots,J, 
\end{align}
where $w \in \mathbb{R}^N$ is a vector of weighting coefficients, $c \in \mathbb{R}^N$ is a vector formed of the (unknown) coefficients of $\{ p_i \}_{i=1}^{\bar{N}} \in \mathcal{R}[x]$ and $\{ p_i \}_{i=\bar{N}+1}^{{N}} \in \Sigma[x]$,  $a_{i,j} (x) \in \mathcal{R}[x]$ are given scalar constant coefficient polynomials,  $p_i(x) \in \Sigma[x]$ are sum-of-squares polynomial (SOSP) variables. 

 The gist of the idea behind sum-of-squares programming is that if there exists an sum-of-squares decomposition for $p(x) \in \mathcal{R}[x]$, \textit{i.e.}, if there exist polynomials $f_1(s),\ldots,f_m(x) \in \mathcal{R}[x]$ such that
$$
p(x) = \sum_{i=1}^m f_i^2(x),
$$
 then it follows that $p(x)$ is non-negative. We denote the class of $p$'s as $\Sigma[x]$. Unfortunately, the converse does not hold in general
 ; that is, there exist non-negative polynomials which do not have an sum-of-squares decomposition. An example of this class of non-negative polynomials is the Motzkin's polynomial (\cite{Mot65}) given by
\begin{equation}
p(x) = 1 - 3 x_1^2 x_2^2 + x_1^2 x_2^4 + x_1^4 x_2^2,
\end{equation}
which is non-negative for all $x \in \mathbb{R}^2$ but is not a SOS. This imposes some degree of conservatism when utilizing sum-of-squares based methods. Generally, determining whether a given polynomial is positive is an NP-hard problem (\cite{bovet1994introduction}) (except for degrees less than 4); but, sum-of-squares decompositions provide a conservative, yet computationally feasible method for checking non-negativity. The next lemma gives an intriguing formulation to the sum-of-squares decomposition problem.

\begin{lemma}[\cite{CLR95}] \label{p5}
A polynomial $p(x)$ of degree $2d$ belongs to $\Sigma[x]$ if and only if there exist a positive semi-definite matrix $Q$ (known as the Gram matrix) and a vector of monomials $Z(x)$ which contains all monomial of $x$ of degree $\le d$ such that $p(x)=Z^T(x)QZ(x)$.
\end{lemma}
In (\cite{CTVG99}) and (\cite{Par00}) it was demonstrated that the answer to the query that whether a given polynomial $p(x)$ is sum-of-squares or not can be investigated via semi-definite programming methodologies.
\begin{lemma}[\cite{Par00}]
Given a finite set $\{ p_i \}_{i =0}^{m} \in  \mathcal{R}[x]$, the existence of a set of scalars $\{ a_i \}_{i=1}^{m} \in \mathbb{R}$ such that
\begin{equation}
p_0 + \sum_{i=1}^m a_i p_i \in \Sigma[x]
\end{equation}
is a { linear matrix inequality}
feasibility problem.
\end{lemma}

In the sequel, we need to verify whether a matrix with polynomial entries is positive (semi)definite. To this end, we use the next lemma from (\cite{prajna2004nonlinear}).

\begin{lemma} [\cite{prajna2004nonlinear}] \label{p1}
Denote by $\otimes$ the Kronecker product. Suppose $F(x) \in \mathcal{R}^{n \times n}[x]$ is symmetric and of degree $2d$ for all $x \in \mathbb{R}^n$. In addition, let $Z(x) \in \mathcal{R}^{n \times 1}[x] $ be a column vector of monomials of degree no greater than $d$ and consider the following conditions
\begin{itemize}
\item [(A)] $F(x) \ge 0$ for all $x \in \mathbb{R}^n$
\item [(B)] $v^T F(x) v \in \Sigma[x,v]$, for any $v \in \mathbb{R}^n$.
\item [(C)] There exists a positive semi-definite matrix $Q$ such that $$v^T F(x) v = (v \otimes Z(x) )^T Q (v \otimes Z(x) ), $$ for any $v \in \mathbb{R}^n$.
\end{itemize}
Then $(A)\Leftarrow (B)$ and $(B) \Leftrightarrow (C)$.
\end{lemma}

Furthermore, we are often interested in checking positivity of a matrix with polynomial entries $F(x) \in \mathcal{R}^{n \times n}[x]$ inside a set $\Omega \subset \mathbb{R}^n$. It turns out that if the set is semi-algebraic
, Putinar's Positivstellensatz~\cite[Theorem 2.14]{Las09} can be used. 

\begin{corollary} \label{cor:psatz}
For $F(x) \in \mathcal{R}^{n \times n}[x]$, $\omega \in \mathcal{R}[x]$ and $\Omega = \{x \in \mathbb{R}^n \mid \omega(x) \ge 0\}$, if there exists $N(x) \in \Sigma^{n \times n} [x]$ such that
\begin{equation} \label{eq:psatz}
F(x)  - N(x)\omega(x) \in \Sigma^{n \times n} [x],
\end{equation}
then $F(x)\ge 0,~\forall x \in \Omega$.
\end{corollary}

If the coefficients of $F(x)$ depend affinely in unknown parameters and the degree of $N(x)$ is fixed, checking whether~\eqref{eq:psatz} holds can be cast as a feasibility test of a convex set of constraints, an SDP, whose dimension depends on the degree of the polynomial entries of $F(x)$ and~$N(x)$.

Algorithms for solving sum-of-squares programs are  automated in MATLAB toolboxes such as SOSTOOLS~(\cite{sostools}) and YALMIP~(\cite{YALMIP}), in which the sum-of-squares problem is parsed into an SDP formulation and the SDPs are solved by LMI solvers such as SeDuMi~(\cite{Stu98}).

\section{Details of Numerical Experiments for Flow Structures}\label{app:NEFS}

In the following, we describe the details of the numerical experiments carried out to obtain the flow structures for the plane Couette flow and the plane Poiseuille flow. We begin by describing the linearized Navier-Stokes equation and its corresponding discretization (\cite{FI93}). 

The non-dimensional linearized Navier-Stokes equations governing the evolution of disturbances in steady mean flow with streamwise velocity varying only in the cross-stream direction are
\begin{equation}\label{OSeqs}
\begin{cases}
\left( \partial_t + U \partial_x   \right) \Delta v - \partial_y^2 U \partial_x v = \frac{1}{Re}\Delta \Delta v,\\
\left( \partial_t + U \partial_x   \right) \eta +\partial_y U \partial_z v = \frac{1}{Re} \Delta \eta,
\end{cases}
\end{equation}
where $U(y)$ is the mean streamwise velocity component, $v$ is the cross-section perturbation velocity, $\eta := \partial_z u - \partial_x w$, the cross-stream component of perturbation vorticity  ($z$ denotes the spanwise direction). Velocity has been non-dimensionalized by $U_0$, the maximum velocity in the channel; length has been non-dimensionalized by $L$, the width of the channel. The Reynolds number is defined as $Re:=\frac{U_0L}{\nu}$, where $\nu$ is the kinematic viscosity. Considering no-slip boundary conditions at $y=\pm 1$, we have $v =\partial_y v = \eta = 0$ at $y =\pm 1$. Recall that for the plane Couette flow $U=y$, and for the plane Poiseuille flow  $U=1-y^2$.

Consider a single Fourier component
\begin{eqnarray}
v &=& \hat{v} e^{ik_x x+ ik_z z},  \\
\eta &=& \hat{\eta} e^{ik_x x+ ik_z z}.
\end{eqnarray}
Physical variables being identified with the real part of these complex form. The field equations can be written in the compact form
\begin{equation}
\partial_t \begin{bmatrix} \hat{v} \\ \hat{\eta} \end{bmatrix}  = \begin{bmatrix} \mathscr{L} & 0 \\ \mathscr{C} & \mathscr{S}  \end{bmatrix} \begin{bmatrix} \hat{v} \\ \hat{\eta} \end{bmatrix},
\end{equation}
in which the Orr-Sommerfield operator $\mathscr{L}$, the Square operator $\mathscr{S}$, and the coupling operator $\mathscr{C}$ are defined as
\begin{eqnarray}
\mathscr{L} &=&  \Delta^{-1} \left( - ik_x U\Delta + i k_x \partial_y^2 U + \frac{\Delta \Delta}{Re}   \right), \\
\mathscr{S} &=& -ik_xU+\frac{\Delta}{Re}, \\
\mathscr{C} &=&-ik_y\partial_y U,
\end{eqnarray}
with $K^2 = k_x^2+k_y^2$ and $\Delta = \partial_y^2 - K^2$. Moreover, we have
\begin{eqnarray}
\hat{u} &=& \frac{-i}{K^2} \left( k_y \hat{\eta} - k_x \partial_y \hat{v} \right), \\
\hat{w} &=& \frac{i}{K^2} \left(   k_x \hat{\eta} + k_y \partial_y \hat{v}     \right).
\end{eqnarray}
For numerical simulations of the Orr-Somerfield equation~\eqref{OSeqs}, we consider its discrete equivalent for an $N$-level discretization (over space)
$$
\zeta = \begin{bmatrix} \hat{v}_1 & \cdots & \hat{v}_N & \hat{\eta}_1 & \cdots & \hat{\eta}_N   \end{bmatrix}^\prime,
$$
and the initial value problem~\eqref{OSeqs} can be rewritten as
\begin{equation}\label{252rws}
\dot{\zeta} = \mathscr{A} \zeta,
\end{equation}
in which the linear dynamical operator, $\mathscr{A}$, is the discretized form of $\left[\begin{smallmatrix} \mathscr{L} & 0 \\ \mathscr{C} & \mathscr{S}  \end{smallmatrix} \right]$. This means that the infinite dimensional  dynamical system~\eqref{OSeqs}, is approximated as a finite dimensional dynamical systems.  

The discretized operator $\mathscr{A}$ was calculated using the codes available in~\cite[Appendix~A]{SH01} using Chebyshev discretization. For both flows, we considered $N=50$. Then, the state-space form~\eqref{252rws} is a linear system that has to be studied. In the following, we obtain linear matrix inequality conditions to check input-to-state stability (ISS) of a linear system.

Now, consider the following linear dynamical system
\begin{equation} \label{exxxxxxxzx1}
\dot{\zeta} =  \mathscr{A} \zeta + B d,\quad t>0,
\end{equation}
where $\zeta(0)=\zeta_0$, $\zeta \in \mathbb{R}^{2N}$, $d \in \mathbb{R}^{2N}$ and $B = \mathrm{I}_{2N \times 2N}$. 
This is the perturbed version of the discrete system \eqref{252rws}. We are interested in studying the ISS of~\eqref{252rws}. That is, given $d \in \mathcal{L}^\infty$, we have the following inequality for all $\zeta_0 \in \mathbb{R}^{2N}$ 
\begin{equation} \label{eq:ISS}
\| \zeta(t) \|_2 \le \beta\left(t,||\zeta_0||_2\right) + \sigma\left( \|d\|_{\mathcal{L}^\infty_{[0,t)}}\right),~t>0
\end{equation}
where $\beta \in \mathcal{KL}$, $\sigma \in \mathcal{K}$ and $\| \zeta(t) \|_2$ is the Euclidean $2$-norm, \textit{i.e.,} $\| \zeta(t) \|_2 = \sqrt{\zeta^\prime \zeta}$. 
\begin{thm} \label{thm1}
Consider system~\eqref{exxxxxxxzx1}. If there exists an ISS-storage function $V(\zeta)$ and a positive semidefinite function $S$, $c_1,c_2 \in \mathcal{K}$, and a positive scalar $\psi$ satisfying 
\begin{equation} \label{eq:x2}
c_1 \left(\| \zeta \|_2 \right) \le V(\zeta) \le c_2 \left(\|\zeta\|_2 \right),
\end{equation}
and 
\begin{equation} \label{eq:x}
\partial_t V(\zeta) \le -\psi V(\zeta) + S(d),
\end{equation}
then solutions of \eqref{exxxxxxxzx1} satisfy estimate~\eqref{eq:ISS} with $\beta(\cdot) = c_1^{-1} \left( 2e^{-\psi t} c_2(\cdot) \right)$ and $\sigma(\cdot) =   c_1^{-1} \left(\frac{2}{\psi} S(\cdot) \right)$.
\end{thm}
\begin{pf}
Multiplying both sides of~\eqref{eq:x} by $e^{\psi t}$, gives 
$$
e^{\psi t} \partial_t V(\zeta) \le -e^{\psi t}\psi V(\zeta) + e^{\psi t}S(d)
$$
which implies $\frac{d}{dt}\left( e^{\psi t}  V(\zeta) \right) \le  e^{\psi t}S(d)$. Integrating both sides of the latter inequality from $0$ to $t$ yields 
$$
e^{\psi t} V(\zeta(t)) - V(\zeta_0) \le \int_0^t e^{\psi \tau}S(d(\tau)) \,\, d\tau \le \left( \int_0^t e^{\psi t} \,\, d\tau \right) \left( \sup_{\tau \in [0,t)} S\left( d(\tau)  \right)  \right) .
$$
where, in the last inequality, we applied the H\"{o}lder inequality. Then, 
$$
e^{\psi t} V(\zeta(t)) - V(\zeta_0) \le \left( \frac{e^{\psi t} -1}{\psi}  \right) \left( \sup_{\tau \in [0,t)} S\left( d(\tau)  \right)  \right) \le \frac{e^{\psi t} }{\psi}   \sup_{\tau \in [0,t)} S\left( d(\tau)  \right).  
$$
Dividing both sides of the last inequality above by the non-zero term $e^{\psi t}$ and re-arranging the terms gives
$$
 V(\zeta(t))  \le e^{-\psi t}V(\zeta_0) +   \frac{1}{\psi}   \sup_{\tau \in [0,t)} S\left( d(\tau)  \right).
$$
Applying the bounds in \eqref{eq:x2}, we obtain
$$
 c_1(\|\zeta \|_2)  \le e^{-\psi t}c_2(\|\zeta_0\|_2) +   \frac{1}{\psi}   \sup_{\tau \in [0,t)} S\left( d(\tau)  \right).
$$
Since $c_1 \in \mathcal{K}$, its inverse exists and belongs to $\mathcal{K}$. Thus,
$$
\|\zeta\|_2  \le  c_1^{-1} \left( e^{-\psi t}c_2(\|\zeta_0\|_2)  +   \frac{1}{\psi}   \sup_{\tau \in [0,t)} S\left( d(\tau)  \right) \right),
$$
which can be further modified to
$$
\|\zeta\|_2  \le  c_1^{-1} \left( 2e^{-\psi t}c_2(\|\zeta_0\|_2)\right)  +  c_1^{-1} \left( \frac{2}{\psi}   \sup_{\tau \in [0,t)} S\left( d(\tau)  \right) \right).
$$
Noting that $S$ is positive semidefinite, we have 
$$
\|\zeta\|_2  \le  c_1^{-1} \left( 2e^{-\psi t}c_2(\|\zeta_0\|_2)\right)  +  c_1^{-1} \left( \frac{2}{\psi}    S\left( \|d  \|_{\mathcal{L}^\infty_{[0,t)}}\right) \right).
$$
\end{pf}

The following corollary gives sufficient conditions based on linear matrix inequalities to check the conditions of Theorem~\ref{thm1}. 

\begin{corollary}\label{qdxxxxzzsas}
Consider system~\eqref{exxxxxxxzx1}. If there exist symmetric matrices $P$ and $S$, and a positive scalar $\psi$ such that
\begin{equation}\label{oopssx}
P\succ0,~S\succ0
\end{equation}
and 
\begin{equation}\label{oopssx2}
\begin{bmatrix}
\mathscr{A}^\prime P + P \mathscr{A} + \psi P & B^\prime P \\
PB & -S
\end{bmatrix} \preccurlyeq 0,
\end{equation}
then the solutions to \eqref{exxxxxxxzx1} satisfy \eqref{eq:ISS} with for $\beta(\cdot) = \left( \frac{2\lambda_M(P)}{\lambda_m(P)} e^{-\psi t}(\cdot)\right)^{\frac{1}{2}}$ and $\sigma(\cdot) = \left( \frac{2\lambda_M(S)}{\psi \lambda_m(P)} (\cdot)   \right)^{\frac{1}{2}}$. 
\end{corollary}
\begin{pf}
This is a result of applying Theorem~\ref{thm1} by considering $V(\zeta) = \zeta^\prime P \zeta$ and $S(d)=d^\prime S d$.
\end{pf}

In order to the find the maximum ISS amplification, we solve the following optimization problem
\begin{eqnarray}
&\text{minimize}_{P,S}~(\lambda_1-\lambda_2)& \nonumber \\
&\mbox{subject to}& \nonumber \\
&S \preccurlyeq \lambda_1 I,~P\succ\lambda_2I,~\text{\eqref{oopssx}},~\text{and~\eqref{oopssx2}}.&
\end{eqnarray}
Then, the system satisfies  inequality~\eqref{eq:ISS} with $\beta(\cdot) = \left( \frac{2\lambda_M(P)}{\lambda_2} e^{-\psi t}(\cdot)\right)^{\frac{1}{2}}$ and $\sigma(\cdot) = \left( \frac{2\lambda_1}{\psi \lambda_2} (\cdot)   \right)^{\frac{1}{2}}$. The upper-bound on the maximum ISS amplification is thus $\left( \frac{2\lambda_1}{\psi \lambda_2} (\cdot)   \right)^{\frac{1}{2}}$. For the wave numbers that correspond to the maximum ISS amplification, we obtain the direction in which maximum amplification is attained. To this end, we carry out a singular-value decomposition of $P$ (since $P$ is symmetric the singular values and eigenvalues coincide) and we obtain the eigenvector in $\mathscr{A}$ that corresponds to the maximum singular value.


\section{Induced $\mathcal{L}^2_{[0,\infty),\Omega}$-norms for the Linearized 2D/3C Model} \label{app:calculation}

In (\cite{mjphd04}), the authors calculated componentwise $\mathcal{H}^\infty$-norms for the linearized 2D/3C model by finding the maximum singular values. This result is described as follows. 

\begin{thm} [Thoerem 11, p. 93 in \cite{mjphd04}]
For any streamwise constant channel flows with nominal velocity $U({y})$, the $\mathcal{H}^\infty$ norms of operators $\mathcal{H}_{rs}(\omega,k_z,Re)$ that maps $d_s$ into $u_r$, $\{r=x,y,z;~s=x,y,z\}$, are given by
\begin{equation} \label{sdsds}
\begin{bmatrix}
\| \mathcal{H}_{xx} \|_\infty (k_z) & \| \mathcal{H}_{xy} \|_\infty (k_z) &  \|\mathcal{H}_{xz} \|_\infty (k_z) \\ \| \mathcal{H}_{yx} \|_\infty (k_z) &\| \mathcal{H}_{yy} \|_\infty (k_z) &  \|\mathcal{H}_{yz} \|_\infty (k_z) \\ \| \mathcal{H}_{zx} \|_\infty (k_z) & \|\mathcal{H}_{zy} \|_\infty (k_z) &  \|\mathcal{H}_{zz} \|_\infty (k_z) 
\end{bmatrix} =
\begin{bmatrix}
h_{xx} (k_z) Re & h_{xy} (k_z) Re^2  &  h_{xz} (k_z) Re^2  \\ 0 & h_{yy} (k_z) Re  &  h_{yz} (k_z) Re \\ 0 & h_{zy} (k_z) Re  &  h_{zz} (k_z) Re
\end{bmatrix},
\end{equation}
where $k_z$ represent the wavenumber in $x_z$ (spanwise direction).
\end{thm}

We are interested in studying the induced $\mathcal{L}^2$-norms from inputs $d_x,d_y,d_z$ to $\boldsymbol{u}=(u_x,u_y,u_z)^\prime$. The following corollary provides the induced norms of interest.

\begin{corollary} \label{app:corollary}
For any streamwise constant channel flows with nominal velocity $U({y})$, we have
\begin{equation}
\frac{\|\boldsymbol{u}\|_{\mathcal{L}^2_{[0,\infty),\Omega}}^2}{\|d_x\|_{\mathcal{L}^2_{[0,\infty),\Omega}}^2} = f_1(k_z) Re^2,
\end{equation}
\begin{equation}
\frac{\|\boldsymbol{u}\|_{\mathcal{L}^2_{[0,\infty),\Omega}}^2}{\|d_y\|_{\mathcal{L}^2_{[0,\infty),\Omega}}^2} = f_2(k_z) Re^2 + g_2(k_z) Re^4,
\end{equation}
\begin{equation}
\frac{\|\boldsymbol{u}\|_{\mathcal{L}^2_{[0,\infty),\Omega}}^2}{\|d_z\|_{\mathcal{L}^2_{[0,\infty),\Omega}}^2} = f_3(k_z) Re^2 + g_3(k_z) Re^4.
\end{equation}
\end{corollary}
\begin{pf}
From~\eqref{sdsds}, we infer that
\begin{equation}
\begin{bmatrix}
\|u_{x} \|_{\mathcal{L}^2_{[0,\infty),\Omega}} \\ \|u_{y} \|_{\mathcal{L}^2_{[0,\infty),\Omega}} \\ \|u_{z} \|_{\mathcal{L}^2_{[0,\infty),\Omega}}
\end{bmatrix}=
\begin{bmatrix}
h_{xx} (k_z) Re & h_{xy} (k_z) Re^2  &  h_{xz} (k_z) Re^2  \\ 0 & h_{yy} (k_z) Re  &  h_{yz} (k_z) Re \\ 0 & h_{zy} (k_z) Re  &  h_{zz} (k_z) Re
\end{bmatrix}
\begin{bmatrix}
\|{d}_x \|_{\mathcal{L}^2_{[0,\infty),\Omega}} \\ \|{d}_y \|_{\mathcal{L}^2_{[0,\infty),\Omega}} \\ \|{d}_z \|_{\mathcal{L}^2_{[0,\infty),\Omega}}
\end{bmatrix}.
\end{equation}
Thus, we have
\begin{multline}
\begin{bmatrix}
\|u_{x} \|_{\mathcal{L}^2_{[0,\infty),\Omega}} \\ \|u_{y} \|_{\mathcal{L}^2_{[0,\infty),\Omega}} \\ \|u_{z} \|_{\mathcal{L}^2_{[0,\infty),\Omega}}
\end{bmatrix}= \\
\begin{bmatrix}
h_{xx} (k_z) Re \|{d}_x \|_{\mathcal{L}^2_{[0,\infty),\Omega}}  + h_{xy} (k_z) Re^2 \|{d}_y \|_{\mathcal{L}^2_{[0,\infty),\Omega}}  +  h_{xz} (k_z) Re^2 \|{d}_z \|_{\mathcal{L}^2_{[0,\infty),\Omega}}  \\  h_{yy} (k_z) Re \|{d}_y \|_{\mathcal{L}^2_{[0,\infty),\Omega}}  +  h_{yz} (k_z) Re \|{d}_z \|_{\mathcal{L}^2_{[0,\infty),\Omega}} \\  h_{zy} (k_z) Re \|{d}_y \|_{\mathcal{L}^2_{[0,\infty),\Omega}}  +  h_{zz} (k_z) Re \|{d}_z \|_{\mathcal{L}^2_{[0,\infty),\Omega}}
\end{bmatrix}.
\end{multline}
Then, multiplying both sides of the above equality by the transpose of vector $\left[\begin{smallmatrix}
\|u_{x} \|_{\mathcal{L}^2_{[0,\infty),\Omega}} \\ \|u_{y} \|_{\mathcal{L}^2_{[0,\infty),\Omega}} \\ \|u_{z} \|_{\mathcal{L}^2_{[0,\infty),\Omega}}
\end{smallmatrix} \right]$ gives
\begin{multline}
\begin{bmatrix}
\|u_{x} \|_{\mathcal{L}^2_{[0,\infty),\Omega}} \\ \|u_{y} \|_{\mathcal{L}^2_{[0,\infty),\Omega}} \\ \|u_{z} \|_{\mathcal{L}^2_{[0,\infty),\Omega}}
\end{bmatrix}^\prime \begin{bmatrix}
\|u_{x} \|_{\mathcal{L}^2_{[0,\infty),\Omega}} \\ \|u_{y} \|_{\mathcal{L}^2_{[0,\infty),\Omega}} \\ \|u_{z} \|_{\mathcal{L}^2_{[0,\infty),\Omega}}
\end{bmatrix} \\= \begin{bmatrix}
h_{xx} (k_z) Re \|{d}_x \|_{\mathcal{L}^2_{[0,\infty),\Omega}}  + h_{xy} (k_z) Re^2 \|{d}_y \|_{\mathcal{L}^2_{[0,\infty),\Omega}}  +  h_{xz} (k_z) Re^2 \|{d}_z \|_{\mathcal{L}^2_{[0,\infty),\Omega}}  \\  h_{yy} (k_z) Re \|{d}_y \|_{\mathcal{L}^2_{[0,\infty),\Omega}}  +  h_{yz} (k_z) Re \|{d}_z \|_{\mathcal{L}^2_{[0,\infty),\Omega}} \\  h_{zy} (k_z) Re \|{d}_y \|_{\mathcal{L}^2_{[0,\infty),\Omega}}  +  h_{zz} (k_z) Re \|{d}_z \|_{\mathcal{L}^2_{[0,\infty),\Omega}}
\end{bmatrix}^\prime \\ \begin{bmatrix}
h_{xx} (k_z) Re \|{d}_x \|_{\mathcal{L}^2_{[0,\infty),\Omega}}  + h_{xy} (k_z) Re^2 \|{d}_y \|_{\mathcal{L}^2_{[0,\infty),\Omega}}  +  h_{xz} (k_z) Re^2 \|{d}_z \|_{\mathcal{L}^2_{[0,\infty),\Omega}}  \\  h_{yy} (k_z) Re \|{d}_y \|_{\mathcal{L}^2_{[0,\infty),\Omega}}  +  h_{yz} (k_z) Re \|{d}_z \|_{\mathcal{L}^2_{[0,\infty),\Omega}} \\  h_{zy} (k_z) Re \|{d}_y \|_{\mathcal{L}^2_{[0,\infty),\Omega}}  +  h_{zz} (k_z) Re \|{d}_z \|_{\mathcal{L}^2_{[0,\infty),\Omega}}
\end{bmatrix}.
\end{multline}
That is, 
\begin{multline}
\overbrace{\|u_{x} \|_{\mathcal{L}^2_{[0,\infty),\Omega}}^2 + \|u_{y} \|_{\mathcal{L}^2_{[0,\infty),\Omega}}^2+\|u_{z} \|_{\mathcal{L}^2_{[0,\infty),\Omega}}^2}^{ \|\boldsymbol{u} \|_{\mathcal{L}^2_{[0,\infty),\Omega}}^2}  \\
= \left( h_{xx} (k_z) Re \|{d}_x \|_{\mathcal{L}^2_{[0,\infty),\Omega}}  + h_{xy} (k_z) Re^2 \|{d}_y \|_{\mathcal{L}^2_{[0,\infty),\Omega}}  +  h_{xz} (k_z) Re^2 \|{d}_z \|_{\mathcal{L}^2_{[0,\infty),\Omega}}  \right)^2 \\ 
+ \left( h_{yy} (k_z) Re \|{d}_y \|_{\mathcal{L}^2_{[0,\infty),\Omega}}  +  h_{yz} (k_z) Re \|{d}_z \|_{\mathcal{L}^2_{[0,\infty),\Omega}}  \right)^2 \\
+ \left( h_{zy} (k_z) Re \|{d}_y \|_{\mathcal{L}^2_{[0,\infty),\Omega}}  +  h_{zz} (k_z) Re \|{d}_z \|_{\mathcal{L}^2_{[0,\infty),\Omega}}  \right)^2.
\end{multline}
In order to see the influence of each $d_x$ on $\|\boldsymbol{u} \|_{\mathcal{L}^2_{[0,\infty),\Omega}}^2$, we set $d_y=d_z=0$ obtaining
$$
\|\boldsymbol{u} \|_{\mathcal{L}^2_{[0,\infty),\Omega}}^2 = h_{xx}^2(k_z)Re^2 \|d_x \|_{\mathcal{L}^2_{[0,\infty),\Omega}}^2. 
$$
It suffices to set $f_1(k_z)= h_{xx}^2(k_z)$. Similarly, we have
$$
\|\boldsymbol{u} \|_{\mathcal{L}^2_{[0,\infty),\Omega}}^2 = h_{xy}^2(k_z)Re^4 \|d_y \|_{\mathcal{L}^2_{[0,\infty),\Omega}}^2 + \left(h_{yy}^2(k_z) +h_{zy}^2(k_z)\right) Re^2 \|d_y \|_{\mathcal{L}^2_{[0,\infty),\Omega}}^2,
$$
$$
\|\boldsymbol{u} \|_{\mathcal{L}^2_{[0,\infty),\Omega}}^2 = h_{xz}^2(k_z)Re^4 \|d_z \|_{\mathcal{L}^2_{[0,\infty),\Omega}}^2 + \left(h_{yz}^2(k_z) +h_{zz}^2(k_z)\right) Re^2 \|d_z \|_{\mathcal{L}^2_{[0,\infty),\Omega}}^2,
$$
wherein $f_2(k_z) = h_{yy}^2(k_z) +h_{zy}^2(k_z)$, $g_2(k_z) = h_{xy}^2(k_z)$, $f_3(k_z) = h_{yz}^2(k_z)+h^2_{zz}(k_z)$ and $g_3(k_z) = h^2_{xz}(k_z)$.
\end{pf}

\bibliographystyle{jfm}
\bibliography{jfm-instructions}

\end{document}